\documentclass[a4paper,fleqn,usenatbib]{mnras}

\usepackage[T1]{fontenc}
\usepackage{ae,aecompl}

\usepackage{graphicx}	
\usepackage{amsmath}	
\usepackage{amssymb}	
\usepackage{array}
\usepackage{pdflscape}
\hypersetup{breaklinks=true}

\newcolumntype{C}[1]{>{\centering}m{#1}}

\title[CMEs on cool stars II]{Stellar coronal mass ejections -- II. Constraints from spectroscopic observations}

\author[P. Odert et al.]{
P. Odert,$^{1}$\thanks{E-mail: petra.odert@uni-graz.at}
M. Leitzinger,$^{1}$
E.~W. Guenther$^{2}$
and P. Heinzel$^{3}$
\\
$^{1}$Institute of Physics/IGAM, University of Graz,
Universit\"atsplatz 5, A-8010 Graz, Austria\\
$^{2}$Th\"uringer Landessternwarte Tautenburg, 07778 Tautenburg, Germany\\
$^{3}$Astronomical Institute, Czech Academy of Sciences, 25165 Ond\v{r}ejov, Czech Republic
}

\date{Accepted XXX. Received YYY; in original form ZZZ}

\pubyear{2020}

\begin{document}
\label{firstpage}
\pagerange{\pageref{firstpage}--\pageref{lastpage}}
\maketitle

\begin{abstract}
Detections of stellar coronal mass ejections (CMEs) are still rare. Observations of strong Balmer line asymmetries during flare events have been interpreted as being caused by CMEs. Here, we aim to estimate the maximum possible Balmer line fluxes expected from CMEs to infer their detectability in spectroscopic observations. Moreover, we use these results together with a model of intrinsic CME rates to infer the potentially observable CME rates for stars of different spectral types under various observing conditions, as well as the minimum required observing time to detect stellar CMEs in Balmer lines. We find that generally CME detection is favoured for mid- to late-type M dwarfs, as they require the lowest signal-to-noise ratio for CME detection, and the fraction of observable-to-intrinsic CMEs is largest. They may require, however, longer observing times than stars of earlier spectral types at the same activity level, as their predicted intrinsic CME rates are lower. CME detections are generally favoured for stars close to the saturation regime, because they are expected to have the highest intrinsic rates; the predicted minimum observing time to detect CMEs on just moderately active stars is already $>$100\,h. By comparison with spectroscopic data sets including detections as well as non-detections of CMEs, we find that our modelled maximum observable CME rates are generally consistent with these observations on adopting parameters within the ranges determined by observations of solar and stellar prominences.
\end{abstract}

\begin{keywords}
stars: activity -- stars: mass-loss -- stars: late-type
\end{keywords}



\defcitealias{Odert17}{Paper I}

\section{Introduction}
Coronal mass ejections (CMEs) may yield an important contribution to mass- and angular momentum loss in young stars. If occurring frequently, they may lead to enhanced atmospheric erosion in young planetary systems \citep{Lammer07a, Khodachenko07a, Cohen12, Garcia-Sage17, Airapetian17, Garraffo17}. Based on the high flare rates in young, magnetically active stars \citep{Maehara12, Balona15, Davenport16} and the close connection between flares and CMEs on the Sun \citep{Priest02, Yashiro06, Compagnino17}, it is hypothesized  that active stars may have very frequent CME events \citep{Aarnio12, Drake13, Osten15a, Cranmer17, Odert17}.

Unfortunately, the occurrence rates and physical parameters of CMEs are not well constrained for stars other than the Sun. On the Sun, various observational techniques like coronagraph imaging, spectroscopy with high spatial and temporal resolution, as well as in-situ measurements of plasma parameters can be applied \citep{Webb12}, which are not feasible for stellar observations with currently available technology. Several studies searched for radio type~II bursts around active stars \citep{Leitzinger09, Leitzinger10, Boiko12, Villadsen16, Crosley16, Crosley18, Crosley18a, Villadsen19}, which are often associated with CMEs on the Sun \citep[e.g.][]{Reiner01a, Gopalswamy05}, but up to now no such bursts have been detected. \citet{Mullan19} suggested that the strong magnetic fields of active stars may be responsible that their CMEs cannot produce type-II bursts because of too large Alfv\'en speeds.

Absorptions detected in X-ray observations during flare events were interpreted as potential stellar mass ejections \citep[e.g.][]{Haisch83, Tsuboi98, Wheatley98, Favata99, Franciosini01, Pandey12}. They could be caused by excess neutral material rising above the flaring region and obscuring the emission, like from erupting prominences. \citet{Moschou17} presented a model of a super-CME aiming to explain prolonged X-ray absorption on Algol detected in 1997 during a superflare \citep{Favata99}. They were also able to give constraints on the CME speed by detailed modeling of the time evolution of this event. In \citet{Moschou19} they applied their model to previous events from the literature. On the Sun, however, X-ray absorption by prominences is rather uncommon \citep{Schwartz15a} and more typically observed at EUV wavelengths. A similar event was also detected in the UV on the active M dwarf EV~Lac \citep{Ambruster86}. In some pre-cataclysmic binary systems, transient eclipses of the white dwarf component have been interpreted as obscuration by mass ejected from the late-type star \citep{Bond01, Parsons13}. In several stellar flares, pre-flare dips were observed in optical photometry which could be caused by prominence eruptions, but could also be related to opacity effects \citep{Giampapa82a, Andersen83, Doyle88}.

The most direct means of detecting mass ejections on stars is the Doppler signature of moving material in spectroscopic observations. However, due to the lack of spatial resolution, only the most massive events may be detected in integrated light. Most often, such events were found in Balmer lines on K/M dwarfs \citep{Houdebine90, Gunn94a, Fuhrmeister04, Vida16, FloresSoriano17, Honda18} and on M-type weak-line T Tauri stars \citep{Guenther97}. Chromospheric lines, such as Balmer lines, probe the cool, neutral material of erupting prominences which often form the cores of solar CMEs \citep{Gopalswamy15e}. One slow event was also found in the FUV, namely in the \ion{O}{vi}\,(1032\AA) line \citep{Leitzinger11a}, which has also been detected in spectroscopic CME observations on the Sun \citep{Kohl06}. \citet{Bourrier17c} investigated the Lyman-$\alpha$ line of the late-M-type planet host star TRAPPIST-1. They discovered an absorption feature in the blue wing of the line during one observing epoch and suggested a filament eruption as a possible interpretation. On the Sun, Ly$\alpha$ emission has been detected in a hot CME core \citep{Heinzel16}. Recently, \citet{Argiroffi19a} detected a blue-shifted X-ray line during the decay phase of a large flare on an evolved star and interpreted this as a signature of a slow CME. Although most of the aforementioned events were observed as blue-shifted signals indicating motion towards the observer and away from the star, some fast red-shifted events were also observed occasionally and could be due either to mass flowing towards the star, similar to coronal rain on the Sun \citep[e.g.][]{Antolin12}, or mass ejections moving away from the star and the observer if seen in emission \citep{Bopp73, Bell17}.

Recently, \citet{Moschou19} analyzed most of these previously reported stellar CME events and compiled (or estimated) the corresponding CME and flare parameters. By comparison with the solar flare-CME relations, they found that the CME mass -- flare energy relation seems to extend to stellar events, as previously found by \citet{Odert17} using a smaller sample of events, but the kinetic energy -- flare energy relation lies below the extrapolated solar relation. This suggests that stellar CME speeds may be smaller than what is expected from the Sun, in agreement with simulations of CMEs on stars with stronger magnetic fields \citep{Alvarado-Gomez18}.

Despite information about the plasma velocity is available with this method, only the line-of-sight component can be retrieved. Thus, it is difficult to disentangle potential flare-related mass motions from real mass ejections with higher velocities seen in projection. Line asymmetries are often found during flares on active stars \citep{Berdyugina99, Montes99a, Fuhrmeister08, Fuhrmeister18, Vida19}. Since blue asymmetries are often found at the beginning of flare events and red ones towards the end \citep{Fuhrmeister08, Vida16}, this suggests either chromospheric evaporation or filament motion for the former and subsequent downflowing/backfalling material for the latter. Red asymmetries at the beginning of the flare may also be related to chromospheric condensations \citep{Kowalski18}. On the Sun, spatially-resolved H$\alpha$ observations revealed mainly red-shifted emissions during the impulsive phases of flares, likely related with chromospheric condensations, as well as blue- and red-shifted absorptions related with filament eruptions \citep{Canfield90}. Blue-shifted Balmer emissions are less common and can be explained by excess absorption on the red side caused by downflowing material, whereas an interpretation as chromospheric evaporation is less likely because of fast heating which makes the evaporated plasma invisible in the optical too quickly \citep{Heinzel94a}. However, during the impulsive phase of solar flares some chromospheric material may move upward and be visible as an emission in the blue wings of strong lines like \ion{Mg}{ii} \citep{Tei18}. Also opacity effects during flares have been found to be responsible for line asymmetries on the Sun \citep{Kuridze15}. On the fast-rotating subgiant component of the RS~CVn binary II~Peg, blue-shifted H$\alpha$ emission is caused by warm spots rotating onto the visible hemisphere \citep{Strassmeier19}.

In fast rotating stars, stellar prominences were detected as absorption features moving across the broadened profiles of Balmer and other chromospheric lines \citep[e.g.][]{CollierCameron89, Eibe98, Dunstone06, Leitzinger16}, in some cases even switching to emission when moving off the disk \citep{Donati00, Dunstone06a}. This is similar to H$\alpha$ observations of solar prominences, which appear in emission above the limb and in absorption in front of the disk, in the latter case termed filaments \citep{Parenti14}. In fast rotating stars, stable prominences can form even above their coronae where they are embedded in the stellar wind above helmet streamers \citep{Jardine05}. \citet{VillarrealDAngelo18} estimated typical masses of $10^{16}{-}10^{17}$\,g and lifetimes of about 0.4\,d using a mechanical support model, whereas the maximum values of mass (${\sim}10^{18}$\,g) and lifetime (several days) should occur if a star reaches its zero-age main-sequence \citep{VillarrealDAngelo19}.

In many solar CMEs, a prominence forms the core of a CME, which can be observed in H$\alpha$ at early stages and X-ray ejecta at later stages when heated \citep{Filippov08, Gopalswamy15e}. On the Sun, neutral prominence material may be observed in H$\alpha$ up to several solar radii \citep{Sheeley80, House81, Dryer82, Illing85, Athay86, Mierla11, Howard15} before the emission transitions to Thomson scattering due to gradual photoionization \citep{Howard15, Howard15a}. Thus, depending on the ionization state of the plasma, CME cores and prominences observed in white-light may consist of H$\alpha$ emission, Thomson scattered light, or a combination of both \citep{Jejcic09, Howard15, Howard15a}. At later stages when the CME core is heated, collisional ionization takes over and the plasma becomes fully ionized, although some residual neutral hydrogen may emit optically-thin Ly$\alpha$ radiation \citep{Heinzel16}. Direct observations of H$\alpha$ emission in erupting prominences at several solar radii are rare because of the current lack of suitable instruments, but can be indirectly inferred from white-light images via polarization properties \citep{Mierla11, Dolei14, Howard15}. In the event studied by \citet{Howard15} the emission was dominated by H$\alpha$ up to $6{-}9\,R_{\sun}$. Recent spectroscopic observations revealed neutral prominence material embedded in a CME \citep{Ding17}. In some cases, the neutral material can even be tracked through interplanetary space \citep{Lepri10, Sharma13, Wood16}. Based on the solar-stellar analogy, erupting prominences are thus a promising explanation for the transient Doppler-shifted Balmer emissions observed in several active stars during flares.

Dedicated searches for stellar CMEs in H$\alpha$ spectra of young cool stars in open clusters resulted in non-detections up to now \citep{Leitzinger14, Korhonen17}. Searches for CME signatures in archival data also yielded no detections \citep{Vida19, Leitzinger20a}, but typically suffer from a small number of consecutive spectra making the search for transient variability more challenging. However, even such non-detections may be useful to constrain CME occurrence rates on other stars. The observed CME rates depend on the stars' intrinsic rates, observing time, flux detection limit, event duration, as well as geometric effects. By setting constraints on the last four quantities, one may constrain the intrinsic CME rates. The aim of this study is therefore to estimate the expected stellar CME rates observable in the Balmer lines and compare them with existing observations. Moreover, we aim to use our model to prioritize target stars for future observations. By estimating the maximum expected Balmer signals we obtain constraints on the minimum required observing conditions for a given star. In section~\ref{sec:flux} we estimate the maximum expected fluxes of the prominences/CME cores around stars with different spectral types. In section~\ref{sec:geo}, we infer by how much projection effects could lower the detectable rates. In section~\ref{sec:cmerates}, we aim to predict the maximum observable CME rates and the necessary minimum observing times to detect CME events in optical spectra based on an empirical CME occurrence rate model \citep[][hereafter Paper I]{Odert17}. In section~\ref{sec:app}, we compare these predicted rates with existing observations. Finally, we discuss limitations and neglected effects of our model in section~\ref{sec:disc} and summarize our results in section~\ref{sec:sum}.

\section{Expected CME signals in Balmer lines}\label{sec:flux}
We estimate the maximum possible Balmer line fluxes of CME cores in a similar manner to radiative transfer calculations applied to solar filaments/prominences. The emergent intensity (i.e. at the optical depth $\tau=0$) in direction $\mu$ of a prominence structure (approximated by a 1D slab) is described by the solution of the radiative transfer equation
\begin{equation}
I(0,\mu) = I_0(\tau,\mu)\exp(-\tau/\mu) + \int^{\tau}_0 S(t)\exp(-t/\mu) dt/\mu
\end{equation}
\citep[e.g.][]{Labrosse10, Heinzel15a}, where $I_0$ is the incident radiation illuminating the opposite side along the line of sight and $S$ is the source function of the spectral line. A simple solution can be found assuming a constant source function through the slab and $\mu=1$,
\begin{equation}\label{eq:rt}
I(0) = I_0(\tau)\exp(-\tau) + S\left[1-\exp(-\tau)\right].
\end{equation}
For prominences seen in emission above the limb, the background illumination $I_0=0$, and therefore Eq.~\ref{eq:rt} gives $I(0)=S[1-\exp(-\tau)]$. In the special case of optically thick material ($\tau\gg1$), $I(0)\approx S$, and for the optically thin case ($\tau\ll1$), $I(0)\approx S\tau$. For filaments seen in absorption in front of the star, $I_0$ is the radiation from the stellar disk. If $S$ is dominated by photon scattering, one can approximate $S\approx WI_0$, where $W$ is the geometrical dilution factor. The geometrical dilution factor 
\begin{equation}\label{eq:w}
W = \frac{1}{2} \left[1 - \left(1 - \frac{R_\star^2}{(R_\star+h)^2}\right)^{1/2}\right]
\end{equation}
depends on the stellar radius $R_\star$ and the height of the prominence $h$ above the surface \citep[e.g.][]{Heinzel95}; it is 1/2 at the stellar surface ($h=0$) and decreases with increasing height. For the optically thin case, $I(0)\approx I_0$, i.e. the filament is invisible because the contrast is zero. For the optically thick case, $I(0)\approx WI_0$. The maximum contrast is therefore obtained for optically thick material for both prominence and filament geometries. Note that for a source function dominated by collisional excitation, the emergent intensity for filaments could also switch to net emission \citep{Labrosse10}, however, we will neglect this possibility here, as this would require electron densities ${>}10^{12}$\,cm$^{-3}$ which is not typical for solar filaments \citep{Heinzel87b}.

The emitted flux includes both intrinsic emission from the prominence material (e.g. from collisional excitation), as well as scattering of stellar photons into the line-of-sight. Scattering depends on the distance of the prominence from the star, as well as on its size and optical thickness \citep{Dunstone06a}. Because the scattering depends on the stellar illumination, it changes strongly upon eruption of a prominence due to two effects. First, the increasing distance from the star results in progressively lower illuminating fluxes; second, the prominence is illuminated by a Doppler-shifted stellar spectrum according to its propagation velocity. This leads to the so-called Doppler dimming or brightening effect which can be determined using detailed NLTE models \citep{Heinzel87}. However, no detailed modeling of the Doppler dimming/brightening effect exists for stars other than the Sun and for non-solar plasma parameters, so we include its effect in an approximate way by assuming that the scattering source function is determined by illumination from the stellar continuum close to the considered Balmer line. This is a reasonable approximation for large velocities ($>$1000\,km\,s$^{-1}$), but could either over- or underestimate the illuminating fluxes for lower speeds, depending on the width and shape of the stellar line and whether it is in emission or absorption.

Therefore, we use here a more simple approach to estimate the minimum signal-to-noise ratio ($SNR_\mathrm{min}$) needed to detect CME cores in the Balmer lines. We therefore estimate the maximum possible signals for both emission (off-disk) and absorption (on-disk) features on stars with spectral types F, G, K, and M in H$\alpha$, H$\beta$, and H$\gamma$. We aim hereby to estimate the maximum Balmer signatures that can possibly be produced by CMEs of a given mass, because this yields the minimum $SNR$ and the most optimistic detection scenario. This will allow to constrain which stellar spectral types are best suited for conducting CME searches with this method (and which are not).

Due to their velocities of typically several 100 to 1000\,km\,s$^{-1}$, CME-related spectral features would appear above the continuum on the blue side of the stellar spectral lines according to the line-of-sight components of their velocities (we ignore the potential events moving away from the observer and are not fully covered by the stellar disk which could be then seen as red-shifted features). If the ejections are slow and/or the angle to the line-of-sight large, only blue-wing asymmetries of the stellar spectral lines may be seen. However, such events cannot be unambiguously identified as mass ejections because one cannot distinguish intrinsically slow plasma motions, or sufficiently fast ejections seen in projection. On M dwarfs, asymmetric Balmer line profiles of are often observed, both simultaneously with and outside the flare events \citep{Fuhrmeister18, Vida19}. Therefore, we will estimate $SNR$ values only for events with assumed line-of-sight velocities of several 100 to 1000\,km\,s$^{-1}$, where the peak of the CME's line would appear well separated from the stellar line. This also minimizes the dilution of the signal by the possibly broadened stellar lines during flare events, as CMEs and flares likely appear in close temporal association. Projected velocities higher than the stellar escape velocity are indicative of actual mass ejections, since the true speeds can only be equal or higher than the ones observed spectroscopically. The mass ejection's spectral line width may be attributed to temperature, turbulent velocity, rotation, expansion, acceleration or superposed motions from overlaying regions in more complex geometries. As the latter effect would lead to very broad, but flat signatures that could reach from the stellar line up to the maximum projected velocity, we assume here that the CMEs are rather compact objects with a distinct value of their propagation velocity, as this assumption produces the maximum peak flux of their signals.

To estimate the minimum $SNR$ required to detect stellar mass ejections, the peak flux enhancement (or depression) relative to the continuum flux $\Delta f = |f_\mathrm{peak}-f_\mathrm{cont}|$ produced by the ejected material must be larger than the measurement uncertainty $\sigma$. We assume here that $\sigma$ is dominated by the error of the stellar continuum flux in the wavelength range where the feature appears. Thus,
\begin{equation}\label{eq:snr}
SNR = \frac{f_\mathrm{cont}}{\sigma} \ge \frac{f_\mathrm{cont}}{\Delta f} \equiv SNR_\mathrm{min}.
\end{equation}
To demonstrate the dependence of $SNR_\mathrm{min}$ on spectral type, we consider five main-sequence stars with spectral types F6, G2 (solar-like), K2, M2, and M5.5. For these stars, we adopt the parameters given in Table~\ref{tab:stars} taken from the extended online table\footnote{\url{http://www.pas.rochester.edu/~emamajek/EEM_dwarf_UBVIJHK_colors_Teff.txt}} originally published in \citet{Pecaut13}.

\begin{table}
	\caption{Adopted stellar parameters.}
	\label{tab:stars}
	\begin{center}
		\begin{tabular}{lllrrl}
			\hline
			Spectral &  $M_\star$  & $R_\star$  & $M_V$ & $\log L_\star$ & $T_\mathrm{eff}$ \\
			type     &  ($M_{\sun}$) & ($R_{\sun}$) & (mag) & ($L_{\sun}$) & (K) \\
			\hline
			F6    & 1.25  & 1.36  & 3.70  & 0.43    & 6340 \\
			G2    & 1.02  & 1.01  & 4.79  & 0.01    & 5770 \\
			K2    & 0.78  & 0.763 & 6.19  & $-0.47$ & 5040 \\
			M2    & 0.44  & 0.434 & 10.30 & $-1.57$ & 3550 \\
			M5.5  & 0.12  & 0.149 & 15.51 & $-2.79$ & 3000 \\
			\hline
		\end{tabular}
	\end{center}
\end{table}

Note that in the following sections we will assume that the CME mass is equivalent to the mass of the neutral material observed in Balmer lines. This overestimates the expected fluxes for a given mass because the prominences are partly ionized and there is also some pile-up of coronal material during propagation. However, since we are interested in estimating the \textit{minimum} $SNR$ needed for detection, this assumption is reasonable.

\subsection{Emission features}
The combined flux of the star and the emitting prominence can be written as
\begin{equation}\label{eq:fem}
\begin{aligned}
f &=  f_\star + f_\mathrm{p} = I_\star A_\star + I_\mathrm{p} A_\mathrm{p} = \\
 &= f_\star \left[1 + W \left(\frac{A_\mathrm{p}}{A_\star}\right)\left(1-\exp(-\tau)\right)\right],
\end{aligned}
\end{equation}
where $I_\star$ and $I_\mathrm{p}$ are the intensities of stellar disk and prominence, $A_\star$ and $A_\mathrm{p}$ the areas of star and prominence, respectively. The corresponding stellar background and prominence fluxes are $f_\star=I_\star A_\star$ and $f_\mathrm{p}=I_\mathrm{p} A_\mathrm{p}$, respectively. Equation~\ref{eq:fem} is evaluated at the line center of the erupting prominence, i.e. at a wavelength $\lambda_\mathrm{p}=\lambda_\star(1-|v|/c)$ corresponding to a motion with line-of-sight velocity $v$ relative to the stellar Balmer line at $\lambda_\star$. The prominence intensity $I_\mathrm{p}$ is given by Eq.~\ref{eq:rt} together with the scattering source function $S=W I_\star$ and $I_0=0$ (no background radiation). Note that we assume for simplicity that $S$ is determined by $ I_\star(\lambda_\star)$, i.e. the continuum blueward of the stellar line, although the prominence is actually illuminated by the redward continuum, as it moves away from the star. The blue and red continua are expected to be roughly symmetric around the stellar Balmer lines.

To estimate the $SNR$ we require that $\Delta f$ must be larger than the uncertainty $\sigma$ (Eq.~\ref{eq:snr}). Then, using Eq.~\ref{eq:fem} we obtain
\begin{equation}\label{eq:snrem}
SNR \ge \left[W \left(\frac{A_\mathrm{p}}{A_\star}\right)(1-\exp(-\tau))\right]^{-1}.
\end{equation}
Thus, the $SNR$ depends on the ratio of areas, the dilution factor, and the optical thickness. The mass of the prominence can be inferred via
\begin{equation}\label{eq:mep}
M = m_\mathrm{H} \mathcal{N}_\mathrm{H} A_\mathrm{p},
\end{equation}
where $\mathcal{N}_\mathrm{H}$ is the column density of hydrogen atoms in the prominence \citep{Dunstone06a}, assuming that the total mass corresponds to that of neutral hydrogen, which is a good approximation at low temperatures, but underestimates the mass for hotter, more ionized plasma. The area ratio can be expressed by
\begin{equation}\label{eq:area}
\frac{A_\mathrm{p}}{A_\star} = \frac{M}{m_\mathrm{H} \mathcal{N}_\mathrm{H} A_\star}.
\end{equation}
Another useful relation is the minimum CME mass which can be observed in spectra with a given $SNR$. Thus, by rewriting Eq.~\ref{eq:snrem} and inserting Eq.~\ref{eq:mep} we obtain the minimum detectable mass for emission lines
\begin{equation}\label{eq:mminem}
M \ge \frac{A_\star m_\mathrm{H} \mathcal{N}_\mathrm{H}}{SNR\times W(1-\exp(-\tau))}.
\end{equation}

\subsection{Absorption features}
For prominences seen in absorption while in front of the stellar disk (i.e. filaments), the attenuated flux can be written as the sum of the fluxes of the stellar background and the prominence
\begin{equation}\label{eq:fabs}
\begin{aligned}
f &=  f_\star + f_\mathrm{p} = I_\star\left(A_{\star}-A_\mathrm{p}\right) + I_\mathrm{p} A_\mathrm{p} = \\
 &= f_\star \left[1+(W-1) \left(\frac{A_\mathrm{p}}{A_\star}\right) (1-\exp(-\tau))\right],
\end{aligned}
\end{equation}
where we assume again a scattering-dominated source function. For simplicity we ignore any inhomogeneities of the stellar disk emission. To estimate the $SNR$ we again require that $\Delta f$ must be larger than the uncertainty $\sigma$. Then, using Eq.~\ref{eq:fabs}
\begin{equation}\label{eq:snrabs}
SNR \ge \left[(1-W) \left(\frac{A_\mathrm{p}}{A_\star}\right)(1-\exp(-\tau))\right]^{-1}.
\end{equation}
Rewriting Eq.~\ref{eq:snrabs} to express the minimum mass that can be observed at a given $SNR$ yields
\begin{equation}\label{eq:mminabs}
M \ge \frac{A_\star m_\mathrm{H} \mathcal{N}_\mathrm{H}}{SNR (1-W)(1-\exp(-\tau))}.
\end{equation}

\subsection{SNR estimates}
Leaving the prominence mass as independent variable, then for any given star there are three free parameters in Eqs.~\ref{eq:snrem} and \ref{eq:snrabs}: the hydrogen column density $\mathcal{N}_\mathrm{H}$, the optical depth $\tau$, and the dilution factor $W$; the area is fixed by Eq.~\ref{eq:mep}. As mentioned before, the dilution factor can take values $\leq0.5$, depending on the distance of the prominence from the star. For $W=0.5$, the resulting $SNR$ for both emission and absorption signals are the same; for smaller $W$, the emission signals decrease, while the absorption signals increase because of higher contrast, leading to an opposite effect. The optical depths of the Balmer lines are related to $\tau_{\mathrm{H}\alpha}$ via $\tau_i/\tau_{\mathrm{H}\alpha} = f_i/f_{\mathrm{H}\alpha} \lambda_i/\lambda_{\mathrm{H}\alpha}$, where $f_i$ are the oscillator strengths. Thus, the higher Balmer lines are more likely to be optically thin ($\tau$ is about a factor of 7 lower for H$\beta$ and about 20 for H$\gamma$ compared to H$\alpha$). Only if H$\alpha$ is very optically thick the higher lines may also be. Hydrogen column densities of solar prominences are typically in the range 10$^{18}$ to a few 10$^{19}$\,cm$^{-2}$ \citep{Labrosse10, Parenti14}. For the large stellar slingshot prominences, values in the range $10^{19}-10^{22}$\,cm$^{-2}$ have been found \citep{CollierCameron90, Dunstone06a, Parsons11}.

We explore the effects of varying the three parameters within realistic ranges. The dilution factor $W$ is varied between 0.5 and 0.005 (corresponding to prominence heights of $0-6R_\star$), and the column density between 10$^{18}$ and 10$^{22}$\,cm$^{-2}$, encompassing the range of typical solar and stellar values. For the optical depth, we use a fixed $\tau_{\mathrm{H}\alpha}=10$ and show additionally the properly scaled higher Balmer lines, which correspond to optical depths of 1.38 (H$\beta$) and 0.46 (H$\gamma$). For higher $\tau_{\mathrm{H}\alpha}$, the results for H$\alpha$ do not change significantly anymore due to saturation, whereas the optical depths of the other Balmer lines increase and the curves approach and eventually merge with H$\alpha$. Smaller optical depths than adopted for H$\gamma$ may be possible, but probably unrealistic for the higher column density values. Moreover, due to Eq.~\ref{eq:mep}, some combinations of column density and CME mass result in prominence areas much larger than the star, especially for small $\mathcal{N}_\mathrm{H}$. Even the large slingshot prominences on active, fast rotating stars have sizes not exceeding $\sim$25\% of the stellar disk area \citep{CollierCameron90, Doyle90a, Dunstone06a}. Thus, we set $A_\mathrm{p}=0.3A_\star$ as an upper limit to the prominence area and scale $\mathcal{N}_\mathrm{H}$ accordingly to keep the mass constant. Note that during eruption $A_\mathrm{p}$ may grow temporarily also to larger values due to expansion. However, its effect on the Balmer signals may be compensated partly by the simultaneously decreasing column density and optical depth, so we ignore it here.

\begin{figure}
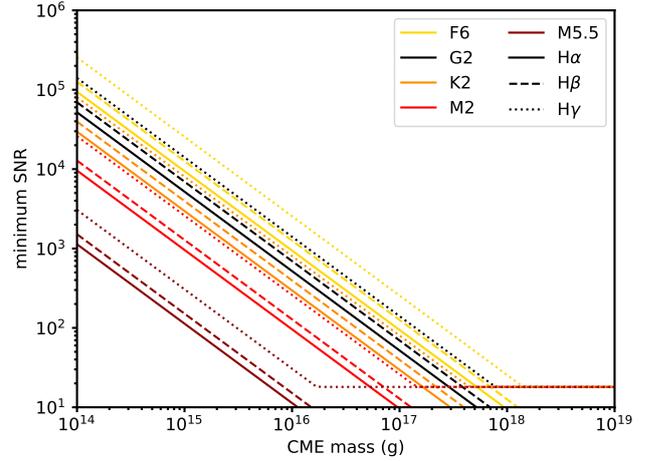

	\centering
	\includegraphics[width=\columnwidth]{{{snr_em_tau10_logN20_wdil0.5}}}
	\caption{Minimum $SNR$ required to detect mass ejections on F, G, K and M dwarfs in the lower Balmer lines (H$\alpha$, H$\beta$, H$\gamma$). Adopted parameters are $W=0.5$, $\mathcal{N}_\mathrm{H}=10^{20}$\,cm$^{-3}$ and $\tau_{\mathrm{H}\alpha} = 10$. For the case of $W=0.5$, the minimum $SNR$ is the same for emission and absorption signals. The horizontal parts of the lines for higher masses indicate where $\mathcal{N}_\mathrm{H}$ was increased so that $A_\mathrm{p}$ does not exceed $0.3A_\star$.}
	\label{fig:snr}
\end{figure}

Figure~\ref{fig:snr} shows the estimated minimum $SNR$ (Eqs.~\ref{eq:snrem} and \ref{eq:snrabs}) as a function of CME mass for the three lowest Balmer lines and five stellar spectral types (Table~\ref{tab:stars}). One can see that for any given CME mass and spectral line the detection becomes more feasible for later spectral types because of the better contrast, and is exceptionally favorable for mid- to late M~dwarfs. This likely explains why most detections with this method up to now occurred on stars later than M3.5 \citep[cf.][]{Moschou19}. For hotter, Sun-like stars the detection will be limited to the most massive events, with masses comparable to those of the large slingshot prominences observed on young fast-rotating stars \citep[e.g.][]{CollierCameron89}. In Fig.~\ref{fig:snr}, we adopt the parameters $W=0.5$ (for which the minimum $SNR$ for emission and absorption signals is the same), $\mathcal{N}_\mathrm{H}=10^{20}$\,cm$^{-3}$ (average value of solar and stellar prominences) and $\tau_{\mathrm{H}\alpha} = 10$. In Appendix~\ref{app:snr}, we show additional plots with variations of these parameters. Specifically, for lower dilution factors the required $SNR$ for emission signals increases, whereas it decreases for absorption lines (cf. Eqs.~\ref{eq:snrem} and \ref{eq:snrabs}). However, the effect for absorptions is not very strong because the flux attenuation dominates the small contribution of the scattering source function. For smaller $\mathcal{N}_\mathrm{H}$ the required $SNR$ decreases, because for a given mass the area grows, whereas it increases for larger $\mathcal{N}_\mathrm{H}$ because the area shrinks. Small values of $\mathcal{N}_\mathrm{H}$ for high CME masses are unrealistic though, because the resulting prominence areas would then become larger than the stellar disk.

\section{Geometrical constraints}\label{sec:geo}
Since stellar mass ejections are transient events with random orientations of their propagation relative to the observer, there are various geometrical constraints which may limit the number of actually observable CMEs. Also, the duration of such events is an important issue. For instance, absorption features can only appear in front of the stellar disk which limits their visibility. Due to expansion, the column densities and thus the optical depths of erupting prominences are expected to decrease with time. If a signal's visibility is much shorter than the exposure time of the spectra, it may be further diluted. These issues are discussed in more detail in this section.

\subsection{CME velocities}
For a successful CME eruption its velocity should overcome the escape velocity $v_\mathrm{esc}$. Since $v_\mathrm{esc}=(2GM_{\star}/r)^{1/2}$ decreases with height the plasma may have to reach a distance of several stellar radii before it becomes gravitationally unbound, depending on the initial acceleration and height of the structure \citep{Vourlidas02, Lewis02}. This also means that if prominences located at several stellar radii from the surface (like those observed on active, fast rotating stars) erupt, less energy is required to become unbound, whereas prominences located at small heights must either be accelerated fast to overcome the high escape speed in the low corona or the energy input must be sufficiently long to bring them to larger heights where the escape speed is lower.

On the Sun, CME speeds were found to be correlated with the energies of the associated flares, similar to their masses and kinetic energies. \citet{Drake13} determined the correlation between CME velocities $v$ and \textit{GOES} flare energies $E$ as $v(E) = 3.6\times10^{-4}E^{0.22}\ \mathrm{km\ s^{-1}}$. However, solar CME speeds  suffer from projection effects, since they are measured in the plane-of-sky and thus tend to be underestimated. \citet{Salas-Matamoros15} determined a similar relation corrected for projection effects by using limb events only. Their relation is in excellent agreement with \citet{Drake13}. Therefore, we assume in the following that the above relation holds for the true CME velocities. We further use a relation between flare energy and CME mass \citep{Drake13} to express $v$ (in km\,s$^{-1}$) as a function of CME mass $M$ (in g), yielding
\begin{equation}\label{eq:vm}
v(M)=1.3\times10^{-3}M^{0.37}.
\end{equation}
Equation~\ref{eq:vm} suggests that CME velocities may become very large for the most massive events (e.g. ${\sim}10^4$\,km\,s$^{-1}$ for $10^{19}$\,g) which is probably unrealistic due to the large energy requirements \citep{Drake13}. Observations of such potential high-speed events, if they exist, would be challenging, because the most massive events are expected to be very rare and the event would travel more than one solar radius within one minute. The line shift at H$\alpha$ in the example above would amount to about 200\,\AA; thus, even strong events with sufficient signal and duration could remain unnoticed if not explicitly searched for. However, for the fastest stellar CME event observed to date \citep{Houdebine90}, the estimated minimum mass ($7.7\times10^{17}$\,g) and line-of-sight velocity (5800\,km\,s$^{-1}$) agree with Eq.~\ref{eq:vm}, which gives 5400\,km\,s$^{-1}$, rather well. This also demonstrates that it may be more difficult to find such signatures with Echelle spectrographs rather than low-resolution instruments, because for such high speeds the light would be diluted over two Echelle orders. For instance, the half width of the Echelle orders for HARPS is 1680\,km\,s$^{-1}$, and 2260\,km\,s$^{-1}$ for ESPRESSO.

On the other hand, recent modeling efforts suggest that the CME velocities may be lower than predicted by Eq.~\ref{eq:vm} for active stars with stronger magnetic fields due to suppressive effects \citep{Alvarado-Gomez18}. This would also reduce the enormous energy requirements for the largest events predicted by Eq.~\ref{eq:vm}. Therefore, we treat the velocities predicted by Eq.~\ref{eq:vm} as upper limits, likely yielding overestimated observable eruption rates. In addition, we perform our calculations also for CME speeds smaller by factors of five and ten than obtained by Eq.~\ref{eq:vm} and show the effect on the predicted CME rates in Appendix~\ref{app:int_obs}.

In stellar spectra, the observed velocity is always a lower limit, because only the line-of-sight component is observed. Therefore it is necessary to consider how many CMEs may not be observable (or observed with such low projected velocities so that they cannot be unambiguously identified as CMEs) due to velocity projection. Considering the visible hemisphere only, we can estimate the average reduction of the line-of-sight speed compared to the propagation speed. In spherical coordinates, the velocity component in the observer's direction is
\begin{equation}\label{eq:vlos}
v_\mathrm{los}=v\cos{\theta}\cos{\phi},
\end{equation}
where $\theta$ is the latitude and $\phi$ the longitude, both counted from -90\degr\ to +90\degr\ in the visible hemisphere. The average value of $\cos{\theta}\cos{\phi}\sim0.4$, i.e. the line-of-sight component is on average only 40\% of the true speed. The projected velocity needs to be high enough to be detected depending on the resolution of the instrument and the width of the stellar line. By assuming an intrinsic CME velocity distribution and a distribution of source locations, one can estimate such correction factors \citep{Leitzinger14}. Hereafter, we will use Eq.~\ref{eq:vm} to assign a true velocity to a given CME mass. Note that on the Sun, erupting prominences have often smaller velocities than their associated CMEs, but in some cases speeds up to 1200\,km\,$s^{-1}$ have been observed \citep{Liu15}. Therefore our adopted velocities provide likely an upper limit to the prominence speeds which are probed by optical spectra. We furthermore assume constant speeds and neglect possible acceleration or deceleration.

\subsection{Source locations}
On the Sun, CMEs can emerge from both active and quiescent regions, in the latter case often related to eruptions of polar crown filaments \citep{Webb12}. During activity minimum, their source regions are located close to the equator, but in maximum, they appear at all latitudes \citep{Yashiro04, Bilenko14}. Their latitude distribution is more similar to those of prominences and streamers than those of spots, flares and active regions \citep{Hundhausen93}. Thus, one cannot generally assume that active stars may have some preferred source location of CMEs. Fast rotating stars often have large polar spots, but spotted regions also reach down to lower latitudes \citep{Strassmeier09}. However, since solar CMEs are not exclusively related to spotted regions, spot locations provide only a weak constraint. Therefore we assume in the following that stellar CMEs may originate from the whole stellar surface.

\subsection{Signal duration}\label{sec:duration}
The signal duration limits the observability of mass ejections because of the required integration time and cadence of the spectra. It depends on several parameters discussed in the previous sections. In Balmer lines it is limited primarily by expansion and ionization. Expansion during the outward propagation may decrease the column density and thus the emitted or absorbed flux of the structure. If assuming roughly self-similar expansion like observed on the Sun, it depends on the initial size, density and ejection/expansion velocity when the point is reached where the signal becomes undetectable. Ionization due to heating \citep{Filippov02} and photoionization \citep{Howard15} decreases the neutral fraction and thus the Balmer signal. On the Sun the Balmer signal dominates up to a few solar radii, but this could be shorter for more active stars where the photoionization rates should be higher due to higher EUV fluxes. If the emission is dominated by scattering of stellar light, the scattered radiation decreases with increasing distance from the star. In addition, Doppler dimming/brightening may reduce/prolong the signal duration \citep{Heinzel87}.

\begin{figure}
	\centering
	\includegraphics[width=\columnwidth]{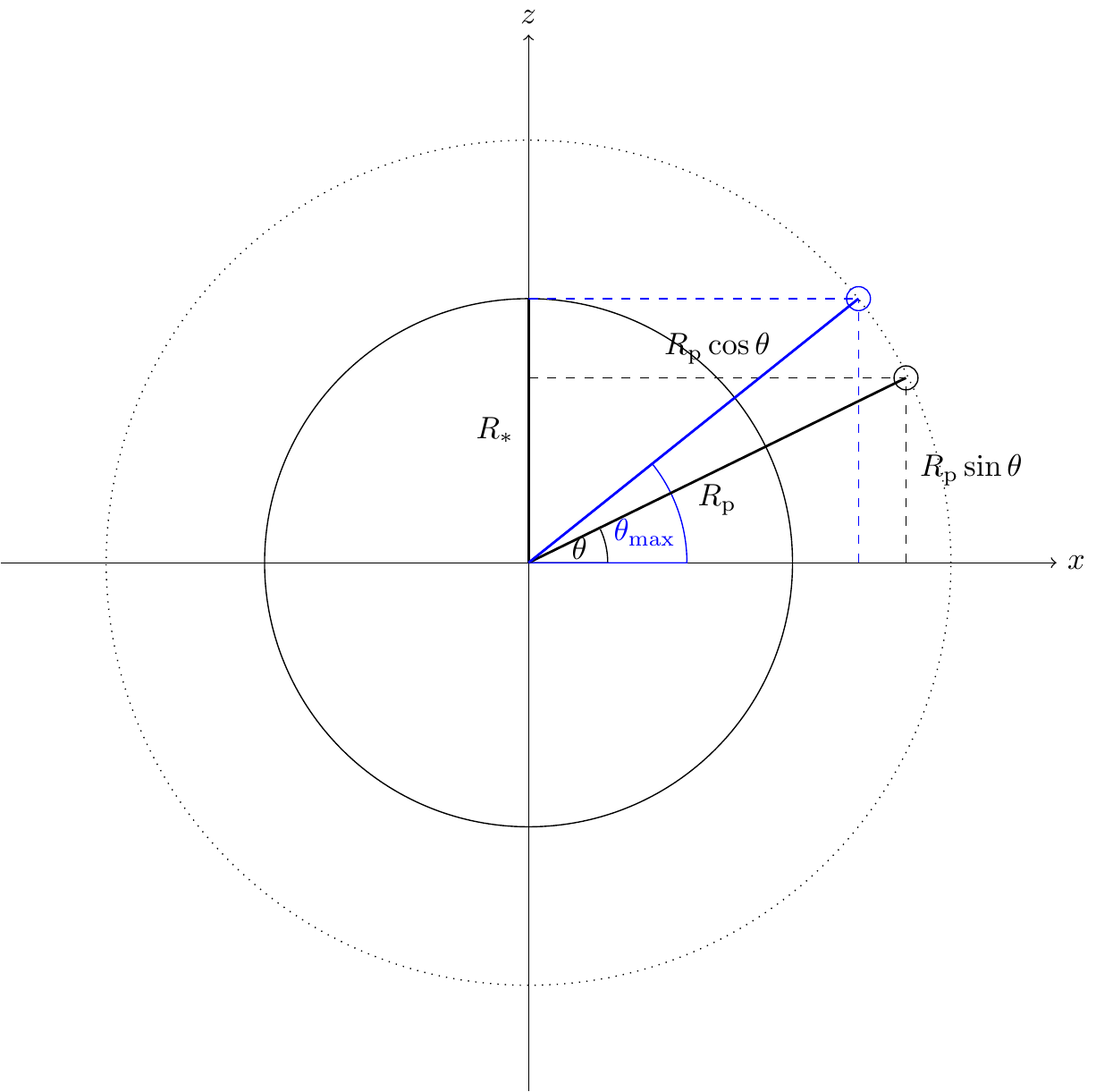}\\
	\includegraphics[width=\columnwidth]{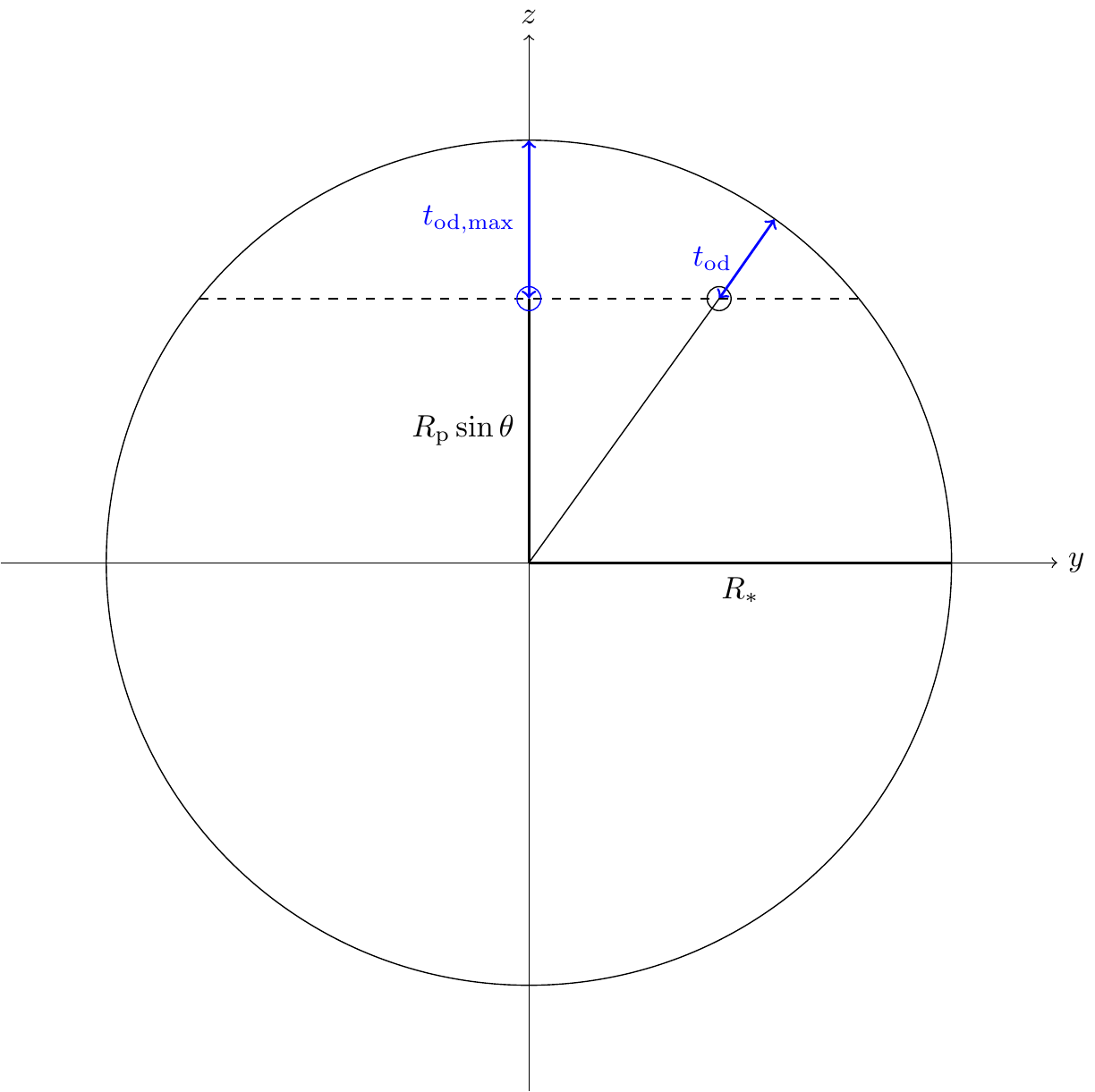}
	\caption{Geometry related to the derivation of the on-disk time. Upper panel: side view ($x{-}z$ plane), lower panel: front view ($y{-}z$ plane). The observer is located in positive $x$-direction. The solid circle represents the stellar radius, small circles the locations of the prominences. The dotted circle (upper panel) has the radius $R_\mathrm{p}$. In the upper panel the prominences lie in the $x{-}z$ plane to define the maximum latitude $\theta_\mathrm{max}$ that a prominence can have to transit the stellar disk. The $z$-axis is the stellar rotation axis (i.e., $i=90\degr$).}
	\label{fig:geo}
\end{figure}

If the typical mass ejections on a star can only be observed as absorption features, the signal duration is further limited by the on-disk time \citep{Leitzinger14}, i.e. the time during which a CME can be seen in projection against the stellar disk. Generally this time can be estimated as the path length in the plane-of-sky from the projected source location to the edge of the stellar disk (on a trajectory passing through the center of the star), $r_\mathrm{pos} = R_{\star} -  R_\mathrm{p} (\cos^2\theta \sin^2\phi + \sin^2\theta)^{1/2}$, divided by the velocity component in the plane-of-sky, $v_\mathrm{pos} = v (\cos^2\theta \sin^2\phi + \sin^2\theta)^{1/2}$ (cf. Fig.~\ref{fig:geo}), yielding
\begin{equation}\label{eq:tod}
t_\mathrm{od} = \frac{r_\mathrm{pos}}{v_\mathrm{pos}} = \frac{R_{\star}}{v} \left[ \left(\cos^2\theta\sin^2\phi+\sin^2\theta\right)^{-\frac{1}{2}} -\frac{R_\mathrm{p}}{R_{\star}} \right],
\end{equation}
where $v$ is the velocity of the ejection. Here, $R_\mathrm{p}$ denotes the initial ``radius'' of the prominence before eruption, i.e. its distance from the center of the star. The maximum on-disk time is always obtained at the observer's meridian ($y=0$) for all $\theta$, i.e. at $\phi=0{\degr}$ (cf. Fig.~\ref{fig:geo}, lower panel), and is given by
\begin{equation}\label{eq:todmax}
t_\mathrm{od,max} = \frac{R_{\star}}{v}\left(\frac{1}{\sin{\theta}} - \frac{R_\mathrm{p}}{R_{\star}}\right).
\end{equation}
Note that these simple estimates assume a point mass and neglect the shape and size of the erupting structure.

\subsection{Correction factor}
Intrinsic CME rates of stars are likely higher than those from observations. First, CME fluxes need to exceed a certain limit to be detected in observations, limiting the minimum mass that can be observed. Second, the random directions of the ejections may prevent the detection of a certain fraction of CMEs in spectra. Here we combine the issues discussed in the previous sections into a mass-dependent reduction factor. For emission signals, the most severe limitation is the velocity projection because the line-of-sight component must be resolvable by the spectrograph and the feature should move out of the stellar line profile sufficiently far to avoid confusion with slower flare related mass motions. Note that the most unambiguous detection of a mass ejection event requires that the line-of-sight velocity is larger than the escape velocity, but slower events can be potential candidates. For an event appearing as absorption feature while still on the disk, the main limiting factor is the on-disk time. We additionally require that for events with sufficient on-disk time the projected velocity must be larger than the chosen limit. As can be seen in Appendix~\ref{app:snr}, the minimum masses for emission and absorption features at a given $SNR$ of the spectrum are not the same for $W<0.5$. Thus, it is possible that for a given star/line/$SNR$ a CME of a given mass can only appear in absorption. The most massive CMEs could be observable in both modes. That means that in principle one may observe an absorption signal while the CME is still in front of the disk and an emission signal later when it has moved off the disk. We assume in the following radial ejections, source locations distributed over the whole stellar surface, and ignore possible signal dilution due to expansion.

\begin{figure}
	\centering
	\includegraphics[width=\columnwidth]{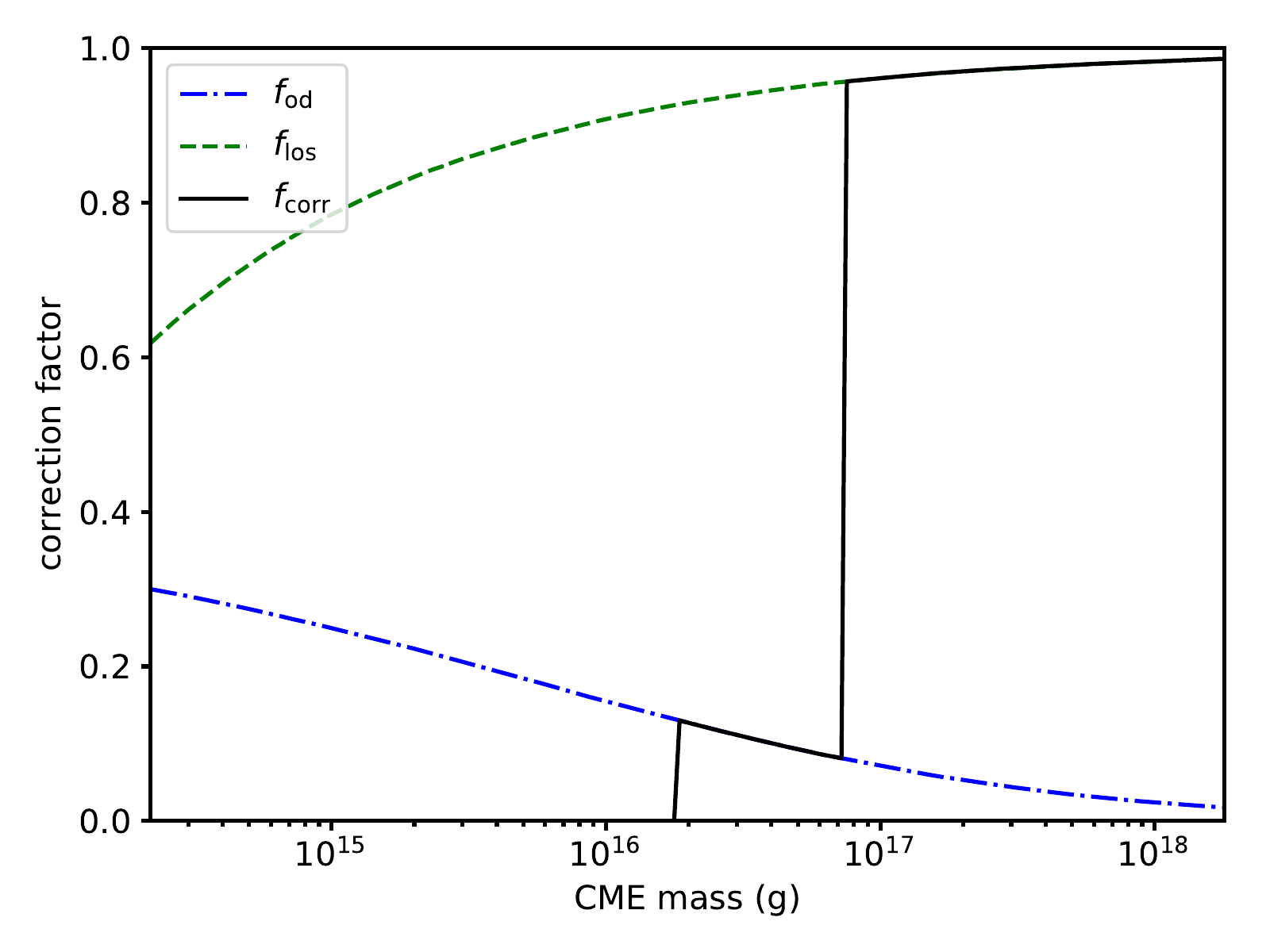}
	\caption{Mass-dependent geometrical correction factors for a K2 star (Table~\ref{tab:stars}), assuming $W=0.2$ (corresponding to $h=0.25R_\star$), $SNR=100$, $t_\mathrm{exp}=5$\,min and $v_\mathrm{lim}=100$\,km\,s$^{-1}$ for observations in H$\alpha$.}
	\label{fig:fgeo}
\end{figure}

We combine the two main geometric reduction factors ($f_\mathrm{los}(M)$ for the velocity correction and $f_\mathrm{od}(M)$ for the on-disk time correction) with the flux limits, $M_\mathrm{min,em}$ (Eq.~\ref{eq:mminem}) and $M_\mathrm{min,abs}$ (Eq.~\ref{eq:mminabs}), to obtain a mass-dependent correction factor $f_\mathrm{corr}(M)$. Note that for a pure scattering source function always $M_\mathrm{min,abs}\le M_\mathrm{min,em}$, where the equality holds for $W=0.5$. For CME masses smaller than $M_\mathrm{min,abs}$, $f_\mathrm{corr}(M)=0$. For masses larger than $M_\mathrm{min,em}$, i.e. the mass range where CMEs could in principle be observed both in emission and absorption, $f_\mathrm{corr}(M)$ is the maximum of $f_\mathrm{los}(M)$ and $f_\mathrm{od}(M)$, because an event is readily detected if it appears either as an emission or absorption feature. This will in most cases be $f_\mathrm{los}(M)$ because it increases with mass (i.e. velocity, Eq.~\ref{eq:vm}) whereas $f_\mathrm{od}(M)$ decreases. The correction factor equals $f_\mathrm{od}(M)$ in the mass range between $M_\mathrm{min,abs}$ and $M_\mathrm{min,em}$ (cf. Fig.~\ref{fig:fgeo}). 

To determine $f_\mathrm{corr}(M)$, we derive $M_\mathrm{min,em}$ and $M_\mathrm{min,abs}$ for the given observational parameters ($t_\mathrm{exp}$, $SNR$, $v_\mathrm{lim}$), spectral line and star using Eqs.~\ref{eq:mminem} and \ref{eq:mminabs}. We estimate CME velocities using Eq.~\ref{eq:vm} and adopt an initial prominence height (and corresponding dilution factor). We then use a Monte Carlo approach to get a statistical estimate for the corrections. We model $n=10^5$ randomly distributed source locations in the visible hemisphere (we ignore the backside like in flare rates) with the longitude $\phi=\pi\times\mathcal{U}(0,1)-\pi/2$ and the latitude $\theta=\arcsin(2\times\mathcal{U}(0,1)-1)$, where $\mathcal{U}$ indicates a uniform distribution\footnote{\url{http://mathworld.wolfram.com/SpherePointPicking.html}}. We calculate $n$ realizations of the line-of-sight component of the velocity (Eq.~\ref{eq:vlos}), and the on-disk time (Eq.~\ref{eq:tod}). The velocity correction factor $f_\mathrm{los}(M)$ is then obtained by the fraction of realizations for which $v_\mathrm{los}{\ge}v_\mathrm{lim}$, whereas $f_\mathrm{od}(M)$ is obtained from the fraction of realizations for which both $t_\mathrm{od}{\ge}t_\mathrm{exp}$ and $v_\mathrm{los}{\ge}v_\mathrm{lim}$. The combined factor is then obtained as described above. An example of $f_\mathrm{corr}(M)$ and its components $f_\mathrm{los}(M)$ and $f_\mathrm{od}(M)$ is shown in Fig.~\ref{fig:fgeo} for a K2 star (Table~\ref{tab:stars}), assuming $W=0.2$ (corresponding to $h=0.25R_\star$), $SNR=100$, $t_\mathrm{exp}=5$\,min and $v_\mathrm{lim}=100$\,km\,s$^{-1}$, for observations in the H$\alpha$ line. One can see that $f_\mathrm{od}$ decreases with CME mass because the associated higher speeds move the CMEs off the disk more quickly. On the other hand, $f_\mathrm{los}$ increases with CME mass because the higher velocities lead to higher line-of-sight components. The discontinuities in $f_\mathrm{corr}$ are related to (from low to high masses) $M=M_\mathrm{min,abs}$ and $M=M_\mathrm{min,em}$. In Appendix~\ref{app:geo} we show how the correction factor changes with changing stellar parameters and observational characteristics.

We note that using a fixed dilution factor, especially $W=0.5$, tends to overestimate the emission and underestimate the absorption signals, since during ejection, the height of the prominence increases and $W$ decreases accordingly. Therefore, the ``effective'' dilution factor during one exposure is smaller. To estimate this effect, we also perform calculations with such an ``effective'' dilution factor
\begin{equation}\label{eq:wav}
\begin{split}
W_\mathrm{eff} &= \frac{1}{2(h_1-h_0)}\times \\
&\times\left(\left[h_1-\sqrt{h_1(h_1+2)} + 2\arctan\left(\sqrt{\frac{h_1}{h_1+2}}\right)\right]\right. \\
&- \left.\left[h_0-\sqrt{h_0(h_0+2)} + 2\arctan\left(\sqrt{\frac{h_0}{h_0+2}}\right)\right]\right),
\end{split}
\end{equation}
which represents the average of Eq.~\ref{eq:w} over the height $h$. Parameter $h_0$ is the initial height of the prominence (taken to be $h_0=0$ if not stated otherwise), and $h_1=h_0+vt_\mathrm{exp}$ is the maximum height reached during the exposure if assuming radial propagation. We show the effect in Fig.~\ref{fig:fgeo_wav}, and together with the variations of the other parameters in Appendix~\ref{app:geo}. For the example shown in Figs.~\ref{fig:fgeo} and \ref{fig:fgeo_wav}, $f_\mathrm{od}$ is larger for the average $W$ at smaller masses, because the average prominence height is in these cases smaller than the fixed value corresponding to $W=0.2$ adopted in Fig.~\ref{fig:fgeo}. On the other hand, $M_\mathrm{min,em}$ increases because the emission signals become fainter for larger heights.

\begin{figure}
	\centering
	\includegraphics[width=\columnwidth]{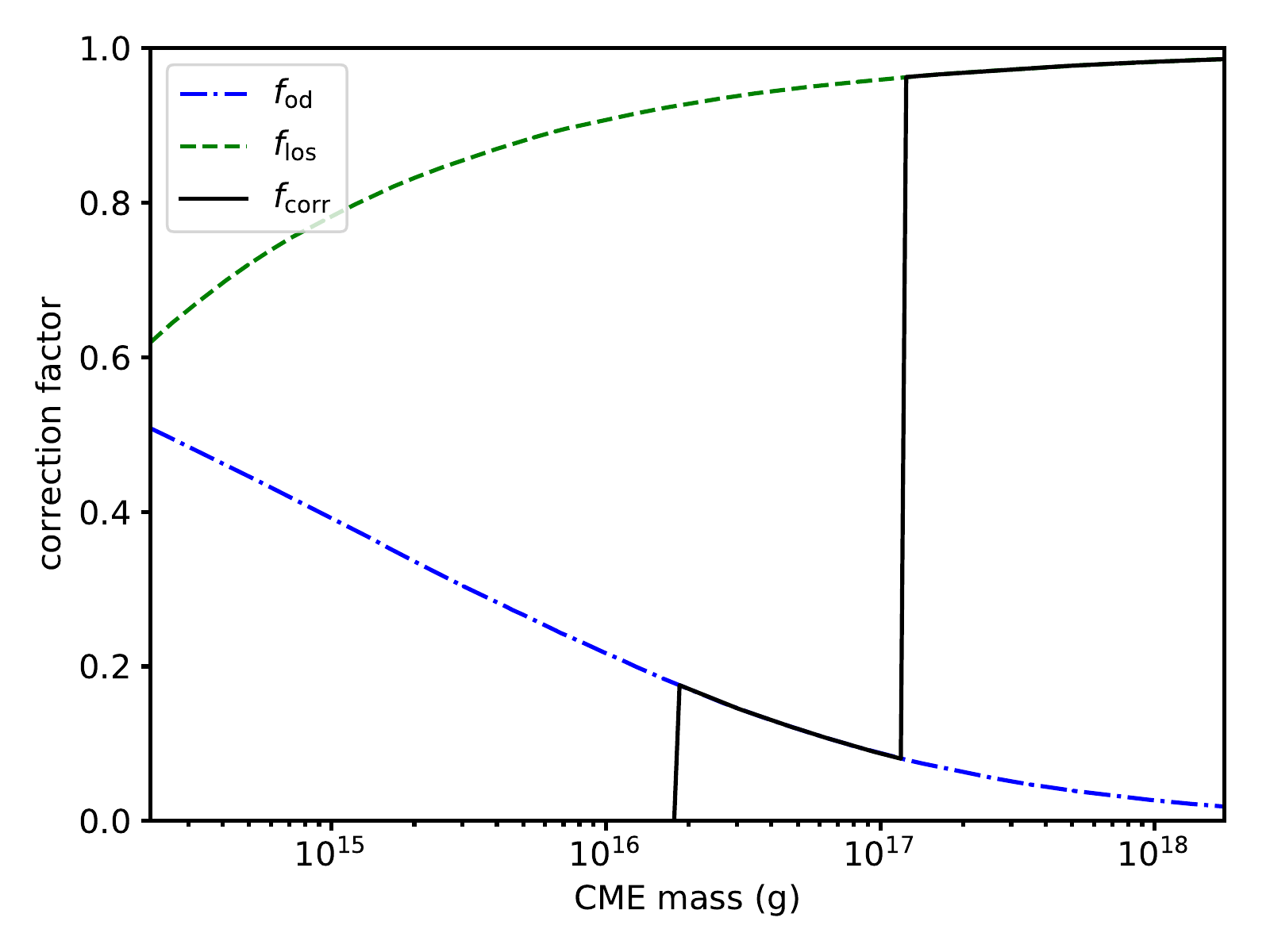}
	\caption{Same as Fig.~\ref{fig:fgeo}, but for the effective dilution factor (Eq.~\ref{eq:wav}).}
	\label{fig:fgeo_wav}
\end{figure}

\section{CME occurrence rates}\label{sec:cmerates}
\subsection{Predicted rates of observable CMEs}\label{sec:rates}
In \citetalias{Odert17} we developed an empirical model based on stellar flare rates and correlations between flare and CME parameters from the Sun to predict stellar CME occurrence rates. To compare these model predictions with optical spectroscopic observations, one needs to include geometrical corrections and flux detection limits. We use Eq.~9 of \citetalias{Odert17}, which gives the predicted differential distribution of CMEs according to their masses, $dN/dM$. It depends on the stellar X-ray luminosity $L_\mathrm{X}$ and the power law index $\alpha$ of the star's XUV flare energy distribution, $dN/dE\propto E^{-\alpha}$. To determine the number of expected CMEs which may be observed with given observational constraints for a given star, we numerically solve
\begin{equation}\label{eq:ntilde}
\tilde{N} = t_\mathrm{obs} \int_{M_0}^{M_\mathrm{max}} \frac{dN}{dM} f_\mathrm{corr}(M) dM,
\end{equation}
where the correction factor $f_\mathrm{corr}$ is calculated as described in Section~\ref{sec:geo}, $M_0$ is the CME mass below which the flare--CME relationship from the Sun is zero, and $M_\mathrm{max}$ is the maximum possible CME mass for a star at a given activity level quantified by its $L_\mathrm{X}$ \citepalias{Odert17}. Note, however, that the statistical correction factors may be more reliable for masses where a large number of CMEs occur, but may not be reliable for the highest masses with low event rates.

\begin{figure}
	\centering
	\includegraphics[width=\columnwidth]{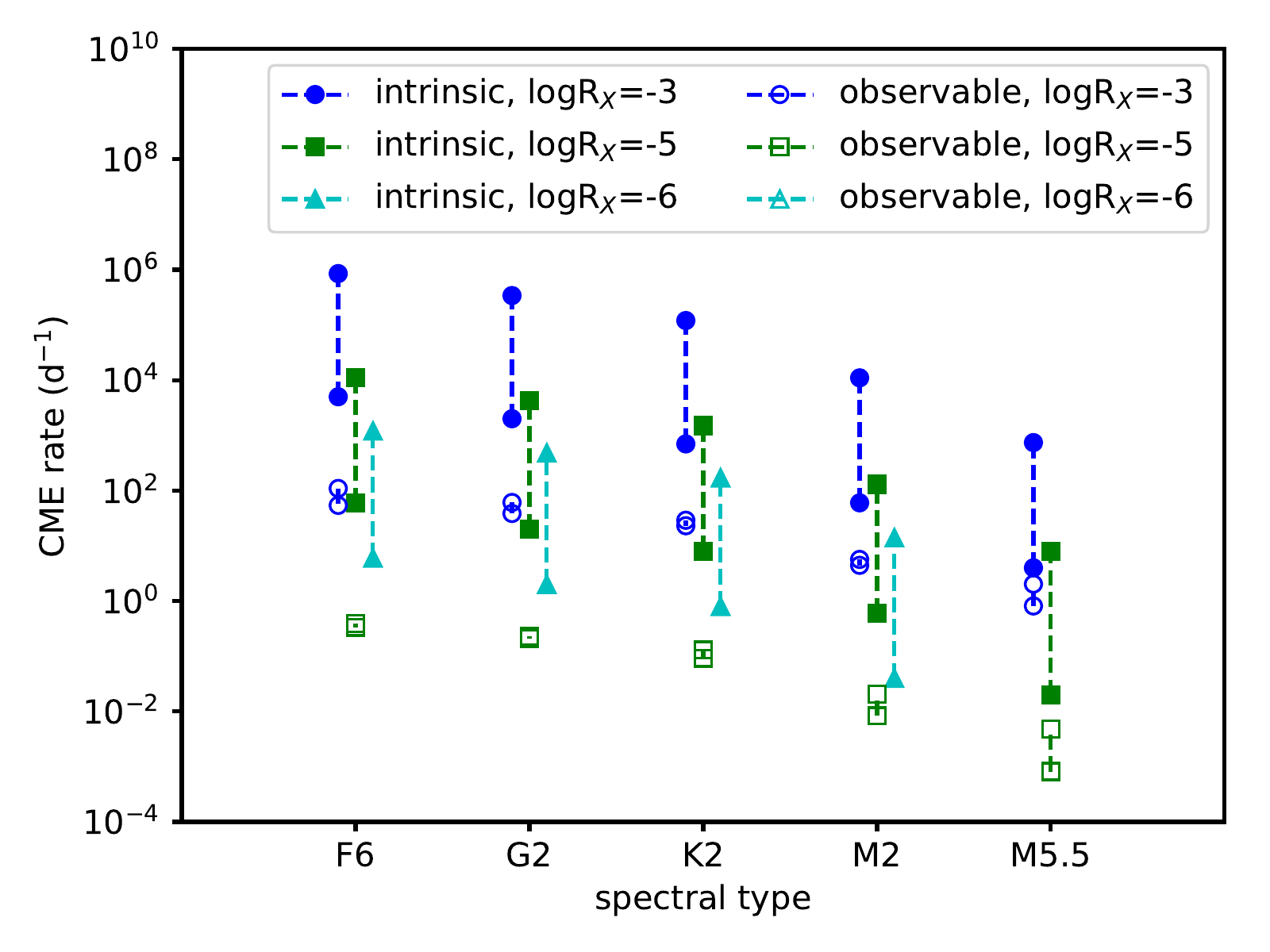}
	\caption{Comparison between intrinsic CME rates \citepalias{Odert17} and those potentially observable in H$\alpha$ if assuming $SNR=100$, $t_\mathrm{exp}=5$\,min, and $v_\mathrm{lim}=100$\,km\,s$^{-1}$. X-ray luminosities were computed from the activity index $\log R_\mathrm{X} = \log(L_\mathrm{X}/L_\mathrm{bol})$, for which we show a highly active case ($-3$, i.e. saturation) as blue circles and two lower activity cases ($-5$, $-6$) as green squares and cyan triangles, respectively. For the latter case, no CMEs are observable anymore. The vertical dashed lines connecting the same symbols correspond to the range of flare power law indices $\alpha=1.5{-}2.5$ \citepalias[cf.][]{Odert17}.}
	\label{fig:int_obs}
\end{figure}

Figure~\ref{fig:int_obs} shows a comparison between the predicted intrinsic and maximum observable rates for stars with spectral types F5--M5.5. We adopt here our default CME and observational parameters, but the effective dilution factor because it is more realistic than using a fixed value. To estimate the X-ray luminosities required for the prediction of intrinsic rates after \citetalias{Odert17}, we adopt the bolometric luminosities of the stars and assume different activity levels represented by the activity index $\log R_\mathrm{X} = \log(L_\mathrm{X}/L_\mathrm{bol})$. For the most active stars, this number is around $-3$, at which $\log R_\mathrm{X}$ saturates \citep[e.g.][]{Pizzolato03, Wright11a}. We note that slightly lower saturation levels are found in studies in which stellar samples are corrected for multiplicity \citep[e.g. $\log R_\mathrm{X}\approx-4$ for M2--M4 stars;][]{Jeffers18}. Thus, $\log R_\mathrm{X}=-3$ represents a maximum value. We also show the results for less active stars ($\log R_\mathrm{X}=-5\ldots-6$), which are expected to have less CMEs. Note that our Sun has $\log R_\mathrm{X}\sim-7\ldots-6$ from cycle minimum to maximum \citep{Peres00}. The vertical dashed lines represent the typically observed range of solar and stellar flare power law indices, 1.5--2.5 \citep[e.g.][]{Guedel03}. The results show that the number of CMEs potentially observable in H$\alpha$ is, as expected, lower that the intrinsic one. The difference is most severe for F5 and decreases with spectral type until M5.5 for which the observable rates lie only slightly below the range of intrinsic rates, especially for the saturated activity case. Thus, for the same intrinsic CME rate distribution, which corresponds to adopting the same X-ray luminosity and flare power law index if using the empirical model of \citet{Odert17}, the detectable CME rates increase towards later spectral types.

We show the effect of varying the observational parameters in Appendix~\ref{app:int_obs}. Reducing the $SNR$ leads to reduced observable rates; conversely, increasing the $SNR$ leads to a higher observable rates, making CMEs observable even on some of the least active stars considered. The magnitude of the effect decreases with later spectral type. Raising the velocity limit slightly reduces the observable rates. However, the effect is negligible for F--K stars, and only significant for M dwarfs. Raising the exposure time slightly lowers the observable rates. Reducing the CME speeds predicted by Eq.~\ref{eq:vm}, which could be relevant for active stars with stronger magnetic fields, lowers the observable rates for M dwarfs, but increases them slightly for F--K stars due to the increased on-disk time. The effect of observing the higher Balmer lines instead of H$\alpha$ slightly lowers the detectable rates due to their smaller optical depths. Finally, lower column densities lead to much higher detectable CME rates, whereas very low rates are obtained for the highest values.

\subsection{Observed occurrence rates}
We model the observed occurrence rates of stellar CMEs as a Poisson process. This requires the assumption that the events occur randomly and are independent. For flares, the occurrence rates seem to be sometimes random \citep{Oskanian71, Lacy76, Pettersen89}, but sometimes with periodic components \citep{Pettersen89}. The latter could depend on rotation or cycle. \citet{Vida16} found that strong flares seem to occur at certain phases. For CMEs, however, it is not yet established if they occur randomly due to the low number of detections. Some CME eruptions could be triggered by a preceeding event which destabilized the overlying magnetic field, like sometimes observed on the Sun \citep{Lugaz17}. However, since we aim to model the \textit{observable} CMEs, likely only the most massive event of such related events would be observed. Furthermore, in all observing campaigns up to now, not more than one event was observed, which indicates that \textit{observable} CME events are rare enough that the detected signals come likely from independent sources and overlapping events are very unlikely, in contrast to flare observations.

Both detections and non-detections of stellar CMEs can be used to constrain their intrinsic rates. In case of successful detection of one or more events, the observed event rate $\dot{N}_\mathrm{obs}=N_\mathrm{obs}/t_\mathrm{obs}$ can be estimated simply as the counts per observing time, with uncertainty $\sqrt N_\mathrm{obs}/t_\mathrm{obs}$. However, since most observations of a given star only found one event, this yields an error of 100\% on the rates. Instead of this number we state the 95\% double-sided confidence intervals in case of detections, which are tabulated in \citet{Gehrels86}. In case of one detected event this interval is 0.0253-5.572. For non-detections an upper limit of $\dot{N}_\mathrm{obs}$ can be estimated as described below. To compare these numbers with model predictions of intrinsic stellar CME rates one has to account for the detection limit of the observations, as well as geometric projection effects.

The occurrence rate of a counting experiment is described by the Poisson distribution
\begin{equation}\label{eq:poisson}
f(k,\lambda) = P(X=k,\lambda) = \frac{\lambda^k\exp(-\lambda)}{k!},
\end{equation}
where $P$ is the probability of occurrence, $\lambda$ is the expected value of the random variable $X$, and $k$ is the number of events. In the present application we have some expected value $N=\dot{N}t$ which represents the number of detectable CMEs within the observing time. It depends on the true CME occurrences within the observing time and a factor accounting for the flux detection limit of the observations and projection effects. Using Eq.~\ref{eq:poisson} the probability to detect no CME for an expected $\tilde{N}$ is
\begin{equation}\label{eq:p0}
P(X=0,\tilde{N}) = \exp(-\tilde{N}).
\end{equation}
We can use Eq.~\ref{eq:p0} to express the probability that $\tilde{N}$ is smaller than some value $\tilde{N}_0$ if no CMEs have been observed
\begin{equation}
P(0|\tilde{N}\le\tilde{N}_0) = \int^{\tilde{N}_0}_0 P(X=0,\tilde{N})~d\tilde{N} = 1-\exp(-\tilde{N}_0).
\end{equation}
Since this represents the confidence limit, we can set $P(0|\tilde{N}\le\tilde{N}_0)$ to some value and find the corresponding $\tilde{N}_0$ which then represents an upper limit to $\tilde{N}$ at the given confidence. Adopting a high confidence limit of $CL=95$\%, a non-detection sets an upper limit of $\tilde{N}_0=-\ln(1-CL)\sim3$ detectable CMEs within the observing time. 

A different observational approach was chosen in the study by \citet{Leitzinger14}. Instead of monitoring a single star they obtained multi-object spectroscopy of a sample of young open cluster stars. By assuming that the coeval stars of similar spectral types should have similar expected CME rates $\tilde{N}$ because of similar activity levels, it is possible to combine observations of $n$ stars to obtain a more stringent upper limit in case of a non-detection. The probability distribution for $n$ stars is given by
\begin{equation}
P(X=0,\tilde{N}) = n\exp(-n\tilde{N}),
\end{equation}
because we assume that observing $\tilde{N}$ CMEs on $n$ stars is equivalent to observing $n\tilde{N}$ CMEs on one star. The additional factor $n$ stems from the normalization so that $\int^{\infty}_0P~d\tilde{N}=1$. Hence, for $n$ stars,
\begin{equation}\label{eq:pn0}
P(0|\tilde{N}\le\tilde{N}_0) = 1-\exp(-n\tilde{N}_0)
\end{equation}
and an upper limit can be found from $\tilde{N}_0=-1/n\ln(1-CL)$.

\subsection{Predicting the minimum observing time}\label{sec:minobs}
\begin{figure}
	\centering
	\includegraphics[width=\columnwidth]{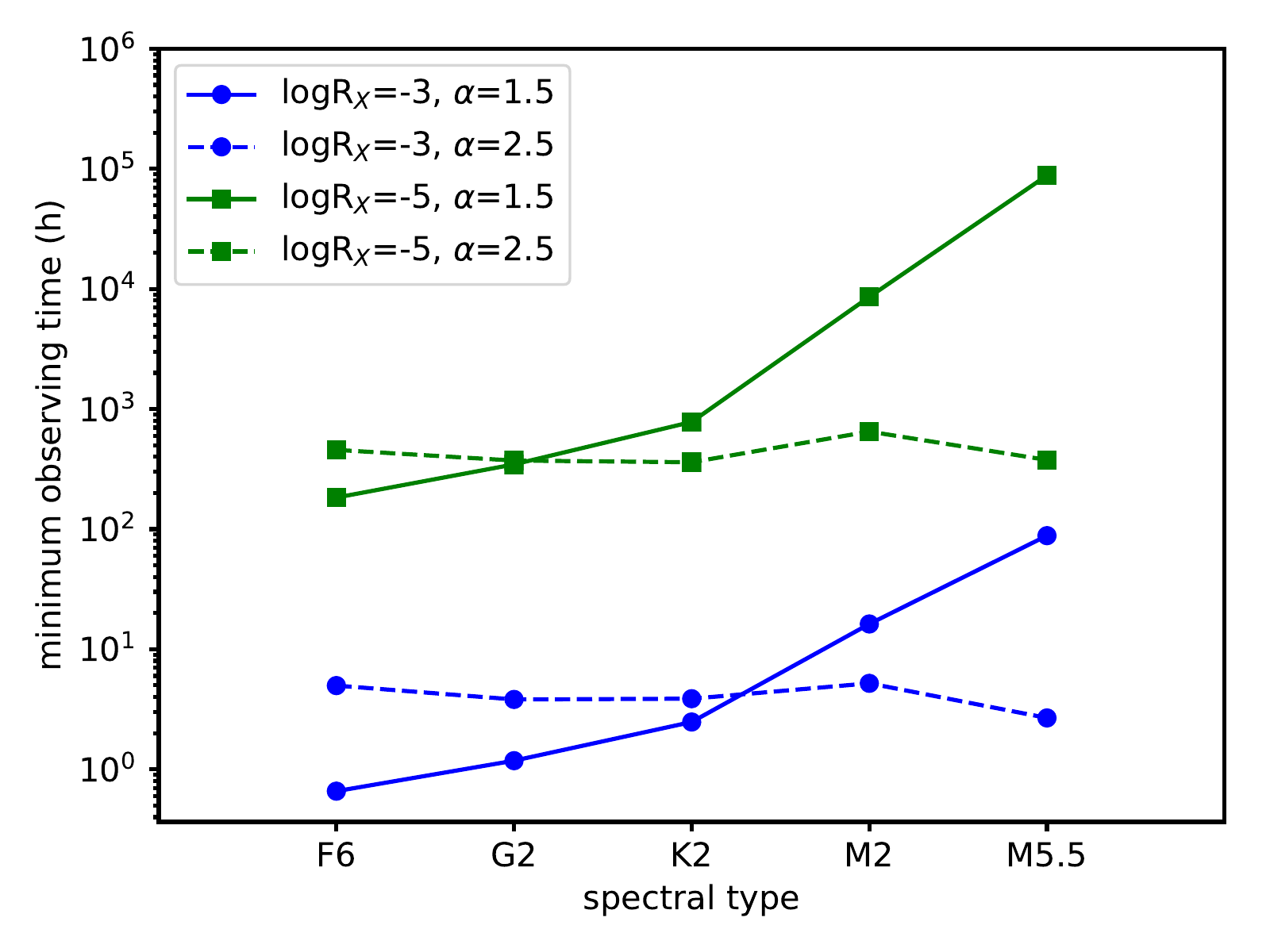}
	\caption{Minimum observing time in hours to detect at least one CME. Two different activity levels of  $\log R_\mathrm{X} = -3$ (blue circles) and $-5$ (green squares) are shown. Same values of the flare power law index $\alpha$ are connected with solid (1.5) and dashed (2.5) lines, respectively.}
	\label{fig:minobs}
\end{figure}

In order to plan observations it is useful to estimate the minimum required observing time for a given star and observational setup to maximize the probability to detect at least one event. Using again the Poisson distribution, the probability to detect at least one event on a single star is
\begin{equation}
P(X\ge1) = 1-P(X=0) = 1-\exp(-N_\mathrm{exp}t_\mathrm{obs}),
\end{equation}
where $t_\mathrm{obs}$ is the total observing time and $N_\mathrm{exp}$ is the expected number of observable events in $t_\mathrm{obs}$. If $P(X\ge1)$ is chosen to be large (e.g. 95\%), one can determine the observing time required to observe at least one event with 95\% probability 
\begin{equation}\label{eq:minobs}
t_\mathrm{obs} \geq \frac{2.996}{N_\mathrm{exp}}.
\end{equation}
The expected observable event rate can be estimated from a model of the intrinsic rate \citepalias[e.g.][]{Odert17} together with a detectability correction factor. In Fig.~\ref{fig:minobs} we show the minimum required observing time for different spectral types and activity levels. The observational parameters correspond to those of Fig.~\ref{fig:int_obs} ($SNR=100$, $t_\mathrm{exp}=5$\,min, $v_\mathrm{lim}=100$\,km\,s$^{-1}$), together with the default CME parameters and effective dilution factor. As expected, lower activity levels ($\log R_\mathrm{X} = -5$) require more observing time than in the very active, saturated case. We do not show here lower activity levels because the estimated observable rates for those are zero for the chosen parameters, and thus the minimum observing time infinite. Another interesting feature is that the flare power law index $\alpha$, i.e. the steepness of the flare energy distribution, strongly affects the minimum observing time. For earlier spectral types, stars with flatter distributions ($\alpha=1.5$) are favored, whereas for M dwarfs, stars with steeper distributions ($\alpha=2.5$) are a better choice. For G--K dwarfs the differences in minimum observing time between the slopes is minimal, with some dependence on activity level. For the steeper distributions, the minimum observing time is rather constant with spectral type, whereas it increases towards cooler stars for the flatter ones. The effect of varying the adopted observational parameters is shown in Appendix~\ref{app:minobs}. Since the minimum observing time is inversely proportional to the expected observable CME rate (Eq.~\ref{eq:minobs}), parameter variations lead to the inverse effects as for the observable rates (cf. Section~\ref{sec:rates}).

\section{Applications}\label{sec:app}
In this section we aim to compare CME occurrence rate estimates from an empirical CME prediction model \citepalias{Odert17} with spectroscopic optical observations, including both data sets with detections and non-detections. We use the sample of 28 late-type main-sequence stars in the young open cluster Blanco-1 which were simultaneously monitored for $\sim$5\,h \citep{Leitzinger14}, as well as observations of two active stars which are known to host prominences, namely HK~Aqr and PZ~Tel, which were monitored for $\sim$18\,h each \citep{Leitzinger16}. Further, we use the $\sim$36\,h of observations from V374~Peg, which include one CME detection \citep{Vida16}. We apply the empirical CME model also to CTTS and WTTS stars in Chamaeleon \citep{Guenther97} which showed during 14\,h of simultaneously observing 36 stars one signature of a CME. Stars with CME detections, as well as stars without such, are important to constrain the CME rates calculated by the empirical model. The constraints are related to the observation parameters $SNR$, spectral resolution, exposure time and total on-source time.

\subsection{Observations of Blanco-1}\label{sec:blanco}
\citet{Leitzinger14} presented observations of the young open cluster Blanco-1 obtained under ESO program ID 089.D-0713(B). Despite its estimated young age (30--145\,Myr) and rich population of low-mass stars they did not detect any CMEs within about 5\,h of simultaneously observing 28 stars, of which 13 are confirmed cluster members. However, four flares were detected, three on confirmed member stars and one on a likely non-member star. The exposure time of the spectra was 180\,s and the spectral resolving power of VIMOS in MOS mode using the orange grism is 2500 which corresponds to a velocity resolution of $\sim$135\,km\,s$^{-1}$. Here, we use this non-detection to infer an upper limit on the detectable stellar CME rates and compare them to the predicted rates. We restrict the analysis to the 13 confirmed cluster members, because they should have similar ages and their X-ray luminosities are known, which are needed as input for the CME prediction model.

Adopting a confidence limit of $CL=95$\%, our non-detection sets an upper limit of $\tilde{N}_0=-\ln(1-CL)\sim3$ observable CMEs per star within $t_\mathrm{obs}$. Assuming that all confirmed 13 cluster members have a similar true value of $\tilde{N}$ we can place the more stringent limit of $\tilde{N}_0=-1/n\ln(1-CL)=0.23$ with $n=13$. However, since the K dwarfs have on average a higher X-ray luminosity which may raise their intrinsic rates, as well as potentially different detectability because of different spectral energy distributions and larger surface areas, we split the sample into 6 K and 7 M dwarfs. The resulting upper limits on $\tilde{N}$ are then 0.5 and 0.43, respectively. This is summarized in Table~\ref{tab:uplim}.

Now we can compare the observed upper limits of $\tilde{N}$ with the intrinsic rates (corrected for detectability). To determine the flux (i.e. mass) limits we use the $SNR$ of the individual spectra given in Table~\ref{tab:blanco}. The geometrical correction is then calculated using an exposure time of 180\,s and a velocity limit of 135\,km\,s$^{-1}$. The total observing time was 4.95\,h. We calculate the expected maximum observable events $N_\mathrm{exp,max}$ per day using Eq.~\ref{eq:ntilde}, the stellar X-ray luminosities given in Table~\ref{tab:blanco} and a range of flare power law indices $\alpha=1.5{-}2.5$. Moreover, we adopt our default CME parameters and the effective dilution factor. The results for the individual stars are given in Table~\ref{tab:blanco}. Using the predicted rates of detectable events, we use Eq.~\ref{eq:p0} to calculate the probabilities that a non-detection occurred despite these non-zero rates. These numbers are also given in Table~\ref{tab:blanco}. We also calculate the predicted averages for the subsamples and compare them to the observational upper limits in Table~\ref{tab:uplim}.

For seven stars, the observational upper limit is larger than the predicted rates, i.e. they are consistent. For two stars, the predictions for the flat power law slope (1.5) are consistent with the observational upper limit, whereas those for the steep slope (2.5) are higher. For the remaining four stars, the predictions for both slopes are higher than the upper limit. The average predicted rates for the subsamples are significantly higher than the upper limits from the observations. The results can be explained if either some of the stars have intrinsically less CMEs than expected, or their true H$\alpha$ fluxes are much smaller than our adopted maximum values, and/or these are diluted too quickly. However, if we allow to vary our default CME parameters and increase the typical column density by a factor of ten, which is still well within the range of stellar values, the obtained observable rates of all stars and subsamples decrease and become consistent with the observations.

\begin{table}
	\caption{Upper limits to CME occurrence rates of stars in the open cluster Blanco-1. The first column gives the considered subsample and the second the number of stars it includes. The upper limits to the number of observable CMEs within the observing time (95\% confidence) are given. The last column gives the expected maximum observable rate for flare power law indices 1.5 and 2.5.}
	\label{tab:uplim}
	\begin{center}
		\begin{tabular}{llcc}
			\hline
			& n & $\tilde{N}$                      & $N_\mathrm{exp,max}$ \\
			&   & ($t_\mathrm{obs}^{-1}$) & ($t_\mathrm{obs}^{-1}$) \\
			\hline
			per star  & 1     & $<$3.0  & see Table~\ref{tab:blanco} \\
			K         & 6     & $<$0.5  & 4.79 - 9.53 \\
			M         & 7     & $<$0.43 & 0.623 - 2.51 \\
			K+M       & 13    & $<$0.23 & 2.54 - 5.75 \\
			\hline
		\end{tabular}
	\end{center}
\end{table}

\subsection{Observations of stars in Chamaeleon}\label{sec:cha}
\citet{Guenther97} simultaneously monitored 18 classical T-Tauri stars (CTTS) and 18 weak-line T-Tauri stars (WTTS) for a total of 14 hours using the multi-object spectrograph FLAIR on the 1.2m UK Schmidt telescope. They detected two flares in two of the WTTS, one accompanied by a likely CME event. We use the same approach as before to set upper limits on observable CMEs for the stars (or subsamples) with non-detections. For the star with one event (DZ~Cha) or subsamples including this star, we calculate the mean rate as $1/t_\mathrm{obs}$, where $t_\mathrm{obs}$ is either the observing time for the star itself (14\,h), or the effective observing time for all $n$ stars of the subsample ($n\times14$\,h). Furthermore, we compute the 95\% double-sided confidence intervals for these rates \citep{Gehrels86}.

We reduce the original sample to 27 stars (10 CTTS, 17 WTTS) for the analysis because of lacking X-ray fluxes for the others. From the non-detection of any events in all but one star within 14\,h we obtain an upper limit on detectable CMEs of $<$5.1 per day for all stars with a non-detection. For DZ~Cha, we obtain a rate of 1.71 (0.04, 9.55) observable CMEs per day. However, if we assume that such young stars may have similar intrinsic rates, we obtain a mean rate of 0.06 (0.0016, 0.354) per day for all 27 stars. If we split the sample into 10 CTTS and 17 WTTS and assume that the intrinsic rates are only similar within these groups, we obtain an upper limit of 0.51 per day for the CTTS and a rate of 0.1 (0.0026, 0.562) per day for the WTTS samples (cf. Table~\ref{tab:ge97} for rates given per $t_\mathrm{obs}$).

Although our empirical CME prediction model \citepalias{Odert17} is based on solar relations we extrapolate it here to the T-Tauri stars of the young ($\sim$2\,Myr) Chamaeleon T1 association. WTTS can be assumed more like main-sequence stars than CTTS although those stars could still have a tenuous disk. CTTS still host disks and are known to have, because of that, variable H$\alpha$ emission. So far very little is known about the CME activity on such stars. By applying our model to pre-main sequence stars one has to be cautious that the flare--CME association rate on those stars might differ significantly from the solar one. We limit the sample to 27 stars with known X-ray luminosities to facilitate comparison with the model. The X-ray luminosities were computed using the count rates compiled by \citet{Guenther97}, a count-to-flux conversion factor \citep{Schmitt04} and a distance of 140\,pc. We assumed a radius of $2R_{\sun}$ for all stars to account for their pre-main sequence state \citep{Guenther97}. As limiting velocity we adopt 120\,km\,s$^{-1}$according to the quoted 2.6\,\AA\ dispersion per pixel, as well as an exposure time of 200\,s \citep{Guenther97}.

The predicted maximum observable rates for all stars are given in Table~\ref{tab:cmestarsg97}. The predicted rates are lower than the observational upper limit for all considered stars, except for two cases if assuming a flat flare distribution ($\alpha{=}1.5$). This means that the non-detection of CMEs on most stars can likely be explained by observational limitations. For DZ~Cha, the predicted rate for the steep flare power law index is below the observationally determined confidence interval, for the flat index it is consistent. Despite the good agreement with our modeling, we caution again that it is not clear up to now how reliable the estimated intrinsic rates based on solar extrapolations are if extrapolated to T-Tauri stars.

\begin{table}
	\caption{CME occurrence rates of stars in Chamaeleon. The columns are similar to Table~\ref{tab:uplim}. The first line corresponds to the upper limit for stars with no detection. For samples including the detected event, the 95\% double-sided confidence limits are given in parentheses.}
	\label{tab:ge97}
	\begin{center}
		\begin{tabular}{llcc}
			\hline
			& n & $\tilde{N}$ & $N_\mathrm{exp,max}$ \\
			&   & ($t_\mathrm{obs}^{-1}$)   & ($t_\mathrm{obs}^{-1}$) \\
			\hline
			per star & 1  & $<$3                 	& see Table~\ref{tab:cmestarsg97} \\
			DZ Cha   & 1  & 1 (0.0253 - 5.572)     	& 0.318 - 0.004\\
			CTTS     & 10 & $<$0.3                	& 0.034 - 0.0015\\
			WTTS     & 17 & 0.059 (0.0015 - 0.328)  & 0.865 - 0.0180\\
			all      & 27 & 0.037 (0.0009 - 0.206) 	& 1.046 - 0.0316\\
			\hline
		\end{tabular}
	\end{center}
\end{table}

\subsection{The fast-rotating M dwarf V374 Peg}\label{sec:v374}
Here we use the observations of the young active M dwarf V374~Peg from the Canada-France-Hawaii Telescope (CFHT) obtained with the ESPaDOnS spectrograph. This data set covers several flares, including one complex event associated with a mass ejection. This CME was preceded by two Doppler-shifted emissions with smaller velocities, possibly failed filament eruptions \citep{Vida16}. The CME event is clearly visible in the first four Balmer lines (H$\alpha$-H$\delta$), but we choose the H$\alpha$ line for computing our predicted rates. We further adopt the X-ray luminosity from \citet{Schmitt04}\footnote{\url{http://www.hs.uni-hamburg.de/DE/For/Gal/Xgroup/nexxus/nexxus.html}} and the stellar radius given in \citet{Vida16}. All model input values including the observational parameters of the used data set (average $SNR$, total observing time, exposure time of the spectra) are summarized in Table~\ref{tab:starscme}. The observationally determined CME rate of 0.67 (0.017, 3.74) per day is consistent with the maximum number of observable CMEs predicted by our model.

\begin{table*}
\caption{Stars with long high-resolution spectral time series. Spectral types were taken from Simbad\protect\footnotemark, references for X-ray luminosities and stellar radii are given in the text. Columns 5-7 give the average $SNR$ of the used data sets, the total observing time in hours, and the exposure time of the spectra. The last three columns show the rates or upper limits determined from the observations, the expected maximum observable CME rate, and the probability for a non-detection within the observing time.}
\label{tab:starscme}
\begin{center}
\begin{tabular}{llllllllll}
\hline
ID & spectral & $\log L_\mathrm{X}$     & $R_\star$    & $SNR$ & $t_\mathrm{obs}$  & $t_\mathrm{exp}$ & $\tilde{N}$                      & $N_\mathrm{exp,max}$             & $P(0)$ \\
   & type     & (erg\,s$^{-1}$) & ($R_{\sun}$) &     & (h)               & (min)            & ($t_\mathrm{obs}^{-1}$) & ($t_\mathrm{obs}^{-1}$) & (\%) \\
\hline
V374 Peg & M3.5Ve & 28.36 & 0.34 & 40 & 35.75 & 5  & 1 (0.0253 - 5.572) 	& 0.27 - 1.42 & 76.3 - 24.2 \\
PZ Tel   & G9IV   & 30.39 & 1.23 & 90 & 19.17 & 5  & $<$3            		& 20.6 - 3.26 & $<$0.1 - 3.82 \\
HK Aqr   & M0Ve   & 29.19 & 0.59 & 60 & 18.33 & 10 & $<$3            		& 0.33 - 0.67 & 71.8 - 51.4 \\
\hline
\end{tabular}
\end{center}
\end{table*}

\subsection{Observations of prominence stars}\label{sec:prom}
Prominences, and possibly also their eruptions, were searched for in a spectral time series obtained under ESO proposal ID 089.D-0709(A) of the young and fast rotating late-type stars HK~Aqr and PZ~Tel by \citet{Leitzinger16}. We refer to this paper for a detailed description of the data, the parameters relevant for the present study are given in Table~\ref{tab:starscme}. Both stars were previously known to host prominences and several of them were identified in the observations. However, no signatures of prominence eruptions/CMEs were found, although the on-source time of each target was $\gtrsim$18\,h each and both stars are very active. The upper limits of detectable CMEs per day from the non-detection are $<$3.75 and $<$3.92 with 95\% confidence. We compare these numbers with the predicted rates calculated with the parameters given in Table~\ref{tab:starscme}. For HK~Aqr, the maximum observable CME rate predicted by our model is consistent with the upper limit from the observations. For PZ~Tel, it is higher by a factor seven for the flat flare power law index, but only marginally higher for $\alpha{=}2.5$.

In this specific data set, however, we consider it more likely that mass ejections are preferably seen in absorption on these stars. This is based on the fact that we see the existing prominences only in absorption when they transit the stellar disk, but do not observe their emission as they rotate off the disk. Higher signal-to-noise would apparently be required to observe their emission signatures \citep{Leitzinger16}. Only if significant Doppler brightening would have occurred during ejection the detection of their emission signals could have been feasible in the given data. The observed prominences of PZ~Tel are located at heights $\gtrsim0.5R_\star$ from the surface \citep{Leitzinger16}. If this is representative for observable prominences on PZ~Tel, this would lead to smaller dilution factors. We find that initial prominence heights ${\ge}1.5R_\star$ could explain the non-detection even for $\alpha{=}1.5$ if the adopted intrinsic CME rates are true. On the other hand, column densities higher than $5\times10^{20}$\,cm$^{-3}$ would also lead to consistent results. Such heights and column densities are not untypical for the known prominence stars.

Another interesting fact is that none of the \textit{observed} prominences were seen to erupt. However, several prominences were re-observed in subsequent nights in case of rotational phase overlap. This indicates that these prominences may have lifetimes in the order of days, which makes observation of their ejection unlikely in the given data set of six nights. In addition, geometrical constraints would allow to observe their ejection only during the times they transit the stellar disk. A rough estimate of the geometrical probability can be obtained as follows (again, ignoring the spatial extent of the structures). The prominences are located at a height $R_\mathrm{p}$ from the center of the star and rotate rigidly with it. Hence, they move on a circle with radius $R_\mathrm{p}\cos\theta$, where $\theta$ is the latitude (cf. Fig.~\ref{fig:geo}, upper panel). The parameter $R_\mathrm{obs}=R_\mathrm{p}\cos\theta$ can be directly measured from the temporal evolution of the radial velocity shifts of the prominence absorption profiles \citep[cf.][]{Leitzinger16}. The circle on which they move is located at a height $z=R_\mathrm{p}\sin\theta$ from the stellar equatorial plane. The fact that the prominences are seen in absorption against the stellar disk restricts their maximum latitude to $\theta_\mathrm{max}=\arctan(R_\star/R_\mathrm{obs})$, corresponding to $z=R_\star$. Note that we ignore the correction for stellar inclination in these calculations, because both HK~Aqr and PZ~Tel have likely an inclination relatively close to 90\degr \citep[cf.][]{Leitzinger16}. The geometric probability can be estimated as
\begin{equation}\label{eq:prob}
p = \frac{2\phi}{2\pi} = \frac{\arcsin\left[\sqrt{1-\left(\frac{R_\mathrm{obs}}{R_\star}\right)^2\tan^2{\theta}}/\left(\frac{R_\mathrm{obs}}{R_\star}\right)\right]}{\pi},
\end{equation}
where $2\phi$ is the longitude range during which a prominence transits the stellar disk. Note that Eq.~\ref{eq:prob} slightly overestimates the geometric probability, because it does not take into account a more restricted longitude range due to the on-disk time (Eq.~\ref{eq:tod}). If inserting typical observed values of $R_\mathrm{obs}$ and choosing some $\theta<\theta_\mathrm{max}$ (e.g., $\theta=\theta_\mathrm{max}/2$), one obtains values of $p$ around $\lesssim30$\%. This indicates, together with the lifetime in the order of days, that it is rather unlikely to have observed the ejection of the visible prominences during the observations.

\footnotetext{\url{http://simbad.u-strasbg.fr}}

\section{Discussion}\label{sec:disc}
Comparison of spectroscopic time series with estimated stellar CME rates observable in Balmer lines suggest that our model tends to predict higher event rates than what is actually observed. This can mainly have two reasons: 1) the intrinsic CME rates of the studied stars are significantly smaller than the predictions of \citetalias{Odert17}; 2) the Balmer line signals are significantly overestimated; or a combination of 1) and 2). As discussed in \citetalias{Odert17}, the intrinsic CME rates could indeed be overestimated for active stars mainly because it is unknown if the solar flare-CME association rate can be extrapolated to stars with much higher than solar activity levels. The stronger magnetic fields of these stars could lead to increased confinement, possibly allowing only the most energetic events to escape from the star \citep{Drake16, Alvarado-Gomez18,Alvarado-Gomez19a}. Furthermore, the modeled intrinsic CME rates of active stars are also higher than present constraints from stellar mass-loss measurements \citepalias[see][]{Odert17}. This indicates that, at least for active stars, the intrinsic CME rates could be lower than predicted by our empirical model. On the Sun, failed filament eruptions are sometimes observed during flares where material first rises, but then decelerates and may eventually fall back to the Sun \citep[e.g.][]{Ji03, Mrozek11}. The second point related to the overestimation of Balmer line signals is also likely true, as we specifically aimed to estimate the \textit{maximum} CME signal in Balmer lines to find an estimate of the \textit{minimum} required $SNR$ level to detect them. Below we discuss these issues in more detail.

\subsection{Estimated fluxes}
There are several processes which could lead to significantly weaker fluxes. First, we assume that the CME masses are equivalent to the neutral hydrogen mass. In reality, the mass of a CME consists of both the cool prominence material and ionized plasma at coronal temperatures. Solar observations of individual events suggest that prominence masses are roughly of the same order or factors of a few lower than CME masses \citep{Kuzmenko17, Lee17d}. CME masses are often larger because additional coronal material is included. For instance, in the event analyzed by \citet{Koutchmy08} the total CME mass was about a factor of five larger than the prominence mass. Typical errors in mass determination for both structures are about a factor of two \citep{Koutchmy08, Vourlidas10}. On the other hand, sometimes the prominence masses are even larger than the CME masses, because some prominence material drains back to the stellar surface and the CME includes then only the remaining part of the prominence matter \citep{Kuzmenko17}. Moreover, prominences themselves are not completely neutral; at their temperatures of about 7500--9000\,K \citep{Parenti14}, they are partially ionized, mainly due to photoionization by the Lyman and Balmer continua for typical prominence densitites. The ionization degree of solar prominences is not very well constrained; prominence models predict a large range between a few per cent and $\gtrsim$1 for the ratio of electron to total hydrogen density \citep{Gouttebroze93}. If similar numbers would be possible for other stars, we may significantly overestimate the neutral masses if the prominences are (almost) completely ionized.

Second, we assume that the structures have the same optical thickness for all masses. A larger range of optical thicknesses would be possible, also depending on the considered spectral line. For instance, prominences on Speedy Mic (masses ${\sim}10^{17}$\,g) have been found to be optically thick at least up to H$\delta$ \citep{Dunstone06a}.

Third, we assumed that the source function is dominated by scattering and neglect possible contributions of collisional excitation. In case of pure scattering, the effect of Doppler dimming/brightening is most important. Depending on the incident stellar spectrum illuminating the prominences, their signals may be enhanced or weakened compared to their appearance in a non-moving state \citep{Heinzel87, Gontikakis97, Gontikakis97a, Labrosse08, Labrosse10}, because moving prominences are illuminated with a Doppler-shifted stellar spectrum. For instance, the Lyman-$\alpha$ line of a prominence is dimmed with increasing radial velocity because the stellar line is an emission line and the prominence ``sees'' only its wings. Higher Lyman-lines and Balmer lines can be either dimmed or brightened depending on the radial velocity. For instance, in the model of \citet{Heinzel87} the H$\alpha$ intensity increases for radial velocities up to 150\,km\,s$^{-1}$, but decreases again for higher speeds. The initial increase can be explained because the solar H$\alpha$ line is in absorption and the incident radiation is that of the wings; the decrease for higher velocities is caused by the decreased population of the second level due to the Lyman-$\alpha$ dimming \citep{Gontikakis97}. For stars other than the Sun, the Balmer lines can also be in emission, e.g. on dMe stars, which will have a different net effect compared to a solar-like spectrum. However, no detailed modeling of Doppler dimming/brightening for stars other than the Sun has been performed yet. We do account for this effect in a simplified way by assuming that the Balmer line source function is determined by illumination from the stellar continuum close to the chosen lines, not from the core of the stellar lines (cf. Section~\ref{sec:flux}). We neglect, however, the possible difference of the illuminating (red) continuum region entering the source function and the blue continuum corresponding to the projected velocity. As the source function of an erupting prominence is determined by integrating the radiation of the stellar-disk continuum from all lines of sight (taking into account the projection of its velocity vector), this results in a complex dependence on the geometry, which is generally only known for the Sun where spatially resolved observations are available. The Doppler dimming/brightening effect also has some dependence on the prominence plasma parameters. An additional signal dilution for moving structures occurs due to the increasing height during propagation, which lowers the incident stellar radiation. Due to this complex behavior, detailed NLTE modeling for different stellar conditions, geometries and plasma parameters would be necessary to obtain improved constraints on the expected Balmer signals, which we plan for future work. For hotter prominences and/or prominences with an extended prominence-to-corona transition region, the added contribution of collisional excitation could also increase their emission as compared to cooler prominences \citep{Labrosse08}.

Another relevant point in this context is that CMEs are expected to occur in close temporal and spatial vicinity to flares. On the Sun, most large CME events are associated with a flare and/or a prominence eruption \citep[e.g.][and references therein]{Webb12}. For small events the association between the different phenomena is not so clear, but may be at least partly due to observational biases. Therefore, for the most energetic events that may be potentially observable on stars, the prominences could be illuminated by the increased flare radiation, which may brighten their signals, at least temporarily. However, the flare may also broaden the stellar Balmer line wings and/or raise the whole continuum level. This could slightly raise the required $SNR$ to detect the CME signal superimposed on the flaring stellar spectrum. However, if such an effect could be visible also depends on the duration of the flare relative to the CME, as well as the area of the flaring region relative to that of the stellar disk.

Comparison with the observational data of prominence stars shows that the non-detection of CMEs can be explained if we consider only absorption features to be observable, based on the fact that the prominences on the considered stars are not visible in emission off-disk. Since stellar prominences are often located at distances of up to several radii from the star, the non-detection of emission signals is consistent with a pure scattering source function because of the strong dilution of the incident radiation from the star \citep{CollierCameron89}. If the typical prominence heights for young active stars are similar to those found for the known fast-rotating prominence stars, adopting such values could indeed result in better agreement with observations, as demonstrated for PZ~Tel (Section~\ref{sec:prom}). The stable locations where prominences can form on stars with different parameters were recently evaluated by \citet{Jardine19}.

\subsection{Geometric effects and duration}
To calculate the effect of velocity projection, we adopted the solar flare energy -- CME velocity relation \citep{Drake13}. As already discussed by these authors, it is possible that the relation cannot be extrapolated to events much more energetic than on the Sun, as these would lead to extremely high kinetic energy losses on active stars. Recent modeling results suggest that CME velocities could be lower than expected due to the stronger magnetic fields on such stars \citep{Alvarado-Gomez18}. A recent analysis of 15 stellar CME events from the literature also suggests lower speeds than what would be expected from the solar kinetic energy -- flare energy relation \citep{Moschou19}. Therefore, we may overestimate the CME velocities using Eq.~\ref{eq:vm}, and therefore underestimate the reduction of observable events due to velocity projection for active stars.

Furthermore, we assume that the CMEs eject radially from the star. Although radial motion is frequently observed in solar events, deflection may occur which alters the direction of propagation. This depends on the size, mass and velocity of a CME (wide, slow, low-mass CMEs are deflected most), as well as on the strength and gradient of the background magnetic field, both global (heliospheric current sheet, coronal holes) and local (active regions). \citet{Kay15a} found that CMEs are preferably deflected towards the heliospheric current sheet, the region of minimum magnetic pressure, and that for stronger background fields the deflection occurs at lower radii. Since deflection is less important for massive and fast CMEs, i.e. those we may actually observe on other stars, it is probably less important for such events. However, the strong magnetic fields and different field geometries on active stars could lead to significantly non-radial motions for lower-mass events which may affect estimations of impact rates on planets \citepalias[e.g.][]{Odert17}. This will depend both on the global magnetic field properties, as well as on the source locations of stellar CMEs. \citet{Kay16} modeled the deflection of CMEs in the young fast-rotating M dwarf V374~Peg (cf.~Section~\ref{sec:v374}). They found that CMEs move towards the astrospheric current sheet, where the less massive ones can even be trapped. This means that around V347~Peg or similar stars, the ejection locations could be more likely around the current sheet. Our model can account for this by specifying a restricted range of source locations. However, the magnetic field structure of the star, as well as its inclination relative to the observer have to be known to evaluate if this deflection would lead to an increase or decrease in observable CME. We note that for applying our model to larger samples of stars, like done in Sections~\ref{sec:blanco} and \ref{sec:cha}, such effects likely cancel out due to the random orientation of the stars relative to the observer. For V374~Peg, the good agreement between our model and the observations does not indicate a significant increase or reduction of observable CMEs because of this effect. However, much longer observations to obtain better statistics would be needed to confirm this.

For our simple estimations, we ignored the spatial extent and shape of the prominence structures. This can have several effects on our estimated event rates. First, large extended structures may not or only be partly optically thick, so their Balmer signals may be smaller (but not necessarily, as this depends also on other plasma parameters). Second, for very large extended structures the simple division into emission and absorption signals may not be valid, as they may partly cover the stellar disk and partly be located off-disk. In such cases, the emission and absorption signals could partly cancel (if the emitting and absorbing parts would have different velocity distributions, complex line profile shapes may result). Third, CMEs expand during propagation. On the Sun, self-similar expansion is observed for many CME events \citep[e.g.][and references therein]{Maricic09}. Expansion leads to a decrease of density with time, which could limit the duration during which a signal would be detectable. Generally, expansion of plasma may also lead to cooling; however, on the Sun, an increase of CME temperatures has been observed during the propagation and expansion of some CMEs, indicating the presence of an additional heating source \citep{Pagano14, Susino16}. If the physical processes involved in flare/CME eruptions on stars are indeed similar to the Sun, expansion may therefore not necessarily lead to a reduced signal.

A further caveat is that prominences/CMEs on the Sun are not homogeneous structures, but exhibit a great amount of finestructure in high-resolution images \citep{Susino18}. Thus, such expanding 3D structures may have complex superimposed velocity components, and may therefore produce very broad lines with small peak enhancements or asymmetric line profiles, in contrast to our assumptions. Even in a rather simple case, expansion could induce non-negligible velocity components perpendicular to the radial propagation direction. On the Sun, the expansion velocity is related with the radial propagation and depends on the CME width \citep{Gopalswamy09a, Gopalswamy12a}. The expansion speeds can be in the order of the radial speeds. Therefore, if the solar relations hold also for CMEs on stars, velocity projection would not be a big issue because at some angle from the line-of-sight the expansion velocity would provide sufficient Doppler shift to be detected. This means that in stellar spectra, the projected components may be dominated by the expansion, which indicates that even CMEs originating far from the disk center could be observed. However, since expansion is isotropic around the radial propagation direction, it broadens the profile, whereas the radial component shifts it away from the stellar line. Thus, the broad CME profiles seen in the events of e.g. \citet{Houdebine90} could be due to combined effects of propagation and expansion, where the maximum line-of-sight velocity would not be the maximum propagation velocity, but added components of radial motion and expansion. On the other hand, if expansion is less relevant, such broad profiles could also stem from a superposition of narrower emission profiles strongly accelerated during the exposure time. In this case the radial velocity component would be closer to the maximum measured value than in the expanding case. In reality, however, both processes may work together.

Expansion leads also to another issue, namely the signal duration. We assume here that any signal lasts long enough to be detected in at least one spectrum of a time series. Therefore, the excess flux must remain sufficiently high at least during a timescale comparable to the exposure time of one spectrum. We aimed to account for this by introducing the effective dilution factor (Eq.~\ref{eq:wav}). In the data sets used here, exposure times are between 3 and 10 minutes. At a velocity in the order of 1000\,km\,s$^{-1}$, a CME travels almost a solar radius within 10\,min. We do account for the time an erupting prominence remains in front of the stellar disk in case of absorption signals, but neglected possible further limits in signal duration. For instance, expansion may reduce the column density, which is important for both emission and absorption features. In a self-similarly expanding shell the density drops as $r^{-3}$ \citep{Howard18} and the column density by $r^{-2}$ if the CME is directed towards the observer (the evaluation of the column density for some non-zero propagation angle would be more difficult to determine). If we adopt such a decrease, the density of the structure could drop by up to a factor of 10 in the example above within the exposure time of one spectrum, and the column density by more than a factor of three. For even higher speeds and shorter exposure times, the CMEs could become optically thin too quickly to be observable. On the other hand, the increasing area of the structure could partly compensate for this. Another process that could also further reduce the Balmer signal duration is, as has been mentioned before, photoionization. However, with increasing distance from the star during propagation, photoionization of hydrogen decreases.

\section{Conclusions}\label{sec:sum}
By estimating the maximum Balmer signals of erupting prominences/CMEs around stars of different stellar spectral types we find that detection on mid- to late-type M~dwarfs is favored. In this case, detection of CMEs would be feasible with moderate $SNR$ values with current telescopes and instruments. Moreover, emission signals have less stringent geometrical limitations compared to absorption signals that are only visible as long as the event is seen against the stellar disk. However, visibility of emissions is restricted to high-mass CMEs, which are rarer. We combine our simple radiative transfer calculations with a model for intrinsic CME rates \citepalias{Odert17} to compare with spectroscopic observations, thereby also taking into account a further reduction of observable events by geometric projection effects and their duration in case of absorption signals. Comparison with upper limits or detections in different spectroscopic data sets of young active stars reveals that our expected observable rates are in most cases consistent with the observations, but in some cases higher. Therefore, either the intrinsic CME rates from the model of \citetalias{Odert17}, or the Balmer signals are overestimated in these cases. The latter seems likely because we specifically aimed at estimating the maximum possible signal to obtain the minimum $SNR$ required for detection. Specifically, we can reproduce the observed CME rates of these stars by increasing the column density by a factor of ten, which is still well within the range of stellar prominence observations. We also estimate the minimum required observing times to detect a stellar CME event in optical spectroscopic observations. The results show that one needs to focus on active stars, as for moderately active stars ($\log R_\mathrm{X}{\sim}-5$) the required observing times would already be at least a few hundred hours, even with our likely overestimated observable CME rates. Although the predicted intrinsic CME rates decrease towards cooler stars for the same activity level because of the correspondingly lower X-ray luminosities, the fraction of observable CMEs becomes larger for later spectral types because their detectability increases.

\section*{Acknowledgements}
We thank the referee for valuable comments that helped to improve the paper. This work was supported by the Austrian Science Fund (FWF) project P30949-N36. PO and ML further acknowledge the Austrian Space Applications Programme of the Austrian Research Promotion Agency FFG (ASAP-14 865972). PH, PO and ML acknowledge support from the Czech Science Foundation, grant 19-17102S.

This study is partly based on observations collected at the European Organisation for Astronomical Research in the Southern Hemisphere under ESO programmes 089.D-0713(B) and 089.D-0709(A). This research has made use of the SIMBAD database, operated at CDS, Strasbourg, France.

\bibliographystyle{mnras}
\bibliography{mybib} 




\appendix

\section{Signal-to-noise ratio}\label{app:snr}
\begin{figure*}
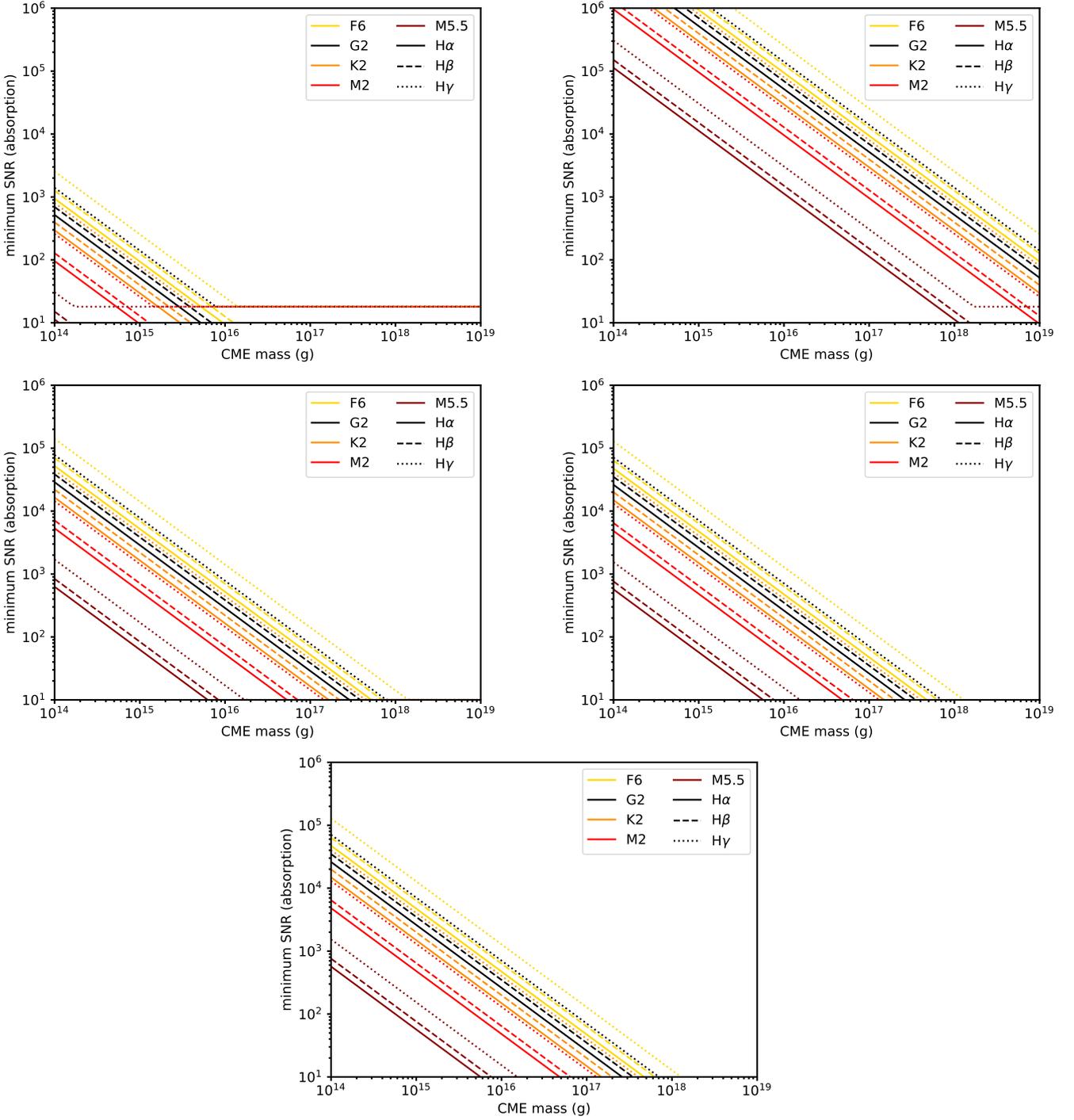

	\centering
	\includegraphics[width=\columnwidth]{{{snr_abs_tau10_logN18_wdil0.5}}}\hfill
	\includegraphics[width=\columnwidth]{{{snr_abs_tau10_logN22_wdil0.5}}}
	\includegraphics[width=\columnwidth]{{{snr_abs_tau10_logN20_wdil0.1}}}\hfill
	\includegraphics[width=\columnwidth]{{{snr_abs_tau10_logN20_wdil0.01}}}
	\includegraphics[width=\columnwidth]{{{snr_abs_tau10_logN20_wdil0.005}}}
	\caption{$SNR$ for absorption features - dependence on parameters. Upper row: $W=0.5$ and column density of $10^{18}$ (left) and $10^{22}$\,cm$^{-3}$ (right); middle row: column density of $10^{20}$\,cm$^{-3}$ and $W=0.1$ (left) and $W=0.01$ (right); lower row: column density of $10^{20}$\,cm$^{-3}$ and $W=0.005$. Horizontal lines for high masses indicate where $\mathcal{N}_\mathrm{H}$ was increased so that $A_\mathrm{p}$ does not exceed $0.3A_\star$.}
	\label{fig:app_snr_abs}
\end{figure*}

In Fig.~\ref{fig:app_snr_em}, we show how the $SNR$ for emission features depends on the different parameters. In Fig.~\ref{fig:app_snr_abs}, we show the same for absorption signatures.

\begin{figure*}
	\centering
	\includegraphics[width=\columnwidth]{{{snr_em_tau10_logN18_wdil0.5}}}\hfill
	\includegraphics[width=\columnwidth]{{{snr_em_tau10_logN22_wdil0.5}}}
	\includegraphics[width=\columnwidth]{{{snr_em_tau10_logN20_wdil0.1}}}\hfill
	\includegraphics[width=\columnwidth]{{{snr_em_tau10_logN20_wdil0.01}}}
	\includegraphics[width=\columnwidth]{{{snr_em_tau10_logN20_wdil0.005}}}
	\caption{$SNR$ for emission features - dependence on parameters. Upper row: $W=0.5$ and column density of $10^{18}$ (left) and $10^{22}$\,cm$^{-3}$ (right); middle row: column density of $10^{20}$\,cm$^{-3}$ and $W=0.1$ (left) and $W=0.01$ (right); lower row: column density of $10^{20}$\,cm$^{-3}$ and $W=0.005$. Horizontal lines for high masses indicate where $\mathcal{N}_\mathrm{H}$ was increased so that $A_\mathrm{p}$ does not exceed $0.3A_\star$.}
	\label{fig:app_snr_em}
\end{figure*}

\clearpage

\section{Geometrical correction factor}\label{app:geo}
In Fig.~\ref{fig:app_geo}, we show how the geometrical correction factor depends on the different parameters. Relative to Fig.~\ref{fig:geo}, we vary one of the default parameters. In Fig.~\ref{fig:app_geo_wav}, we show how the geometrical correction factor depends on the different parameters if assuming an effective dilution factor.

\begin{figure*}
	\centering
	\includegraphics[width=\columnwidth]{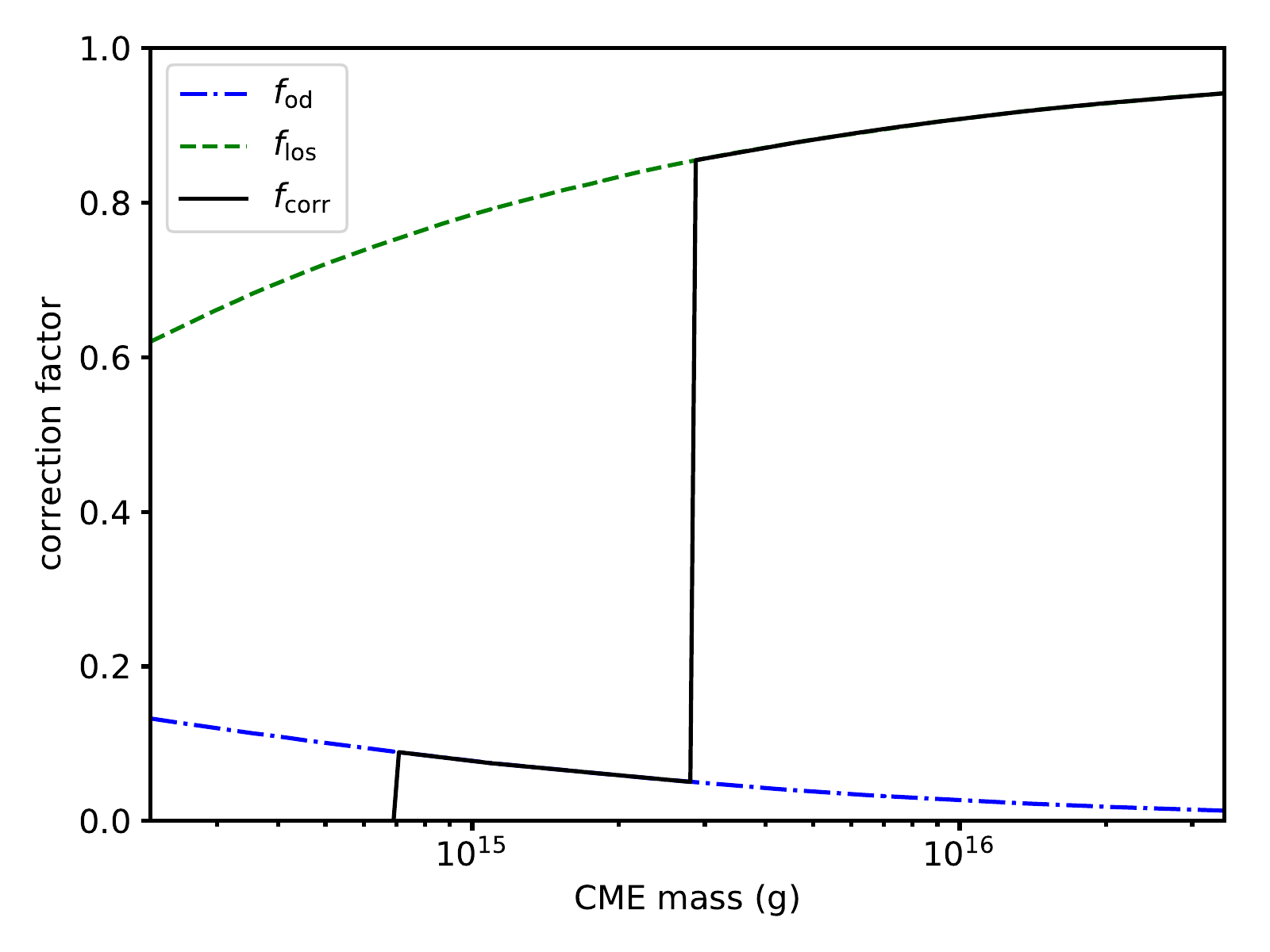}\hfill
	\includegraphics[width=\columnwidth]{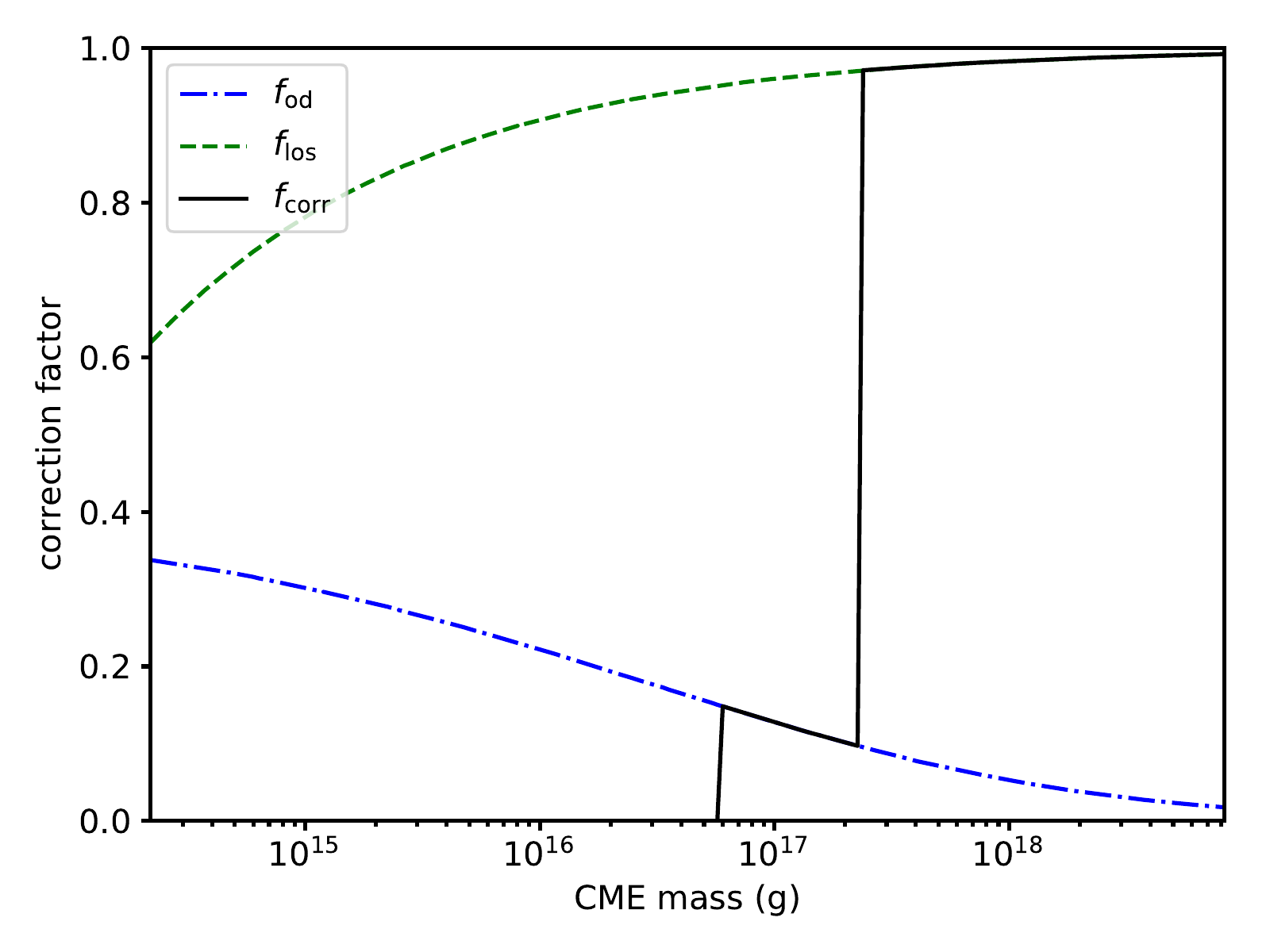}
	\includegraphics[width=\columnwidth]{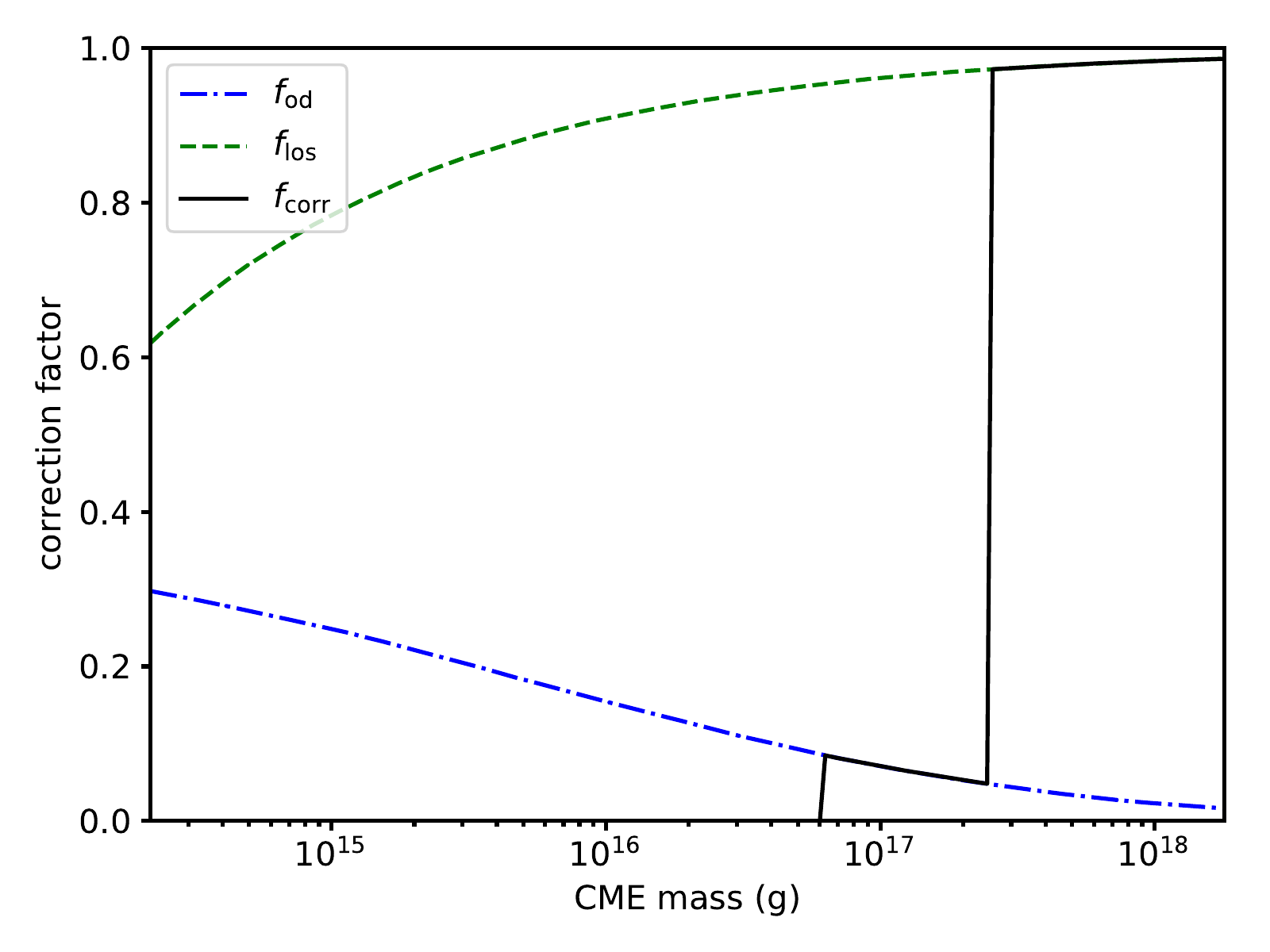}\hfill
	\includegraphics[width=\columnwidth]{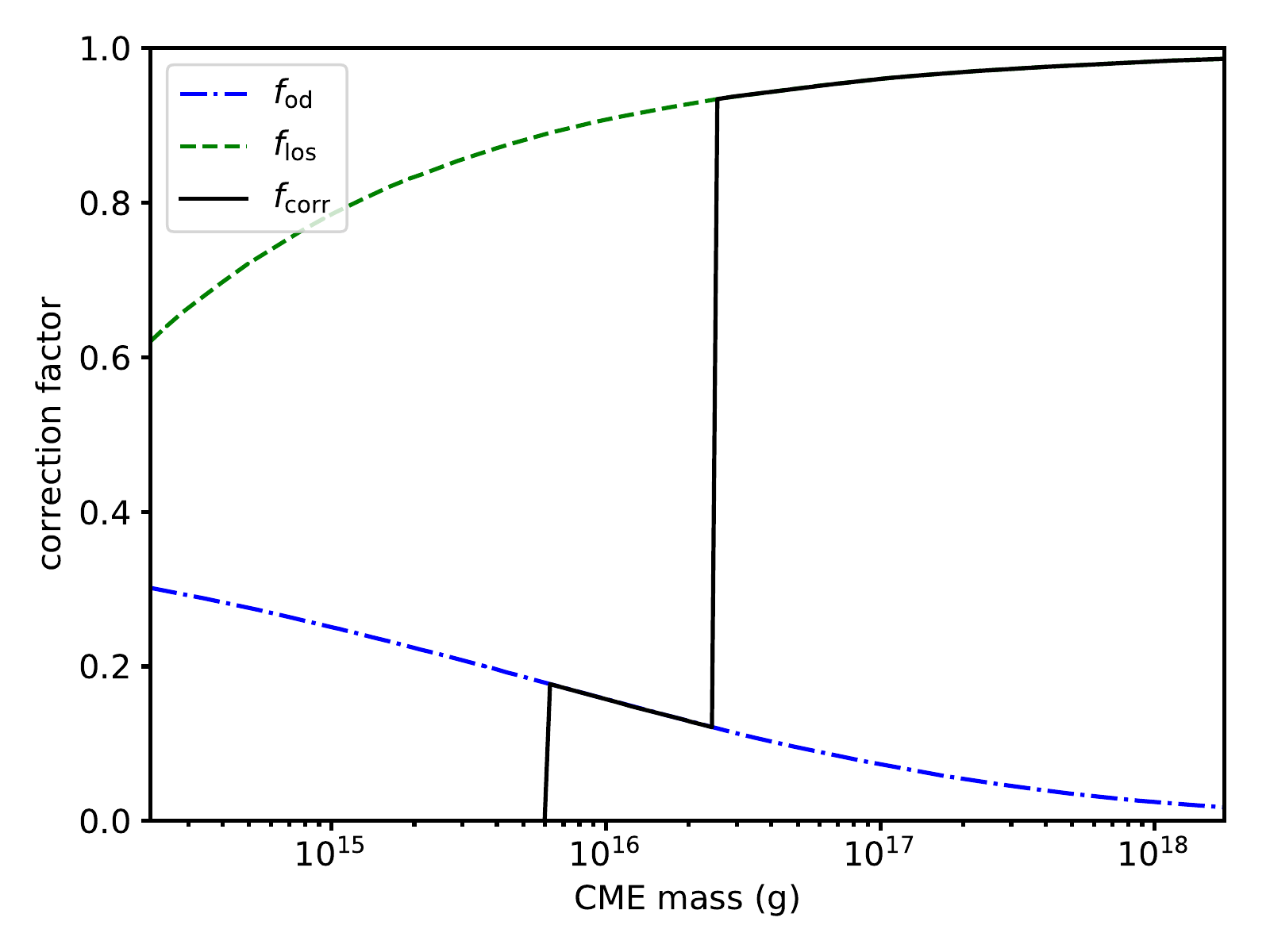}
	\includegraphics[width=\columnwidth]{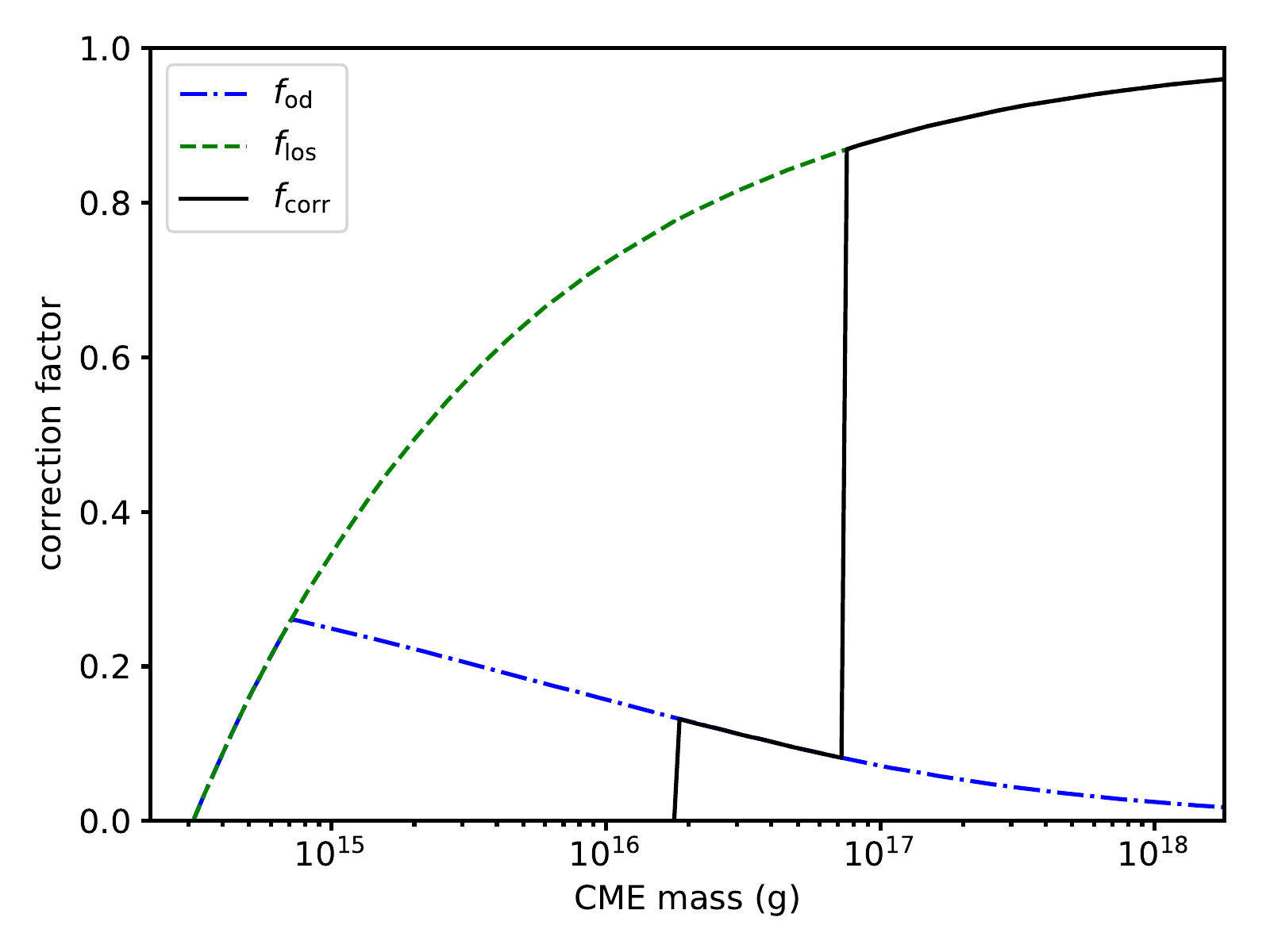}\hfill
	\includegraphics[width=\columnwidth]{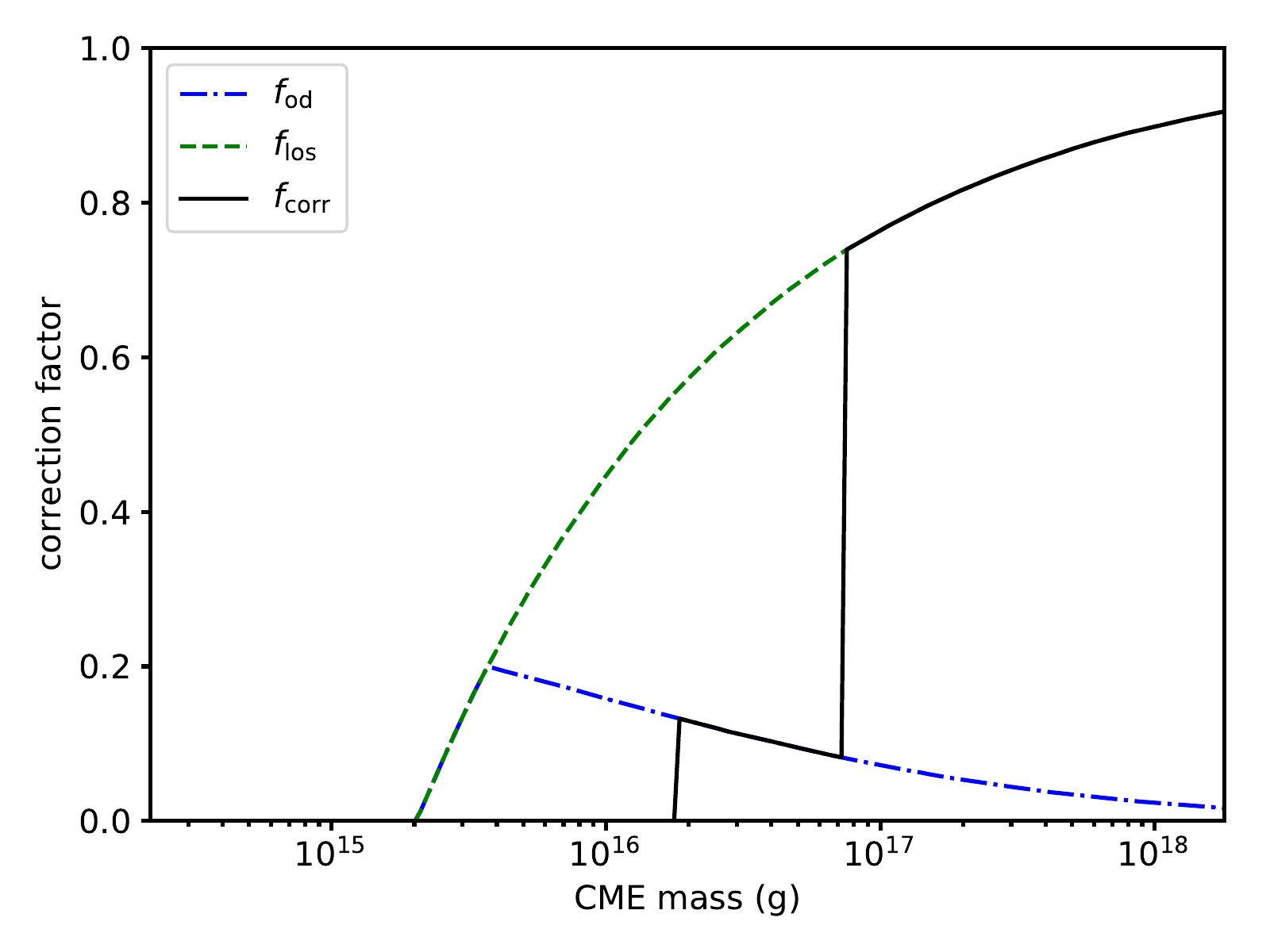}
	\caption{Geometrical correction factor for fixed dilution factor $W=0.2$ - dependence on parameters. Upper row: spectral types M5.5 (left) and F6 (right); middle row: $SNR$ of 30 (left) and 300 (right); lower row: velocity limit of 300\,km\,s$^{-1}$ (left) and 600\,km\,s$^{-1}$ (right).}
	\label{fig:app_geo}
\end{figure*}

\begin{figure*}
	\centering
	\includegraphics[width=\columnwidth]{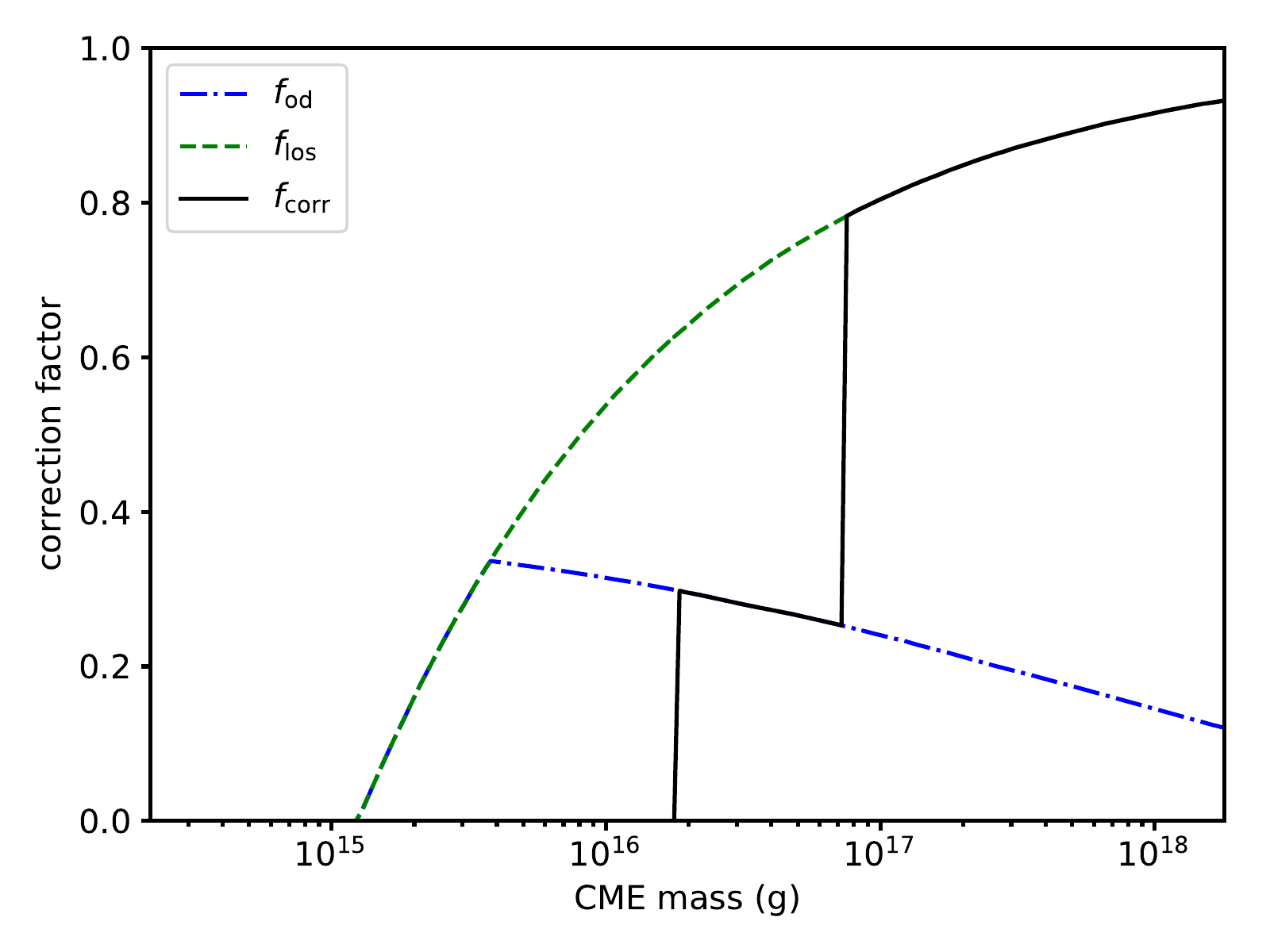}\hfill
	\includegraphics[width=\columnwidth]{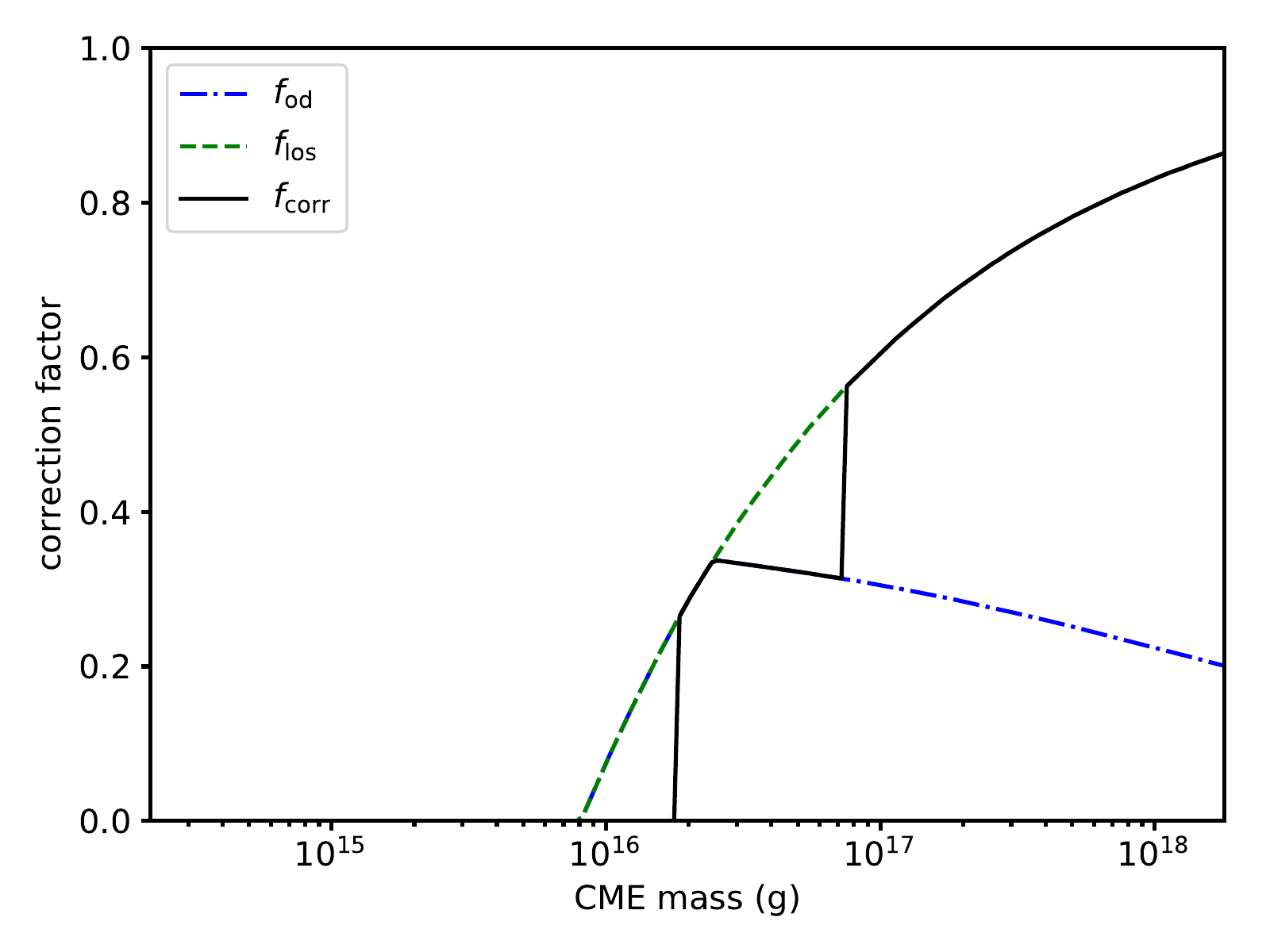}
	\includegraphics[width=\columnwidth]{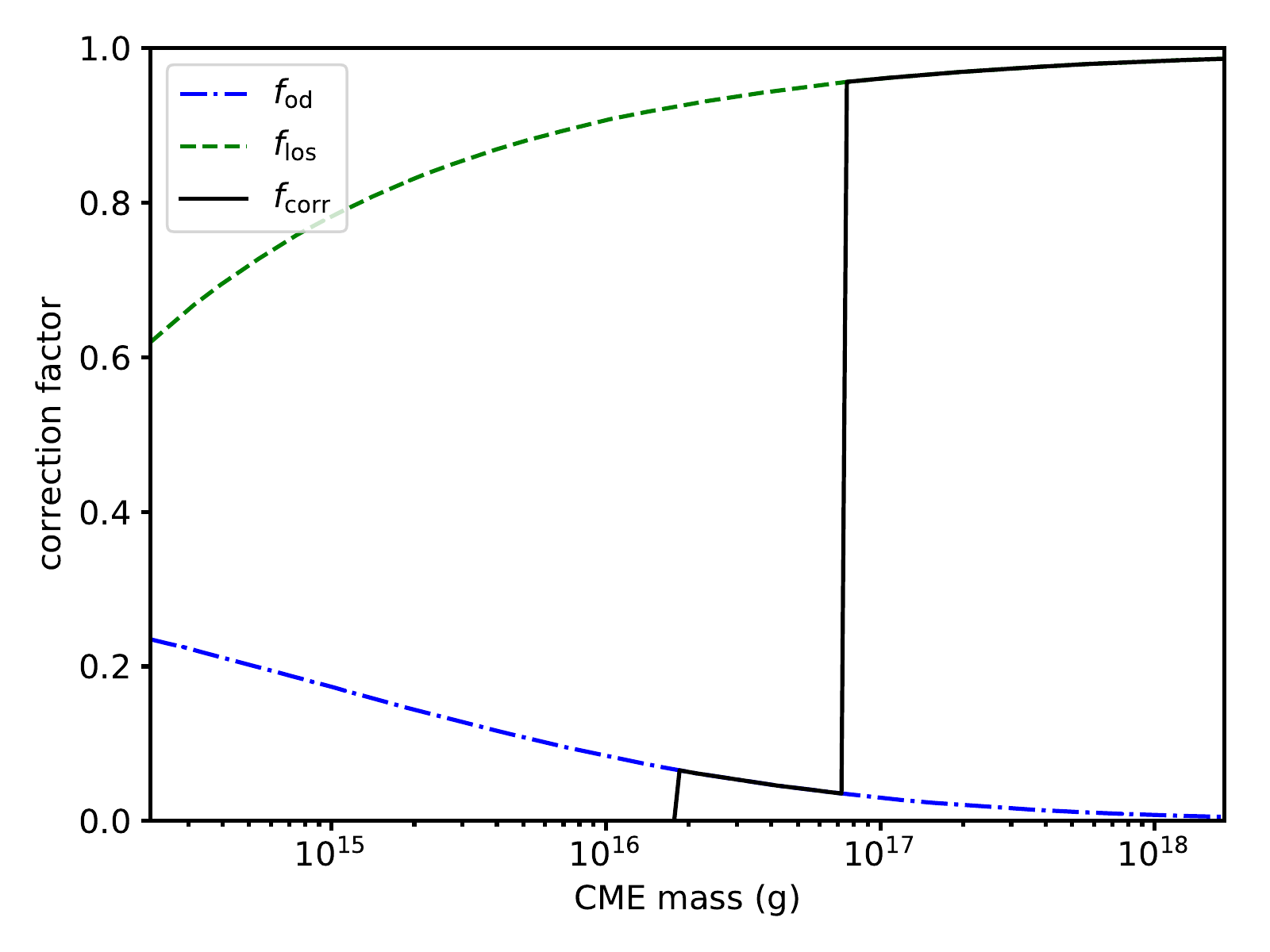}\hfill
	\includegraphics[width=\columnwidth]{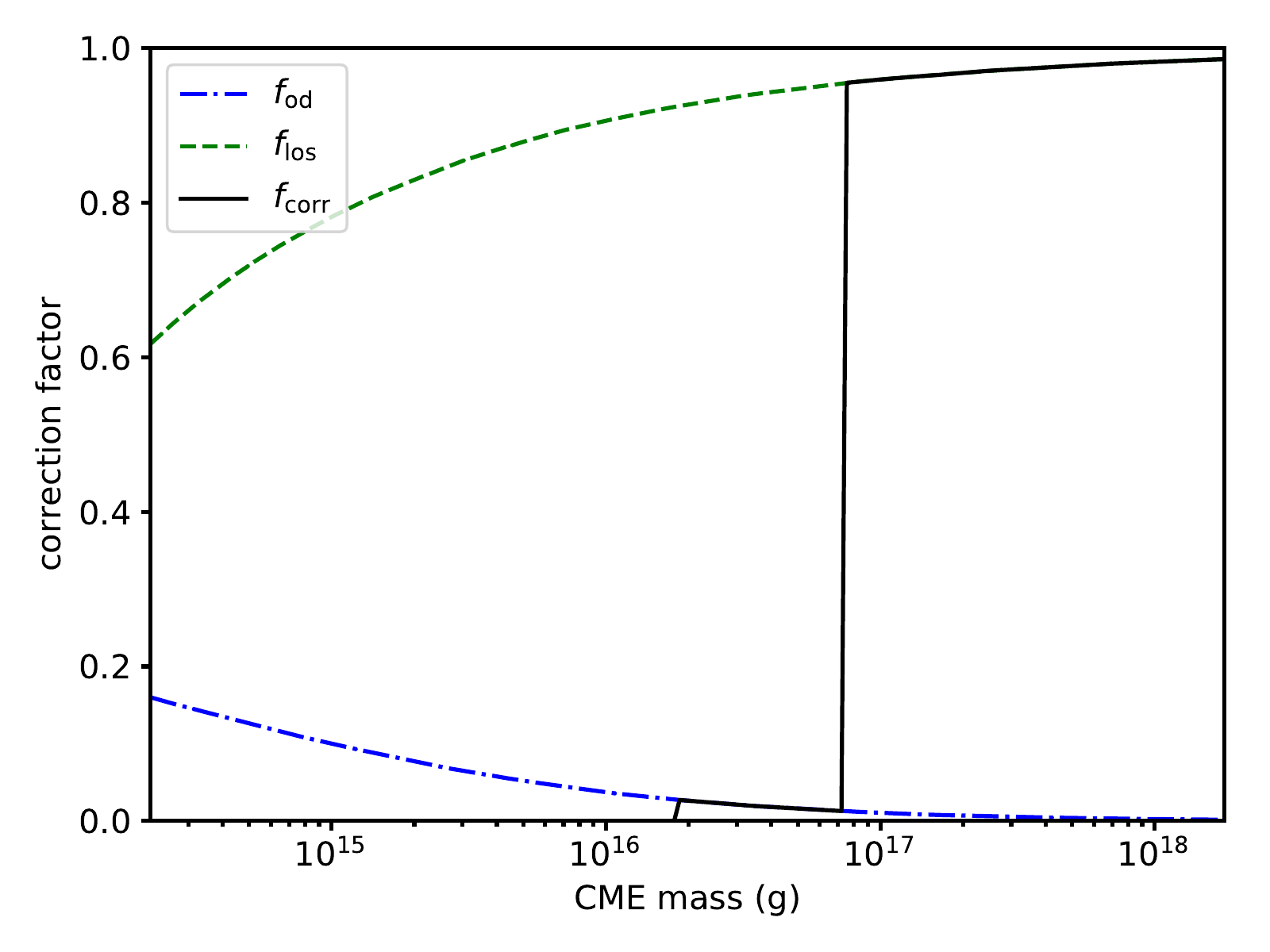}
	\includegraphics[width=\columnwidth]{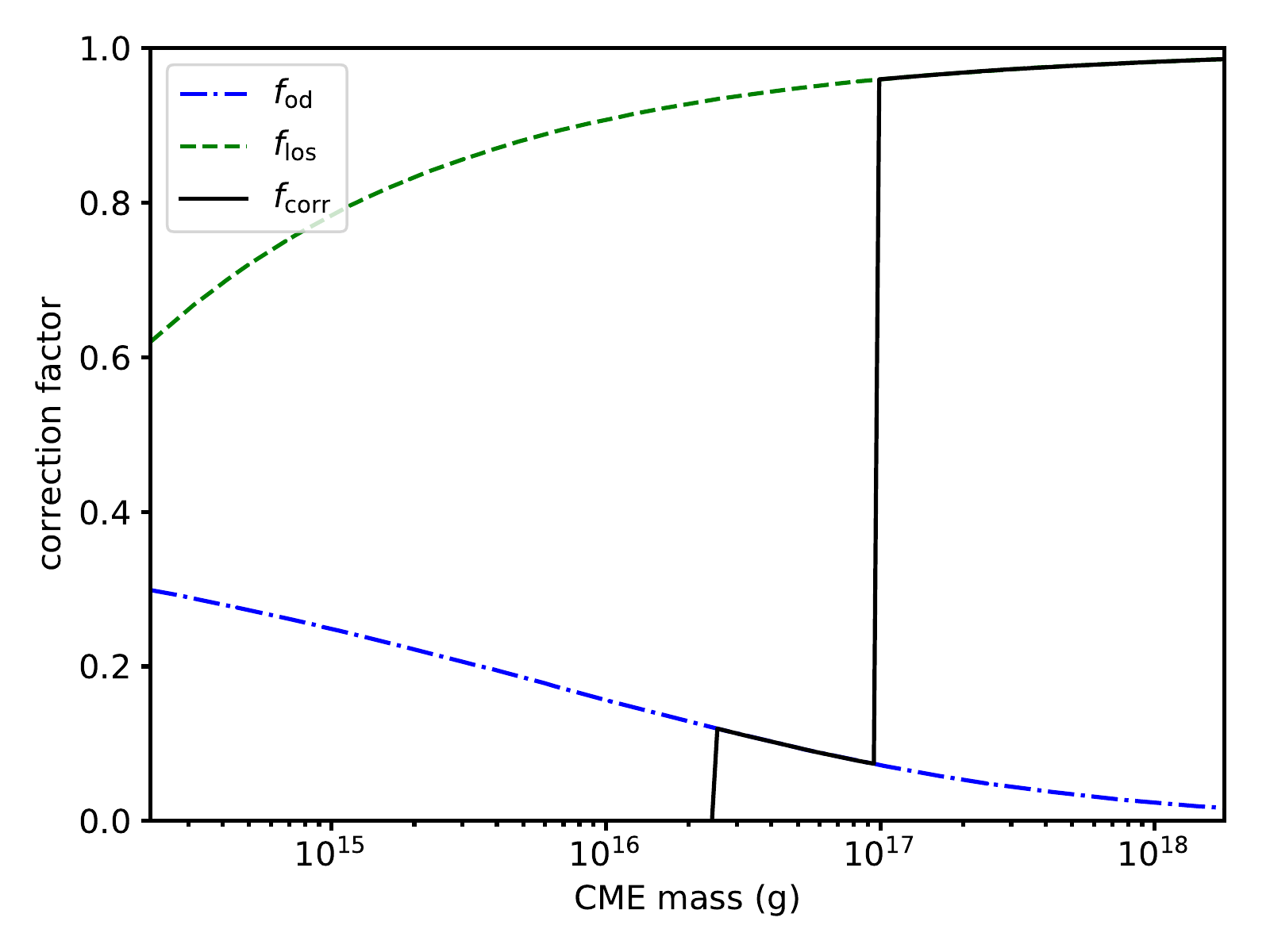}\hfill
	\includegraphics[width=\columnwidth]{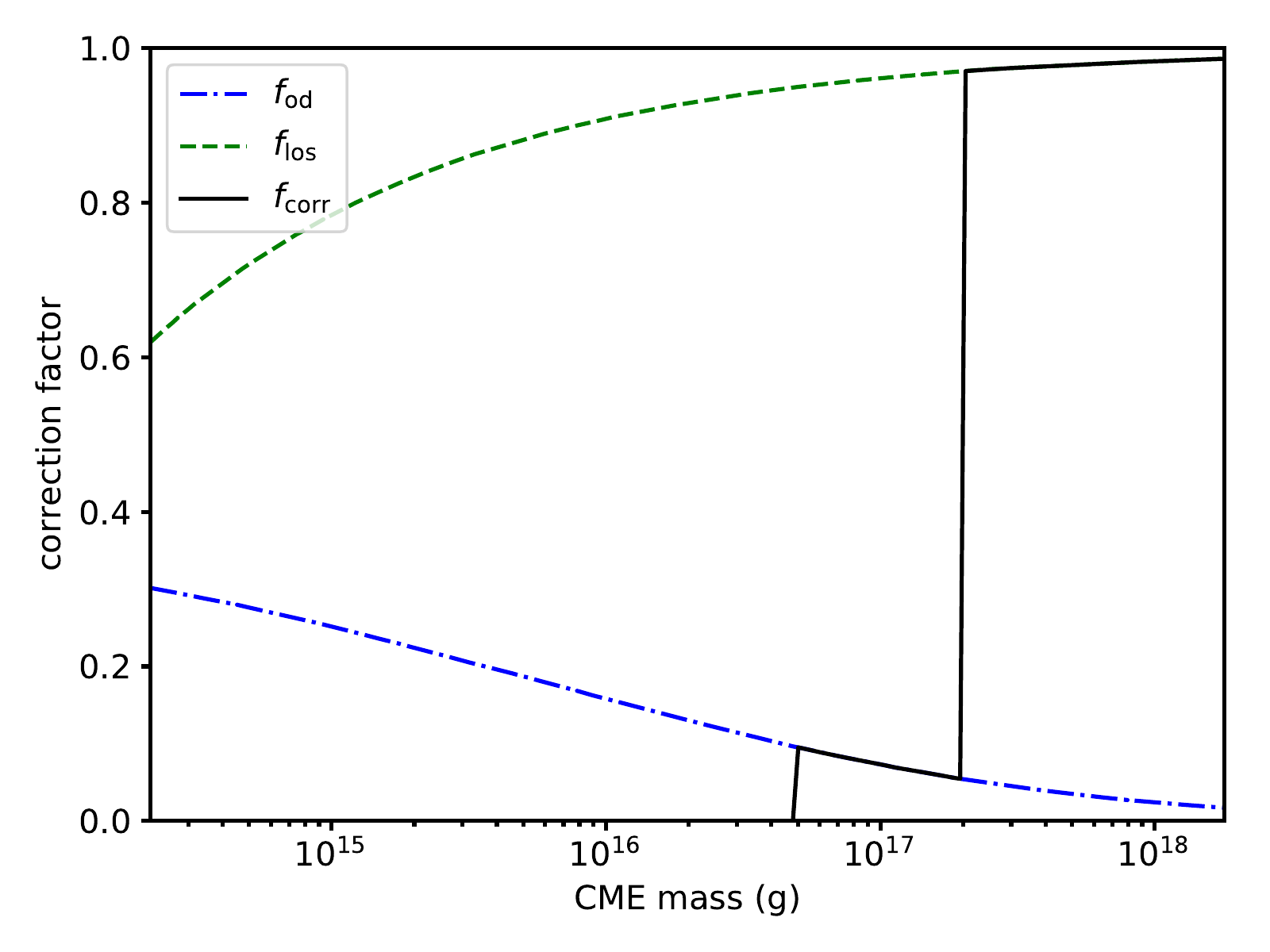}
	\contcaption{Upper row: reduction of CME speeds by factors of 5 (left) and 10 (right); middle row: exposure time of 10~min (left) and 20~min (right); lower row: spectral lines H$\beta$ (left) and H$\gamma$ (right).}
\end{figure*}

\begin{figure*}
	\centering
	\includegraphics[width=\columnwidth]{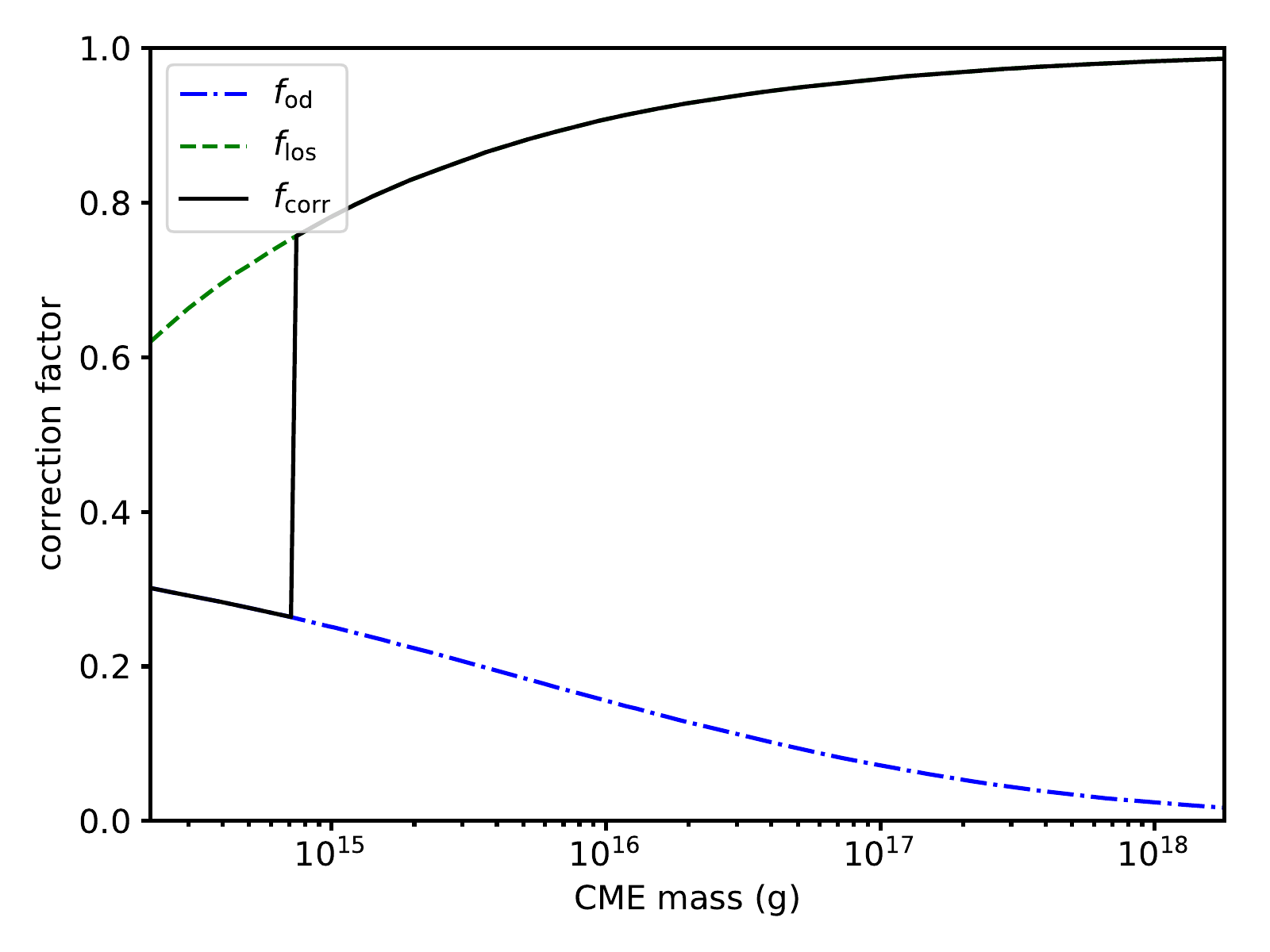}\hfill
	\includegraphics[width=\columnwidth]{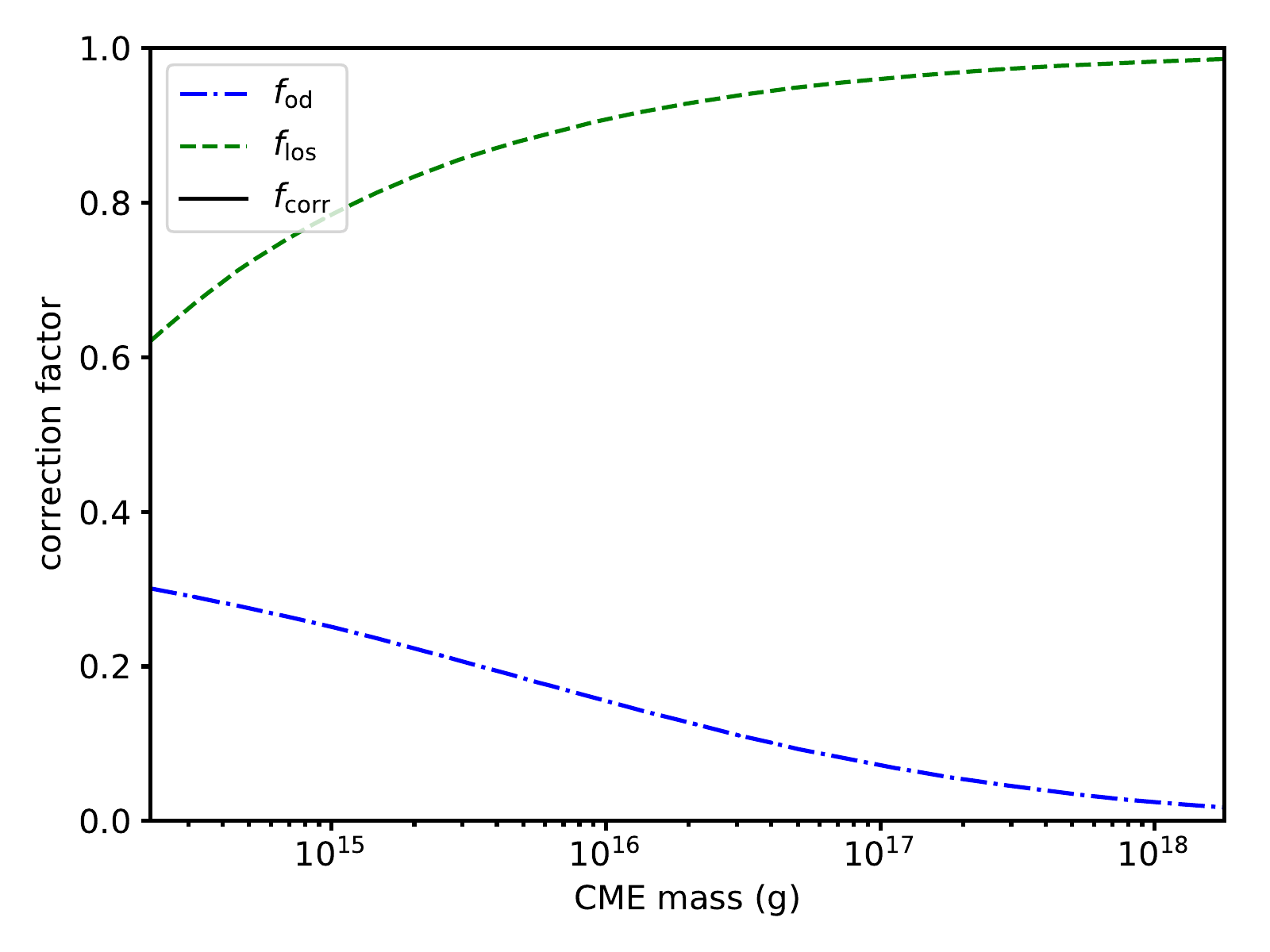}
	\includegraphics[width=\columnwidth]{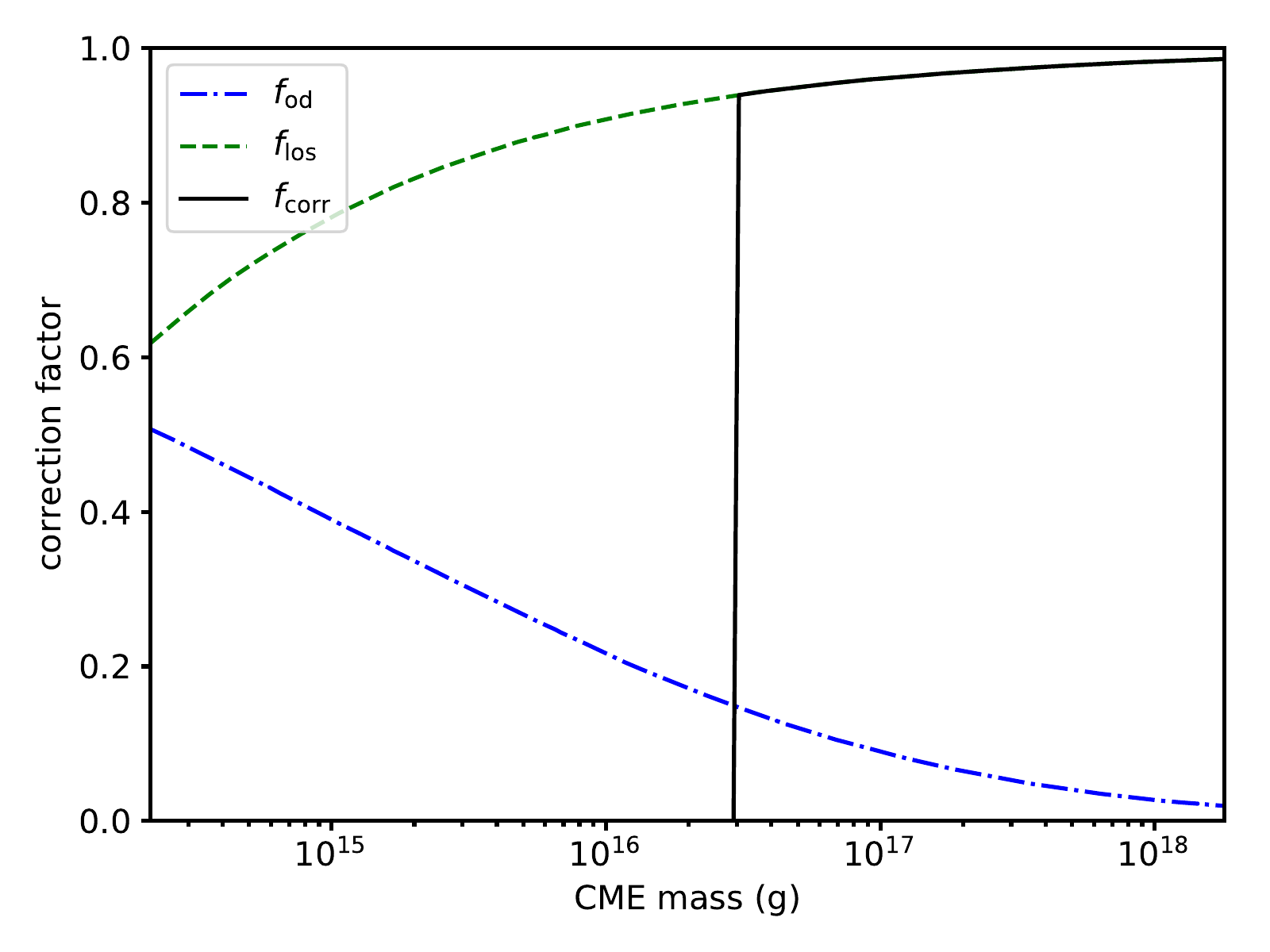}\hfill
	\includegraphics[width=\columnwidth]{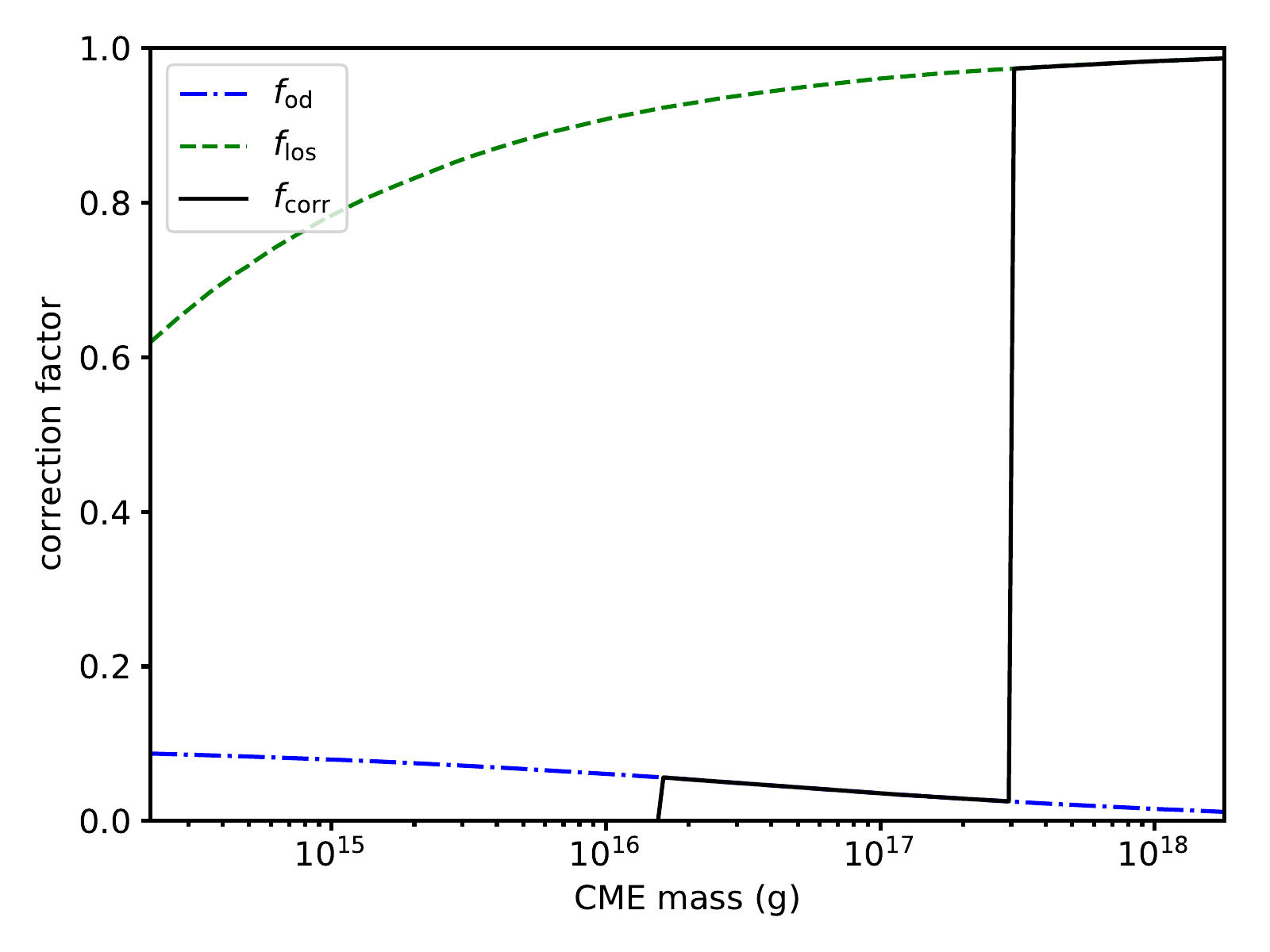}
	\contcaption{Upper row: column densities of $10^{18}$ (left) and $10^{22}$\,cm$^{-3}$ (right); lower row: dilution factor $W=0.5$ (left) and $W=0.05$ (right).}
\end{figure*}

\begin{figure*}
	\centering
	\includegraphics[width=\columnwidth]{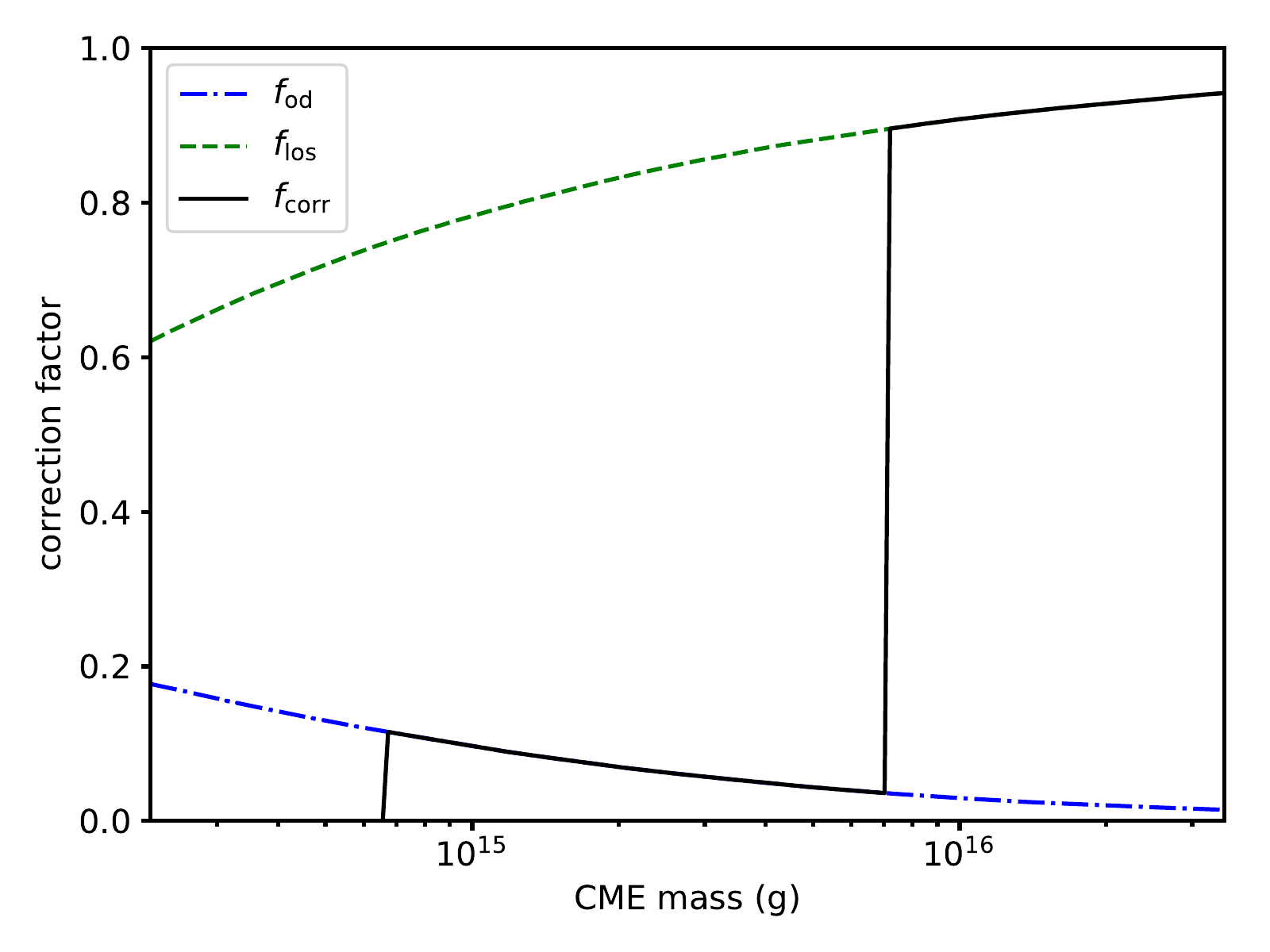}\hfill
	\includegraphics[width=\columnwidth]{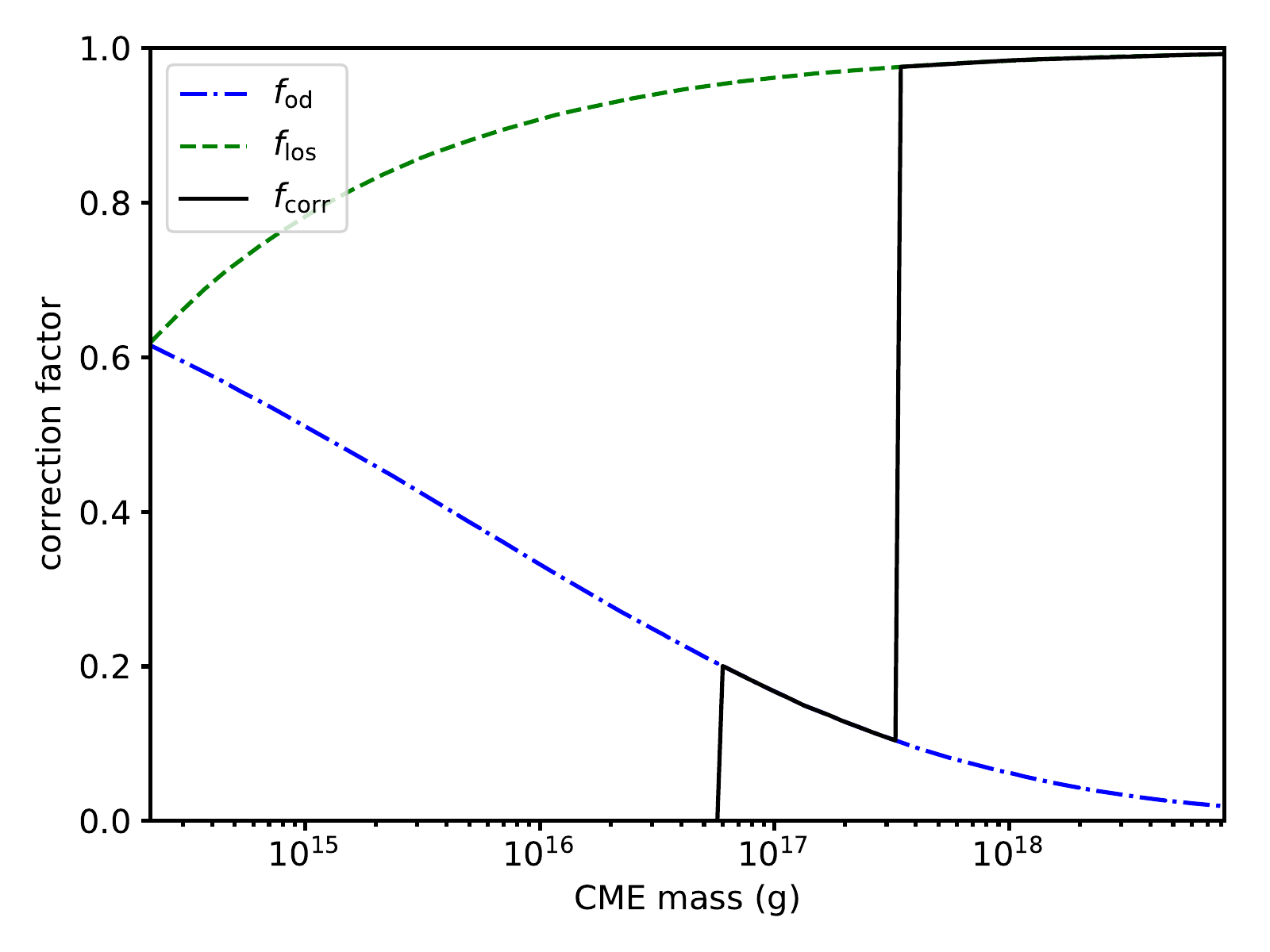}
	\includegraphics[width=\columnwidth]{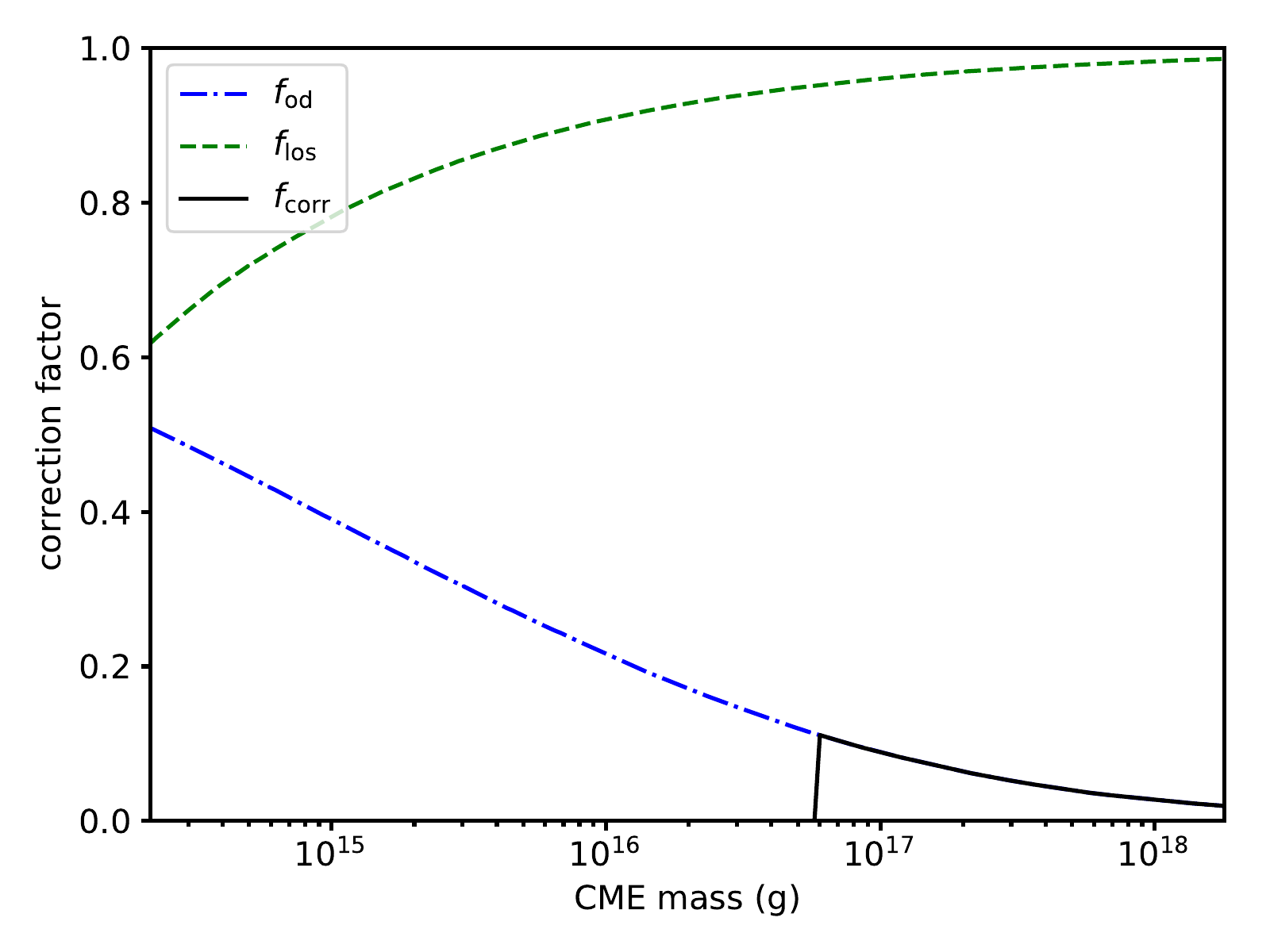}\hfill
	\includegraphics[width=\columnwidth]{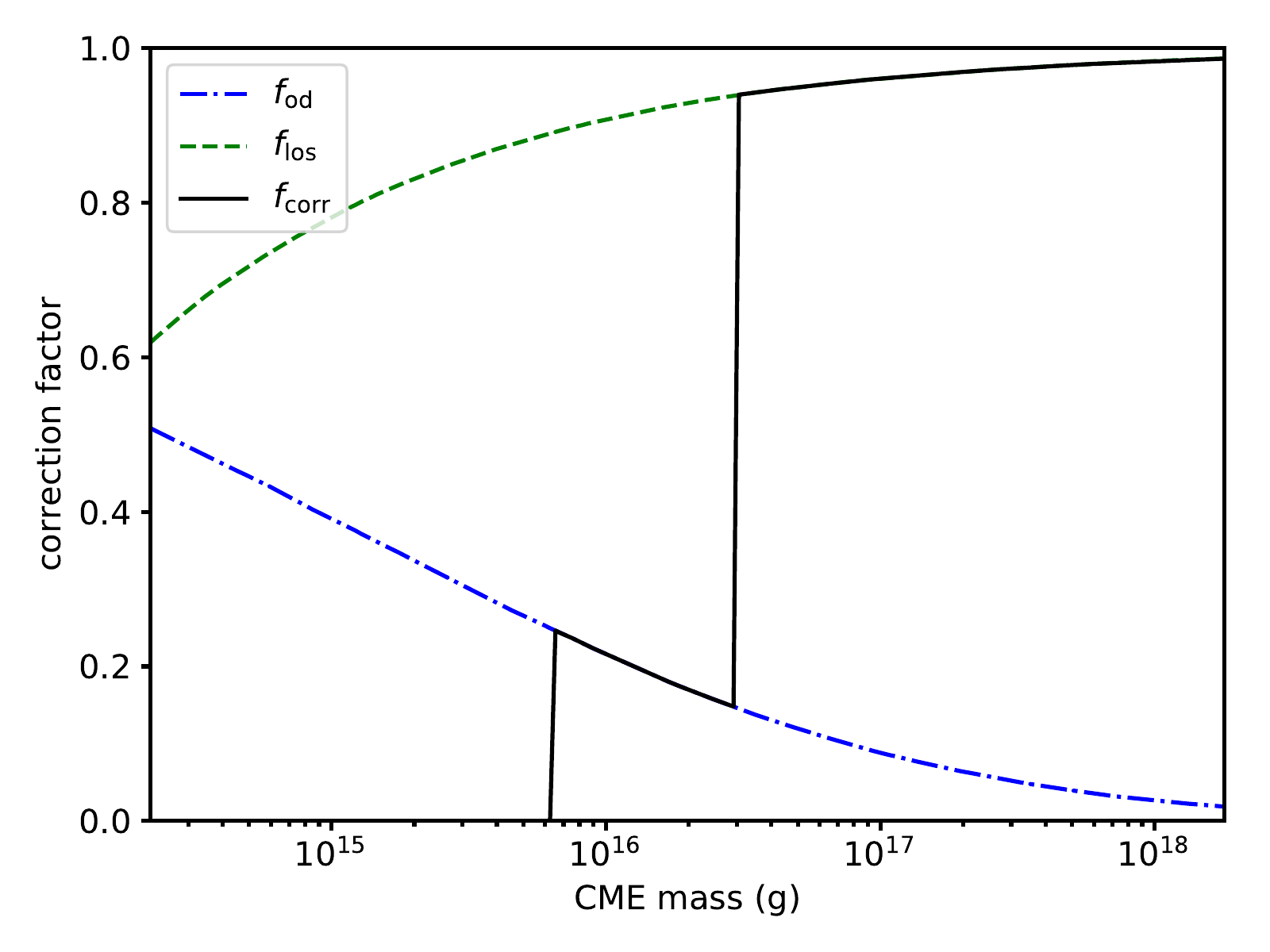}
	\includegraphics[width=\columnwidth]{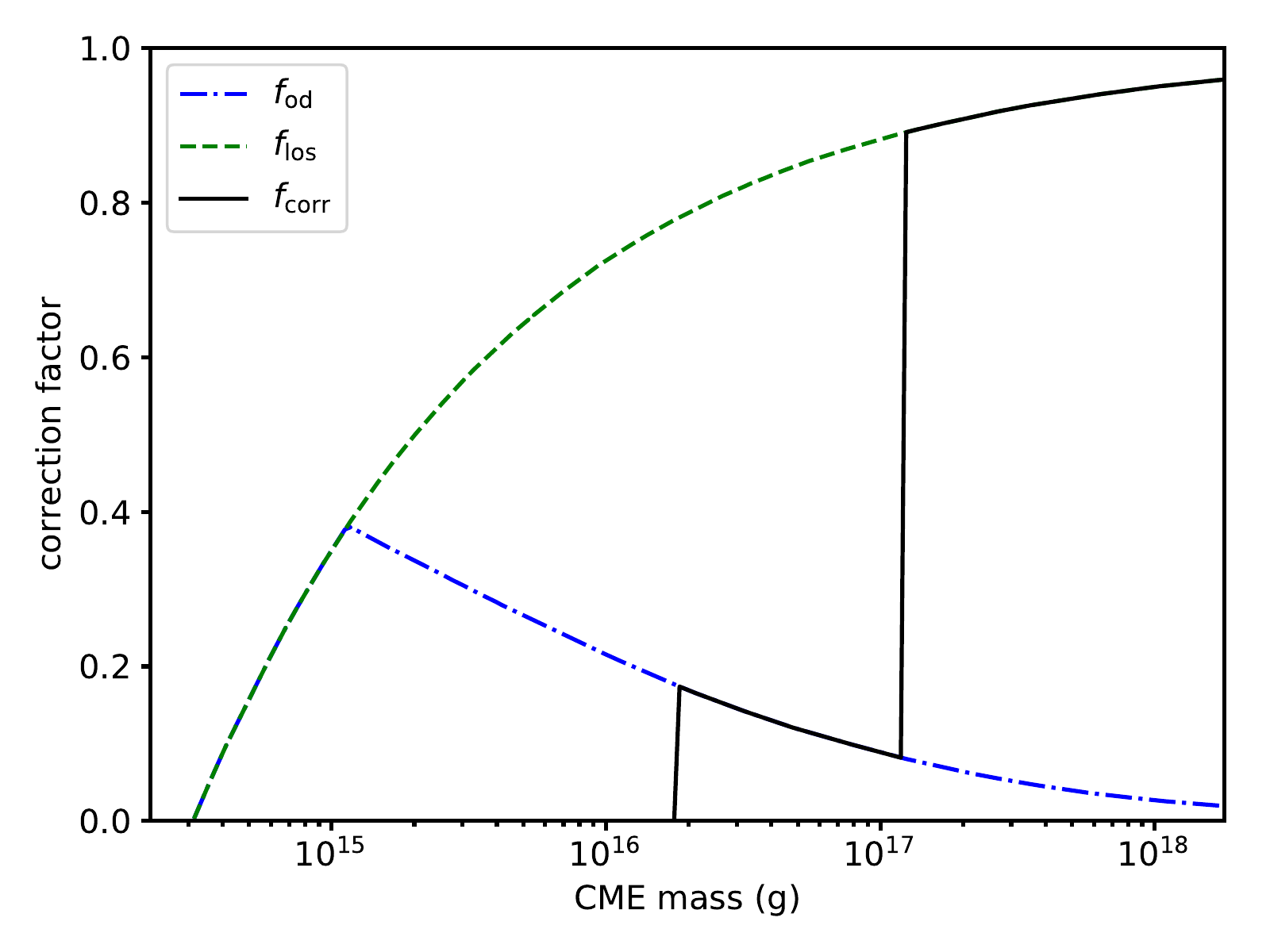}\hfill
	\includegraphics[width=\columnwidth]{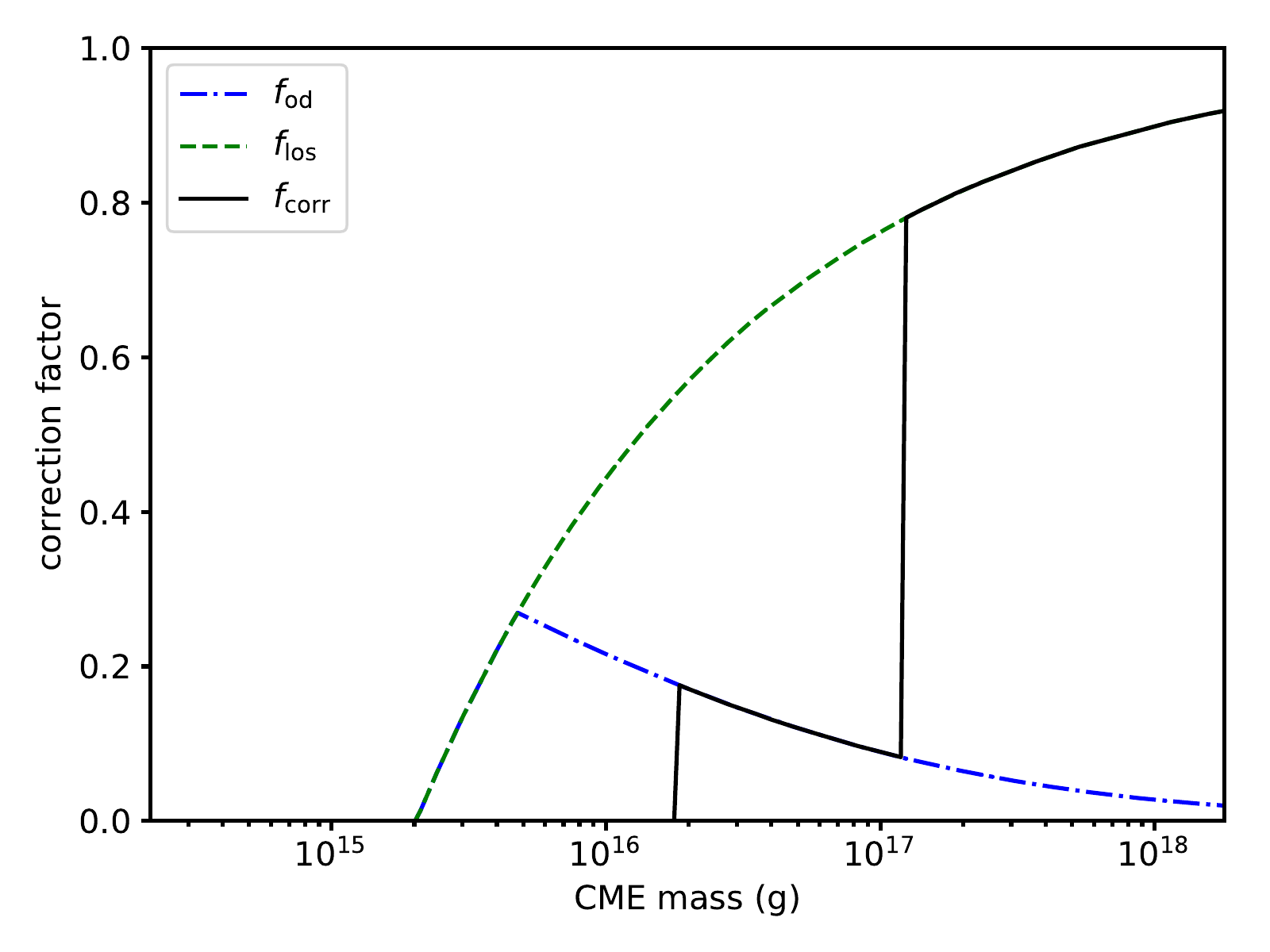}
	\caption{Geometrical correction factor for the effective dilution factor - dependence on parameters. Upper row: spectral types M5.5 (left) and F6 (right); middle row: $SNR$ of 30 (left) and 300 (right); lower row: velocity limit of 300\,km\,s$^{-1}$ (left) and 600\,km\,s$^{-1}$ (right).}
	\label{fig:app_geo_wav}
\end{figure*}

\begin{figure*}
	\centering
	\includegraphics[width=\columnwidth]{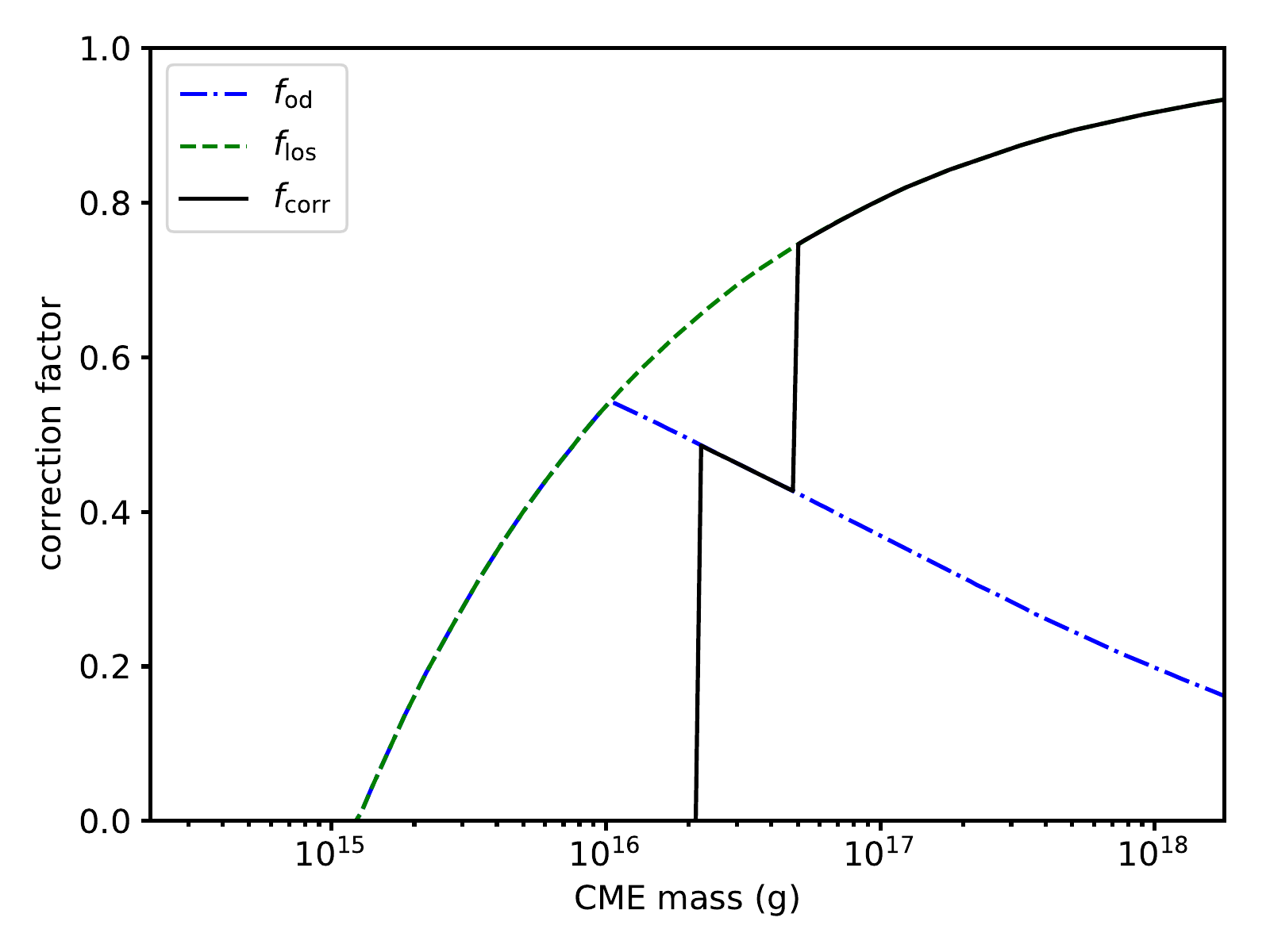}\hfill
	\includegraphics[width=\columnwidth]{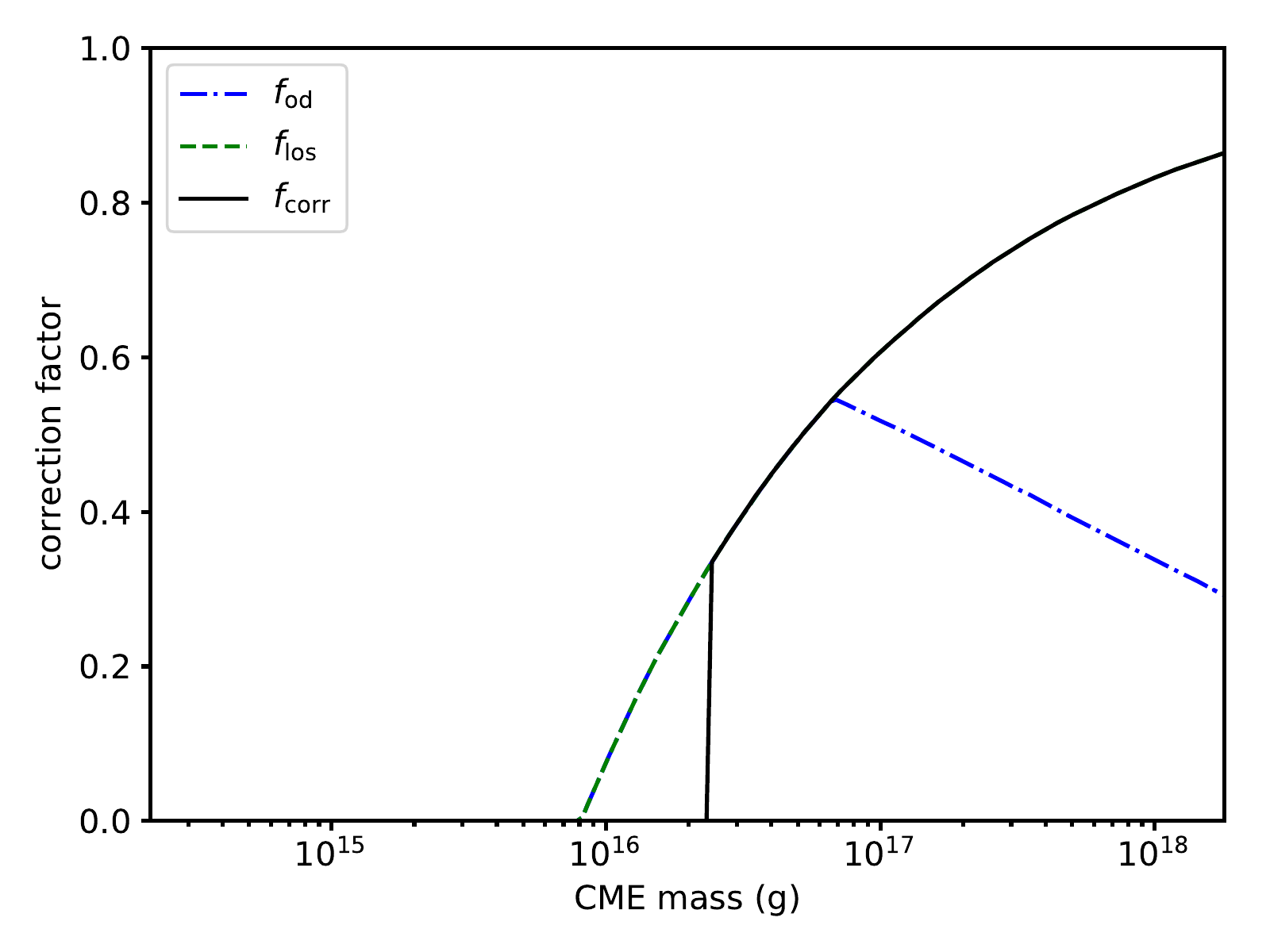}
	\includegraphics[width=\columnwidth]{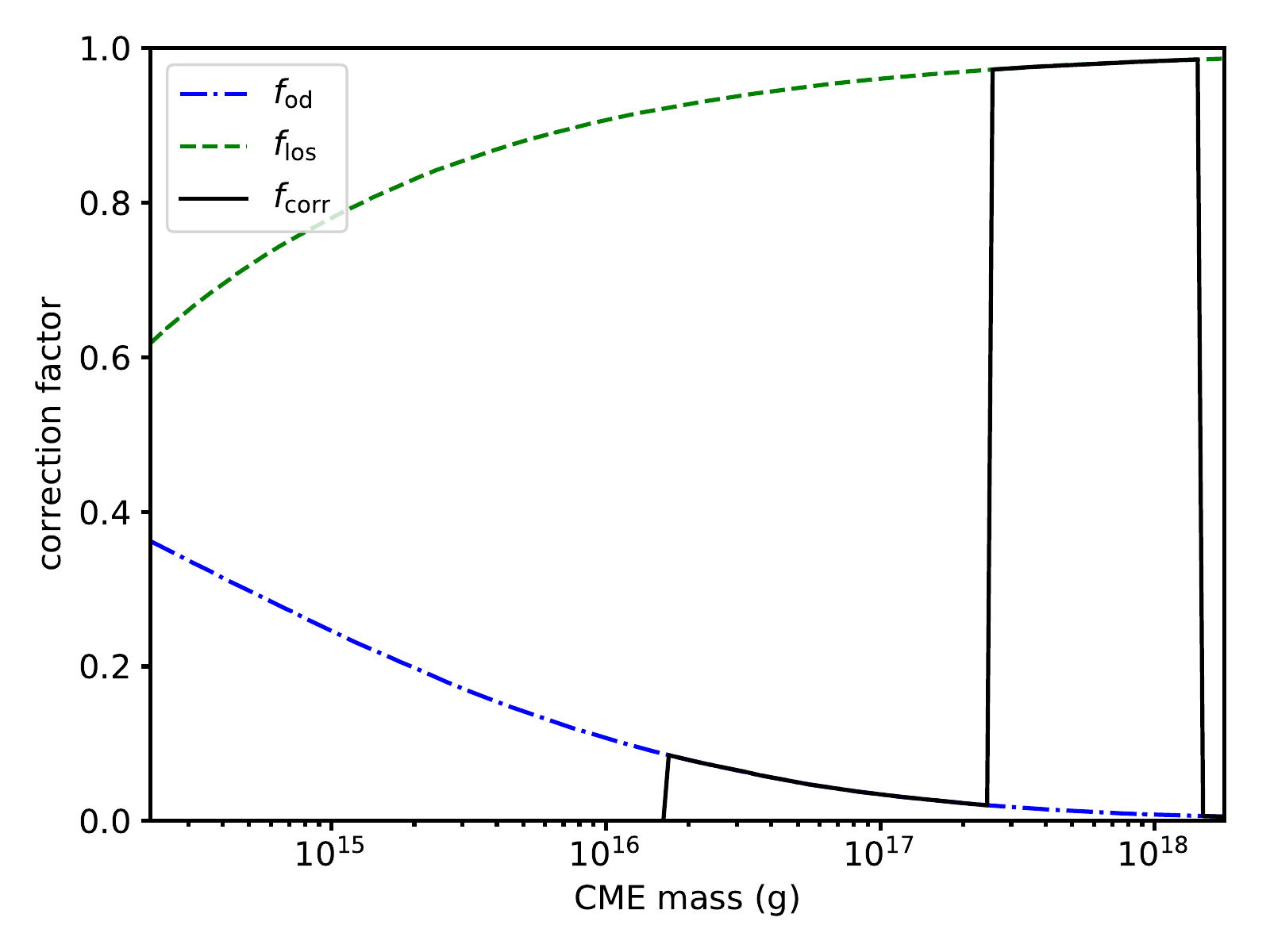}\hfill
	\includegraphics[width=\columnwidth]{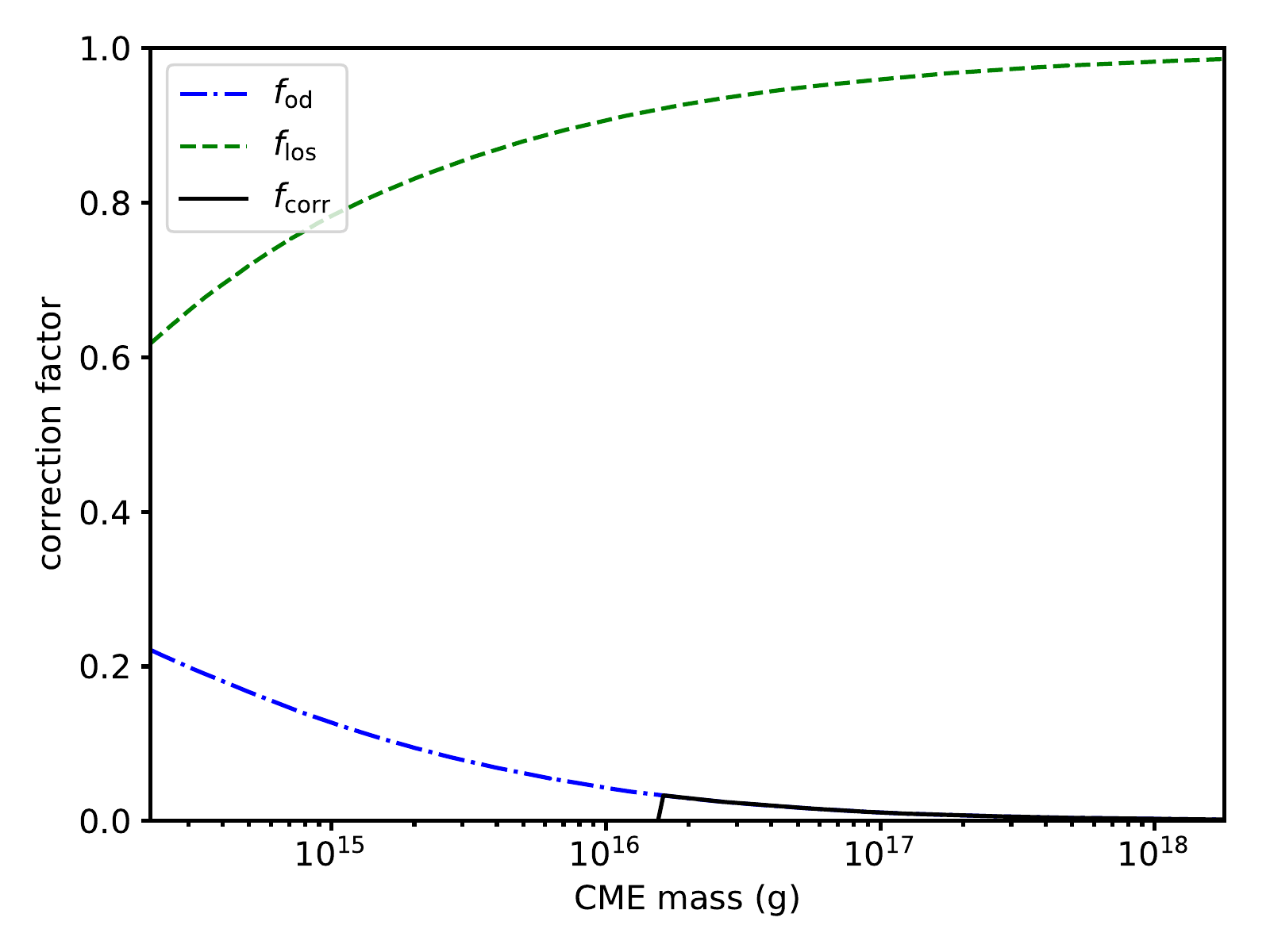}
	\includegraphics[width=\columnwidth]{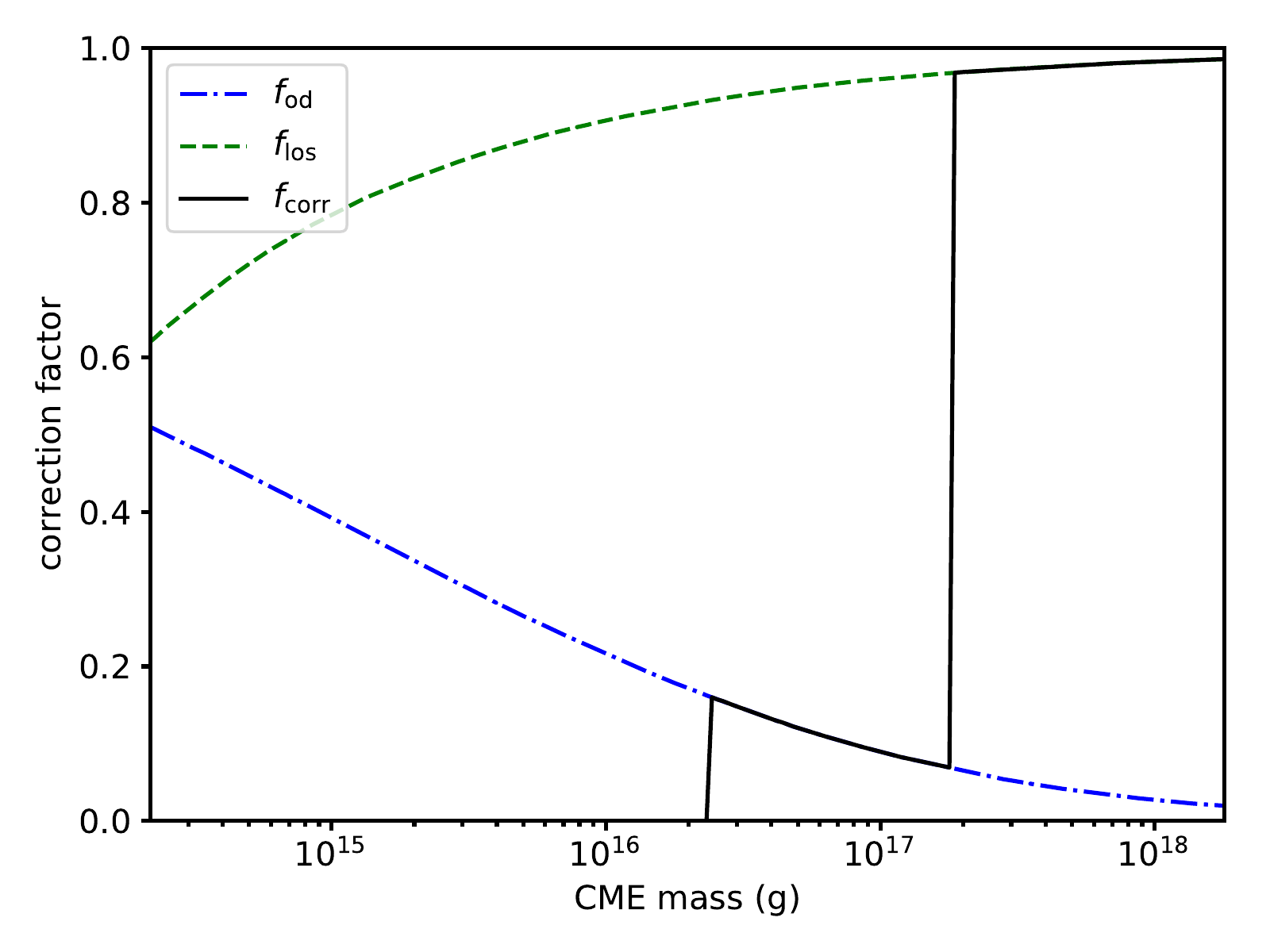}\hfill
	\includegraphics[width=\columnwidth]{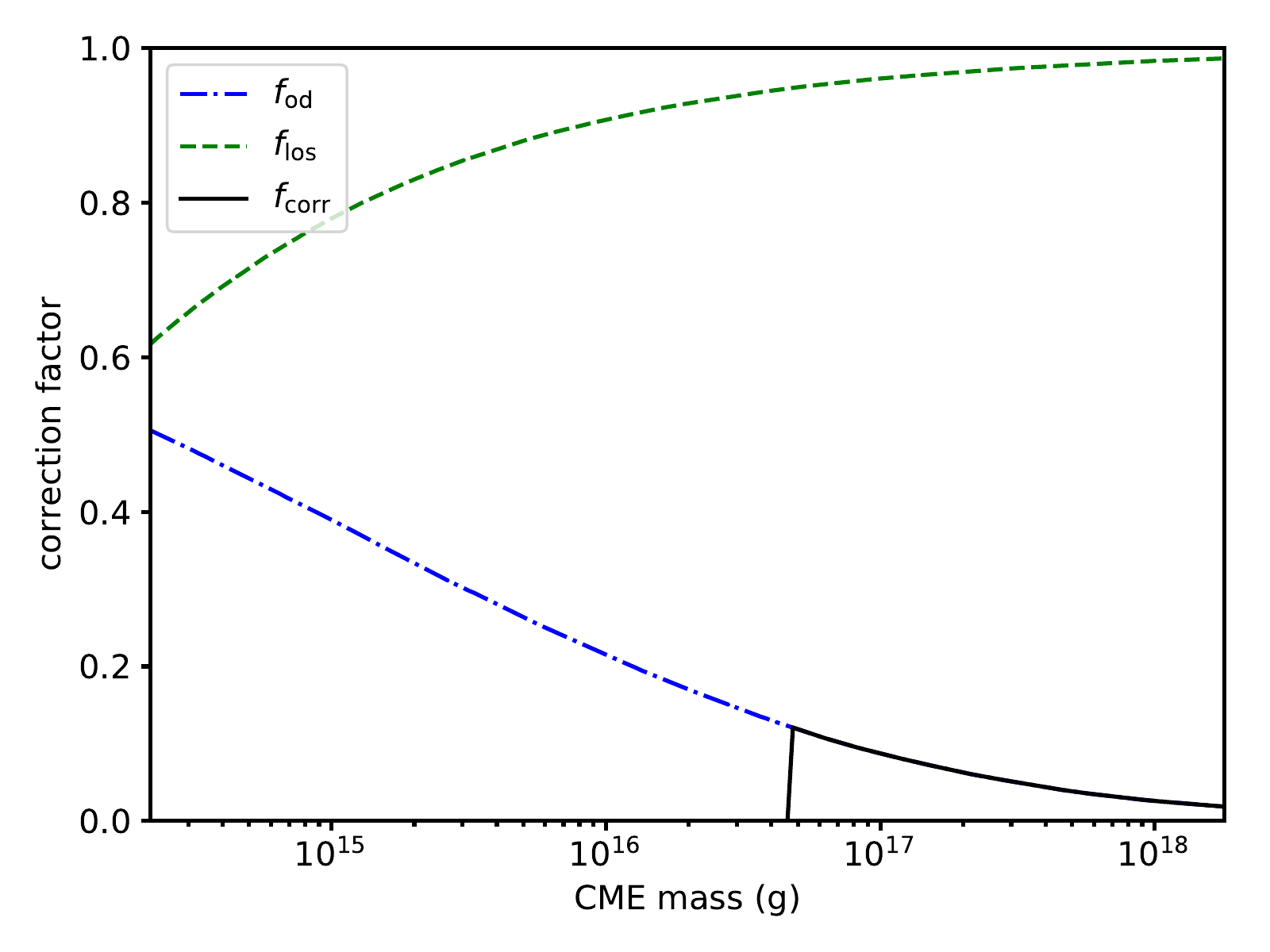}
	\contcaption{Upper row: reduction of CME speeds by factors of 5 (left) and 10 (right); middle row: exposure time of 10~min (left) and 20~min (right); lower row: spectral lines H$\beta$ (left) and H$\gamma$ (right).}
\end{figure*}

\begin{figure*}
	\centering
	\includegraphics[width=\columnwidth]{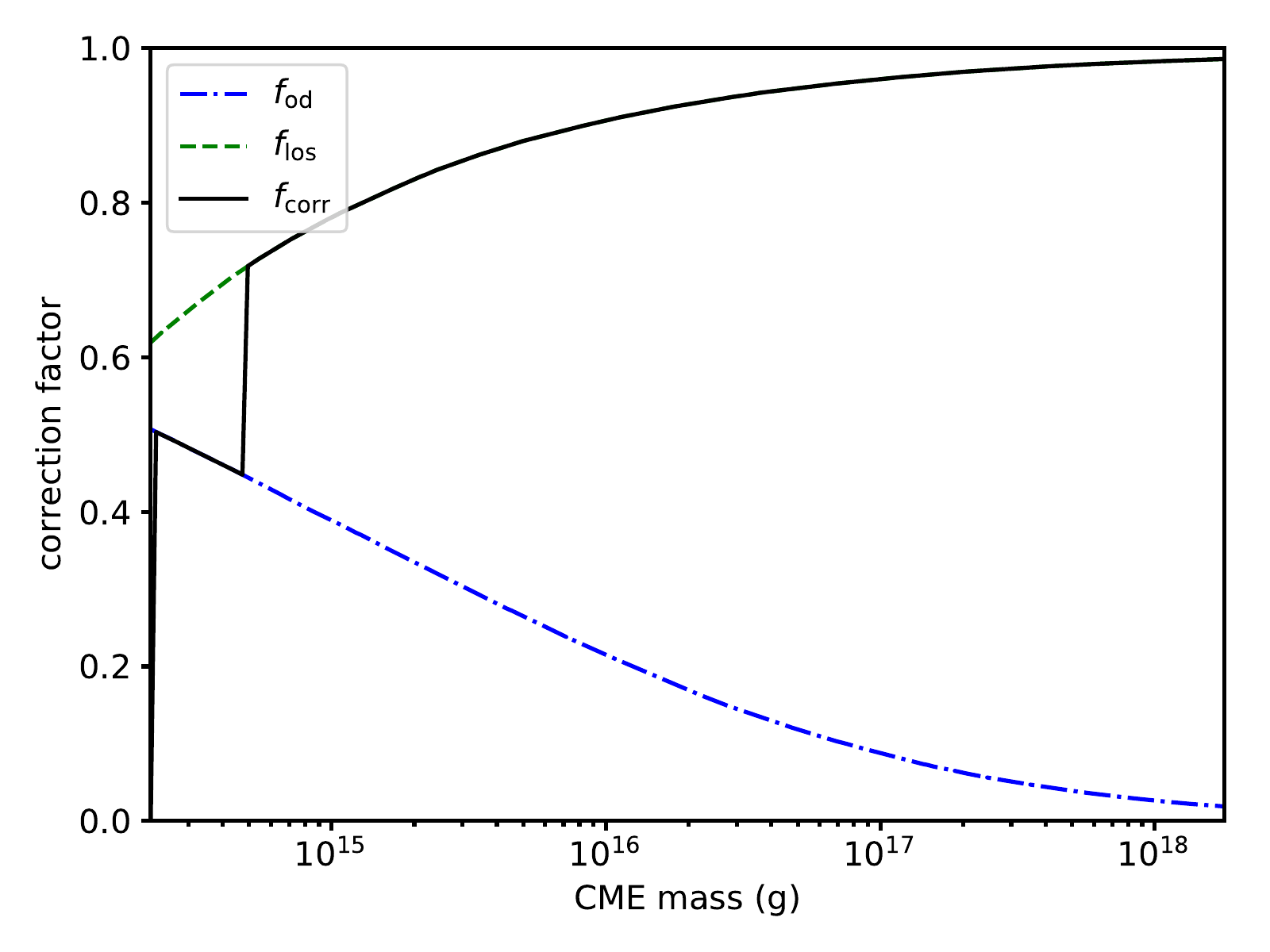}\hfill
	\includegraphics[width=\columnwidth]{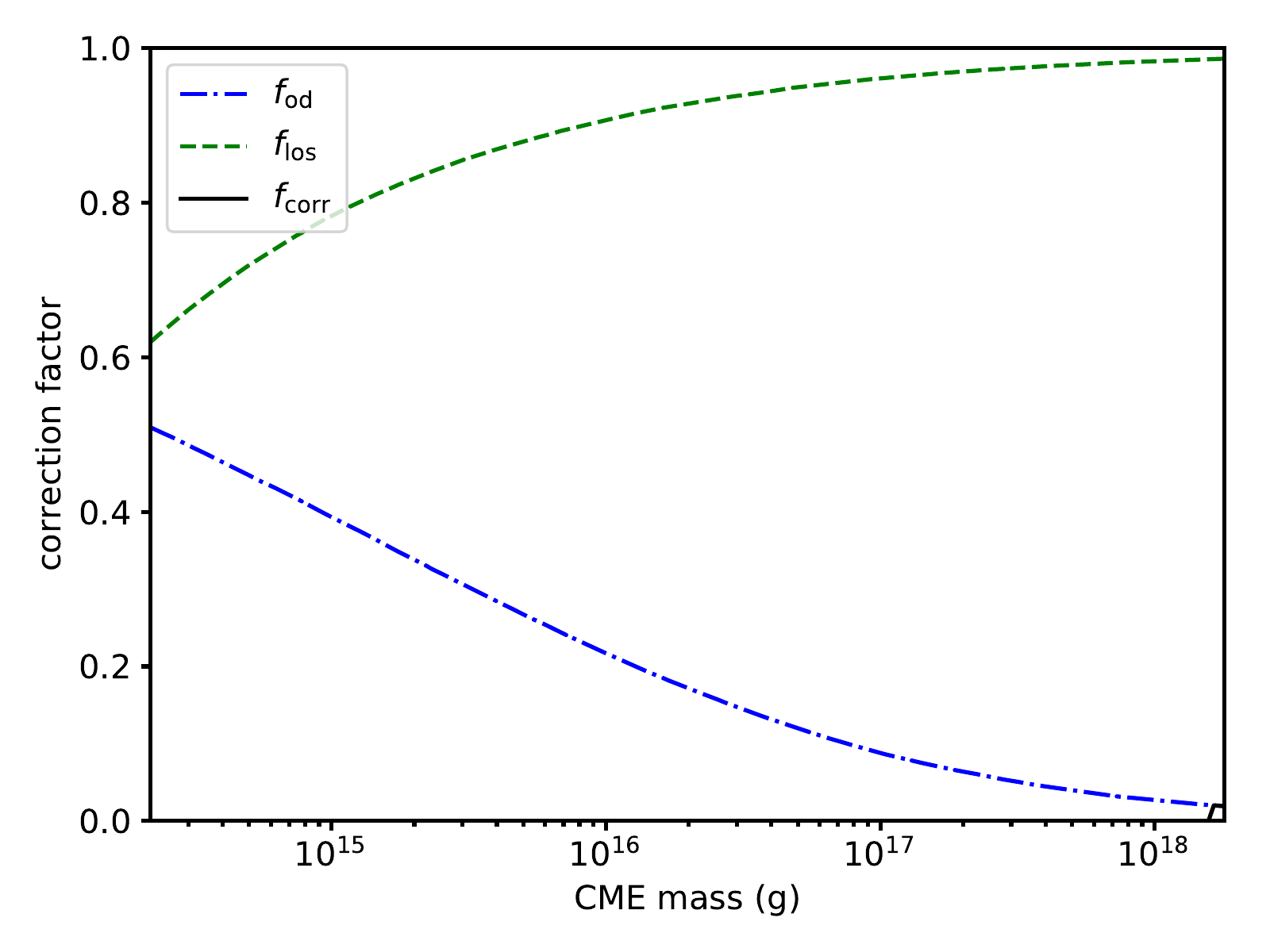}
	\contcaption{Column densities of $10^{18}$ (left) and $10^{22}$\,cm$^{-3}$ (right).}
\end{figure*}

\clearpage

\section{Observable vs. intrinsic CME rates}\label{app:int_obs}
In Fig.~\ref{fig:app_int_obs}, we compare intrinsic CME rates with observable ones for several stellar spectral types. Relative to Fig.~\ref{fig:int_obs}, we vary one of the default parameters.

\begin{figure*}
	\centering
	\includegraphics[width=\columnwidth]{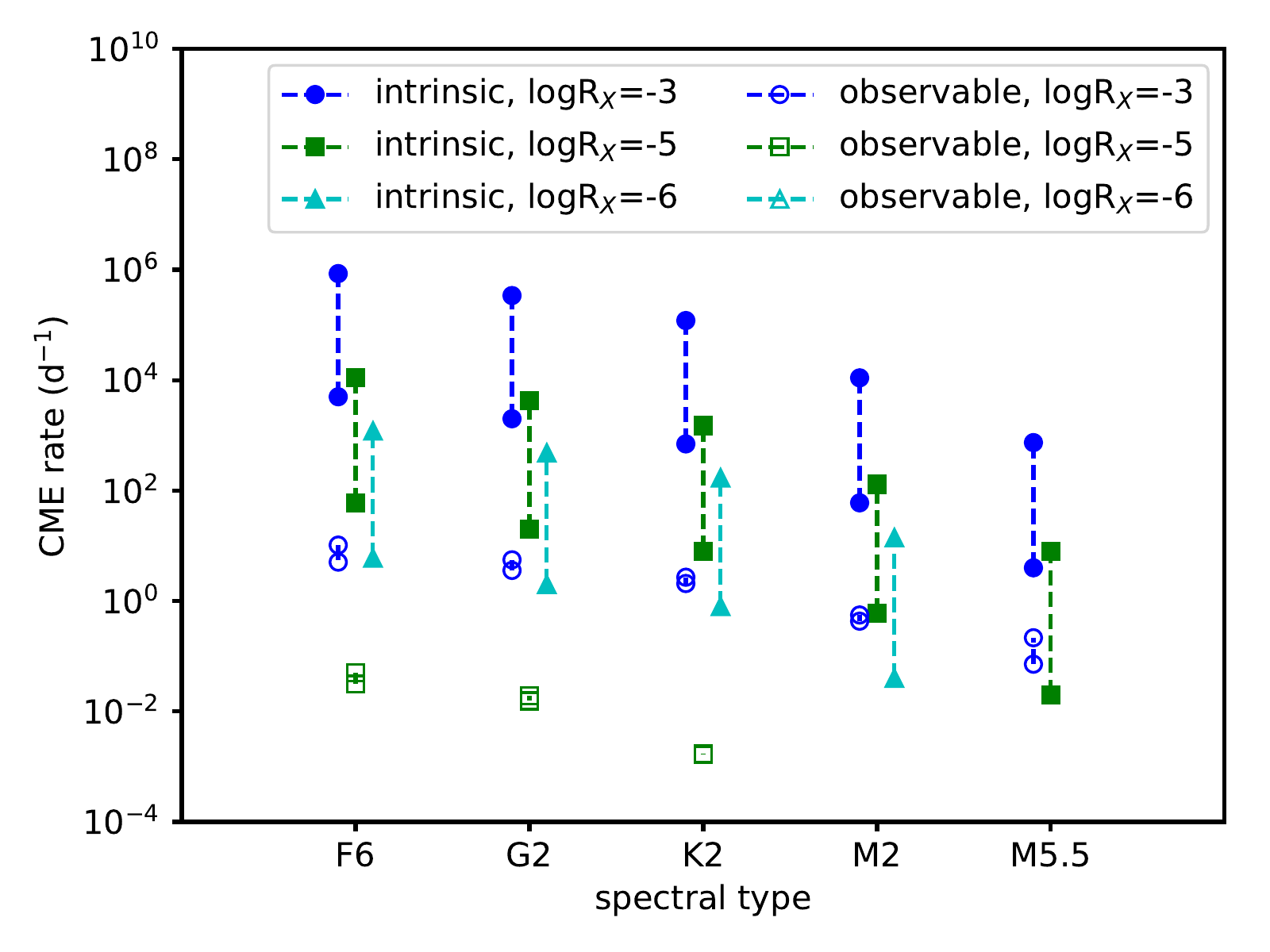}\hfill
	\includegraphics[width=\columnwidth]{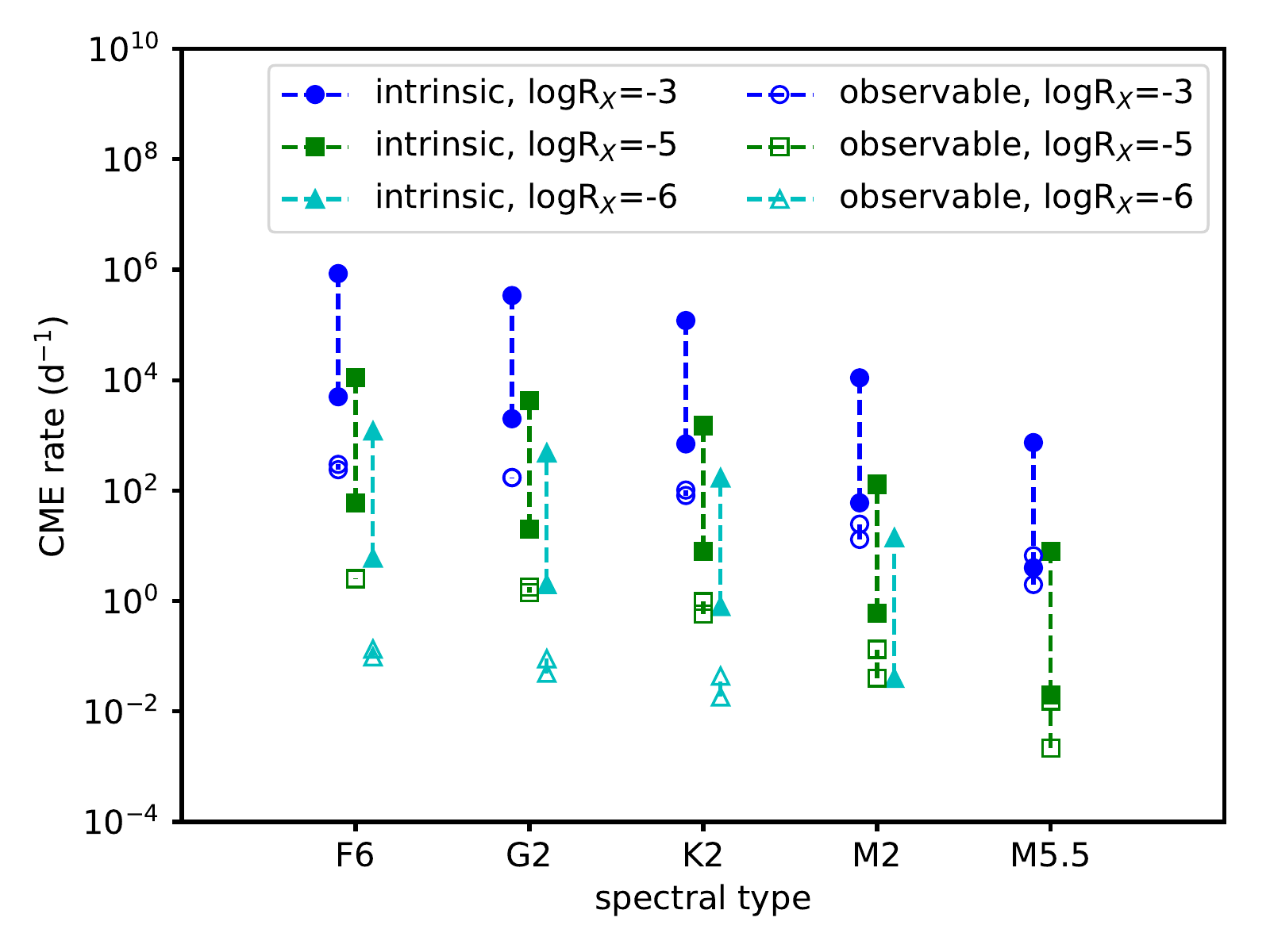}
	\includegraphics[width=\columnwidth]{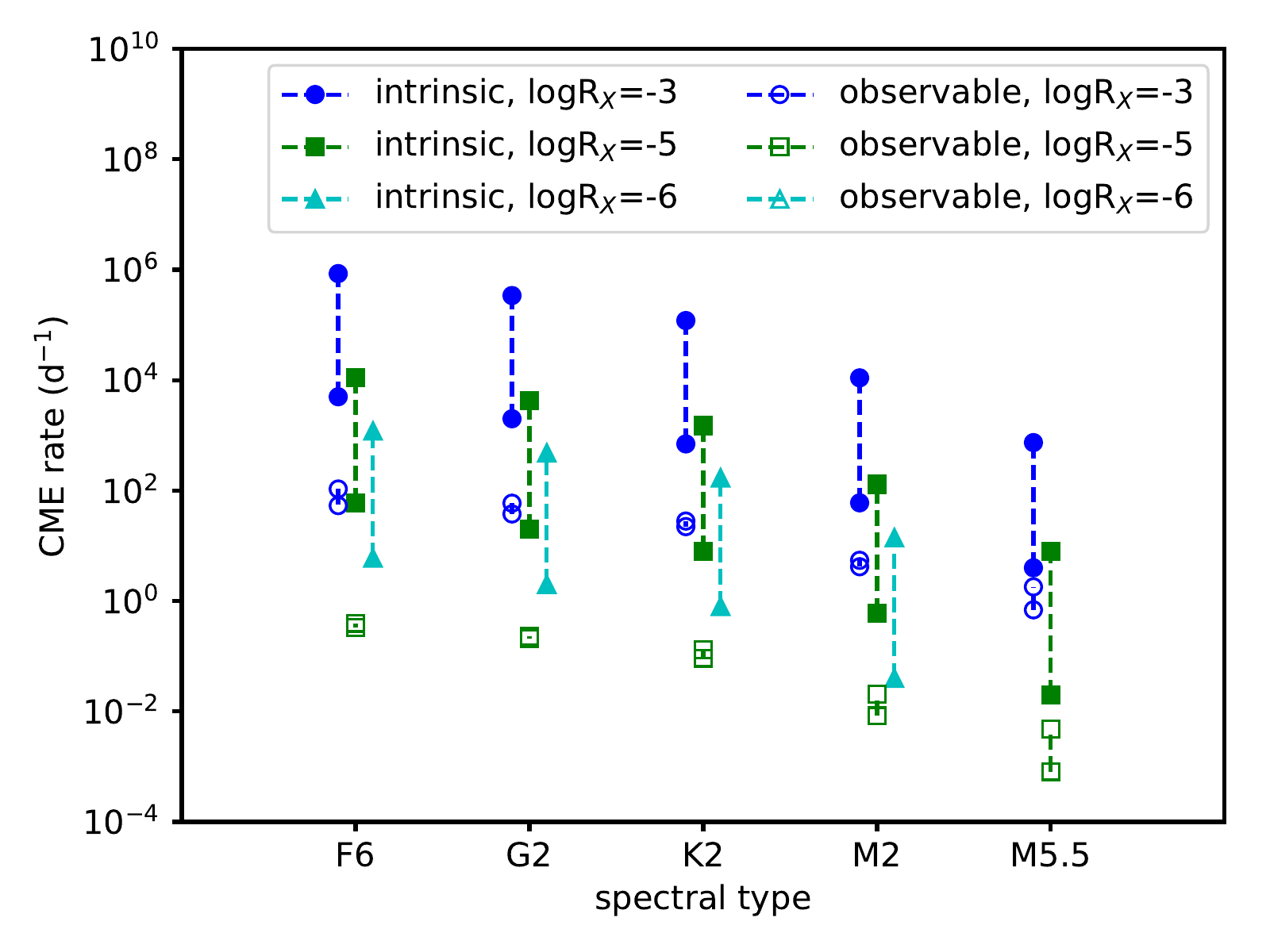}\hfill
	\includegraphics[width=\columnwidth]{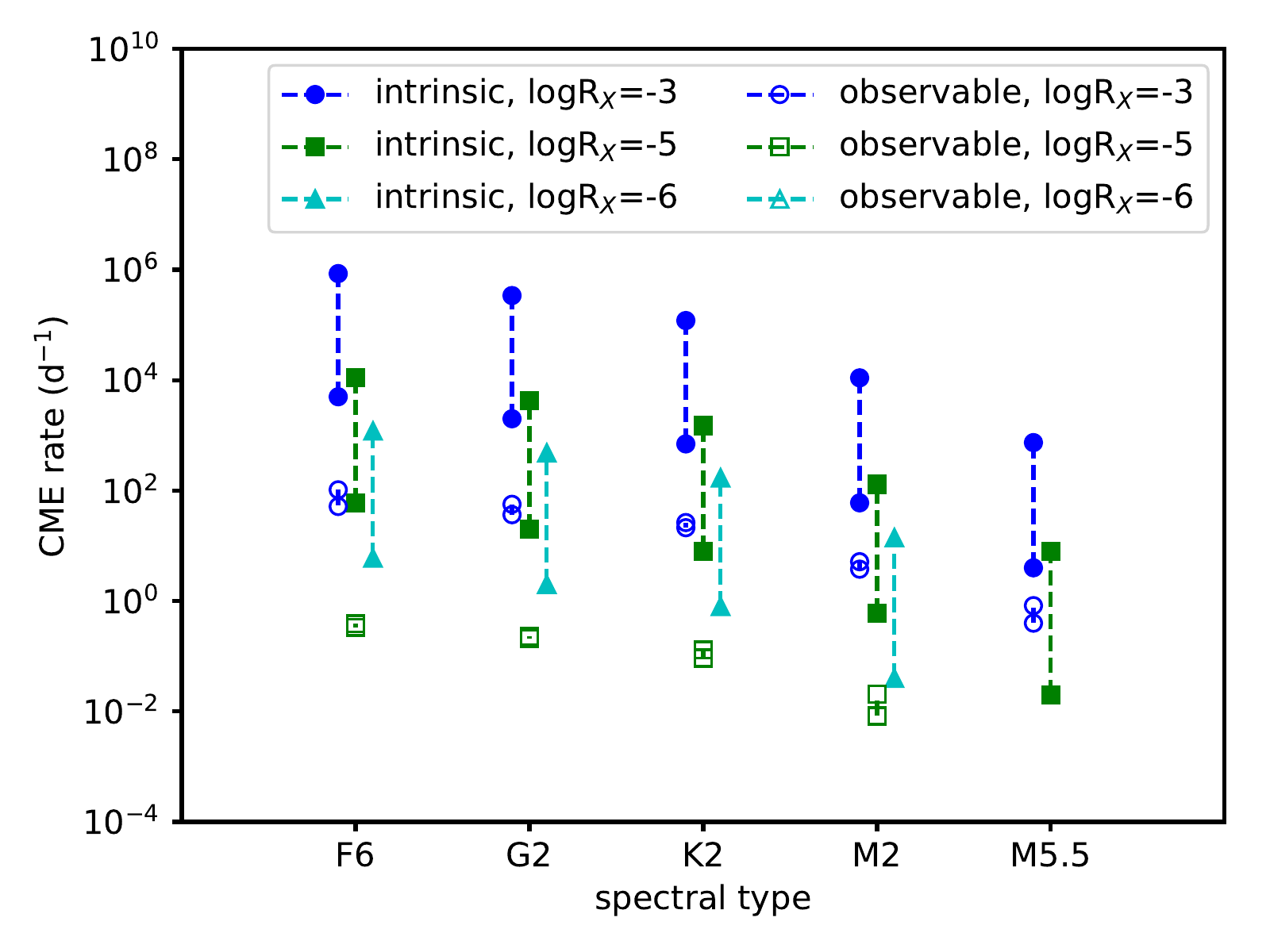}
	\includegraphics[width=\columnwidth]{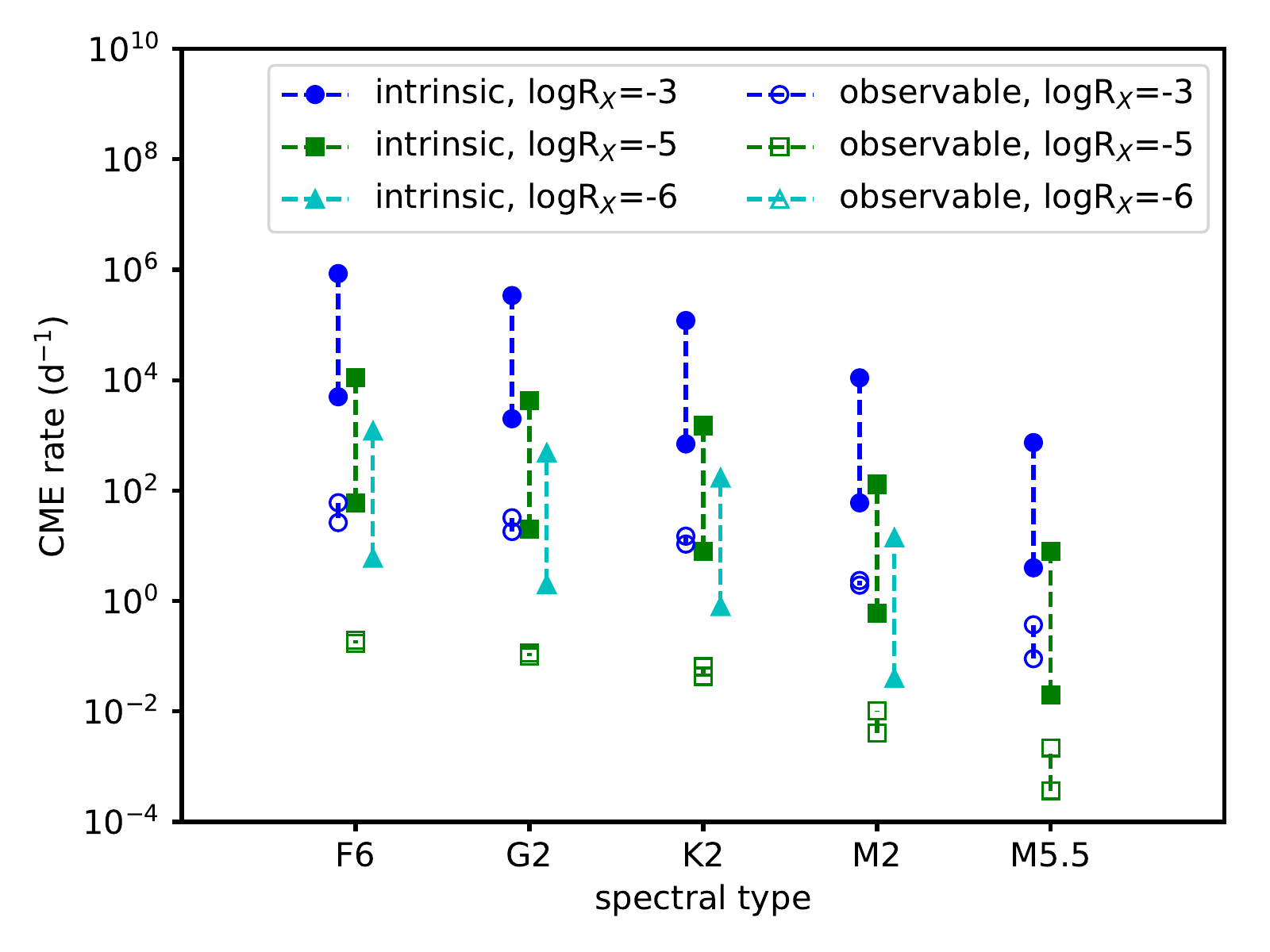}\hfill
	\includegraphics[width=\columnwidth]{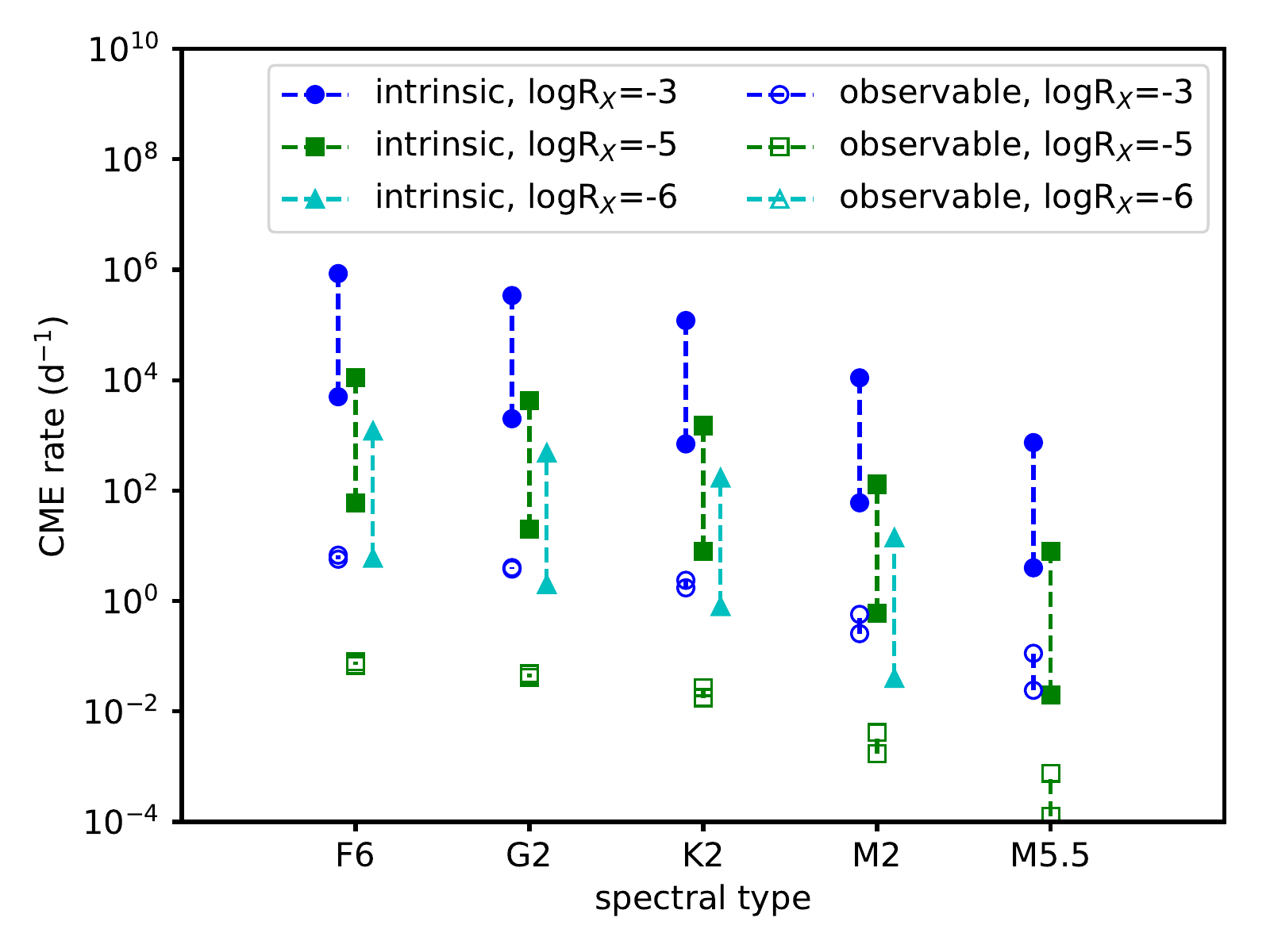}
	\caption{Intrinsic vs. observable CME rates - dependence on parameters. Upper row: SNR of 30 (left) and 300 (right); middle row: velocity limit of 300\,km\,s$^{-1}$ (left) and 600\,km\,s$^{-1}$ (right); lower row: exposure time of 10~min (left) and 20~min (right). Missing symbols indicate observable and/or intrinsic rates of zero.}
	\label{fig:app_int_obs}
\end{figure*}

\begin{figure*}
	\centering
	\includegraphics[width=\columnwidth]{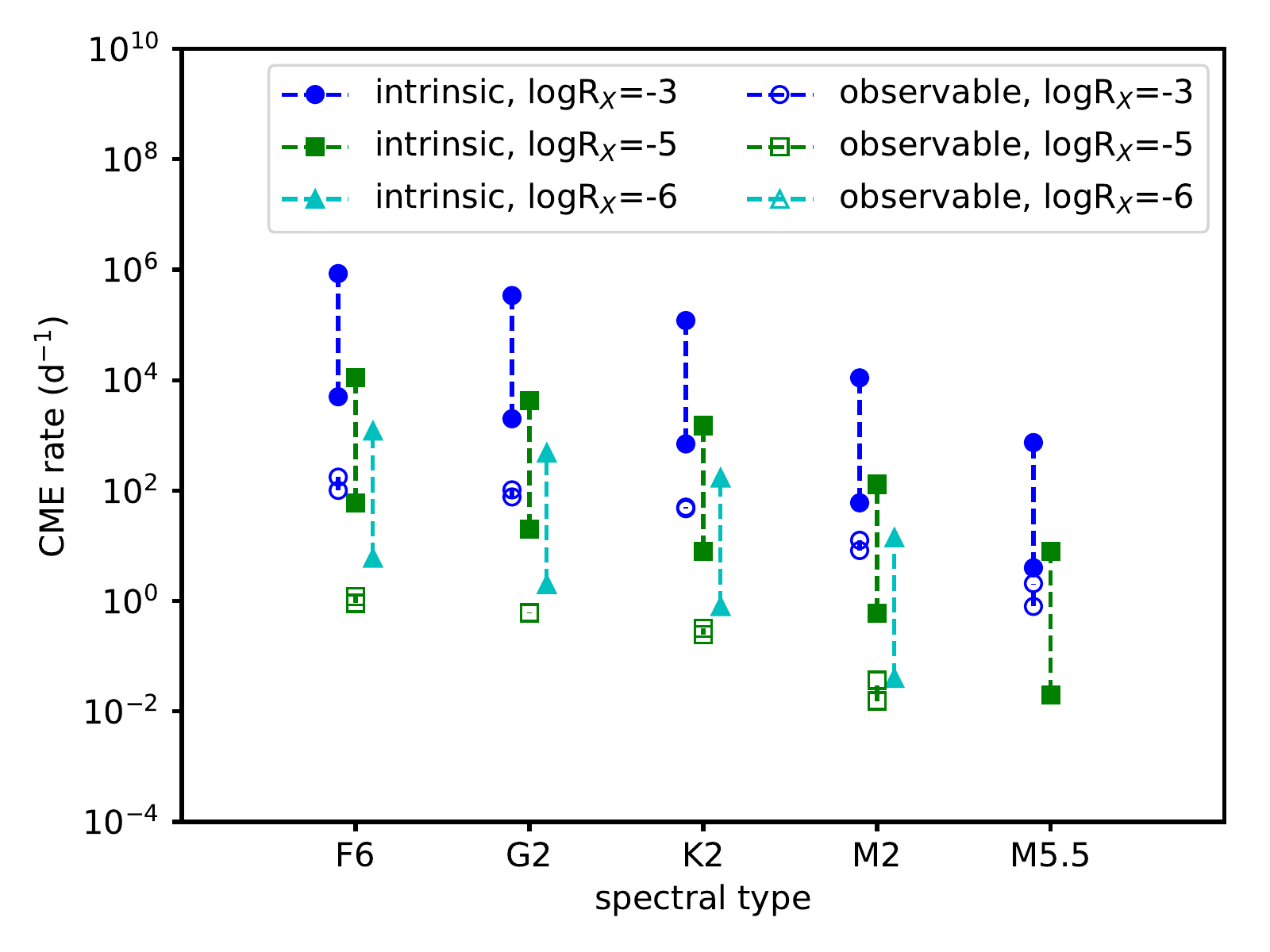}\hfill
	\includegraphics[width=\columnwidth]{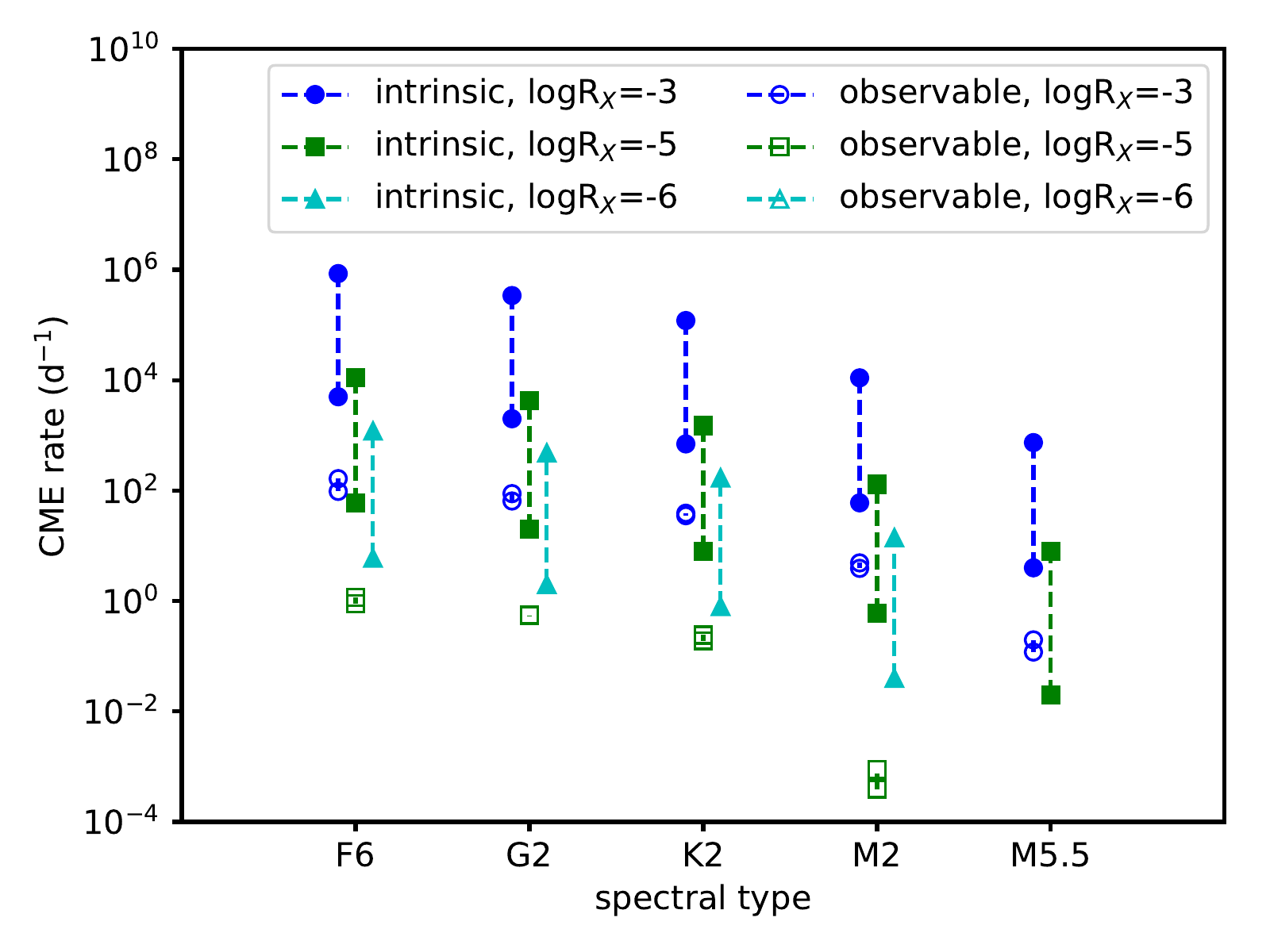}
	\includegraphics[width=\columnwidth]{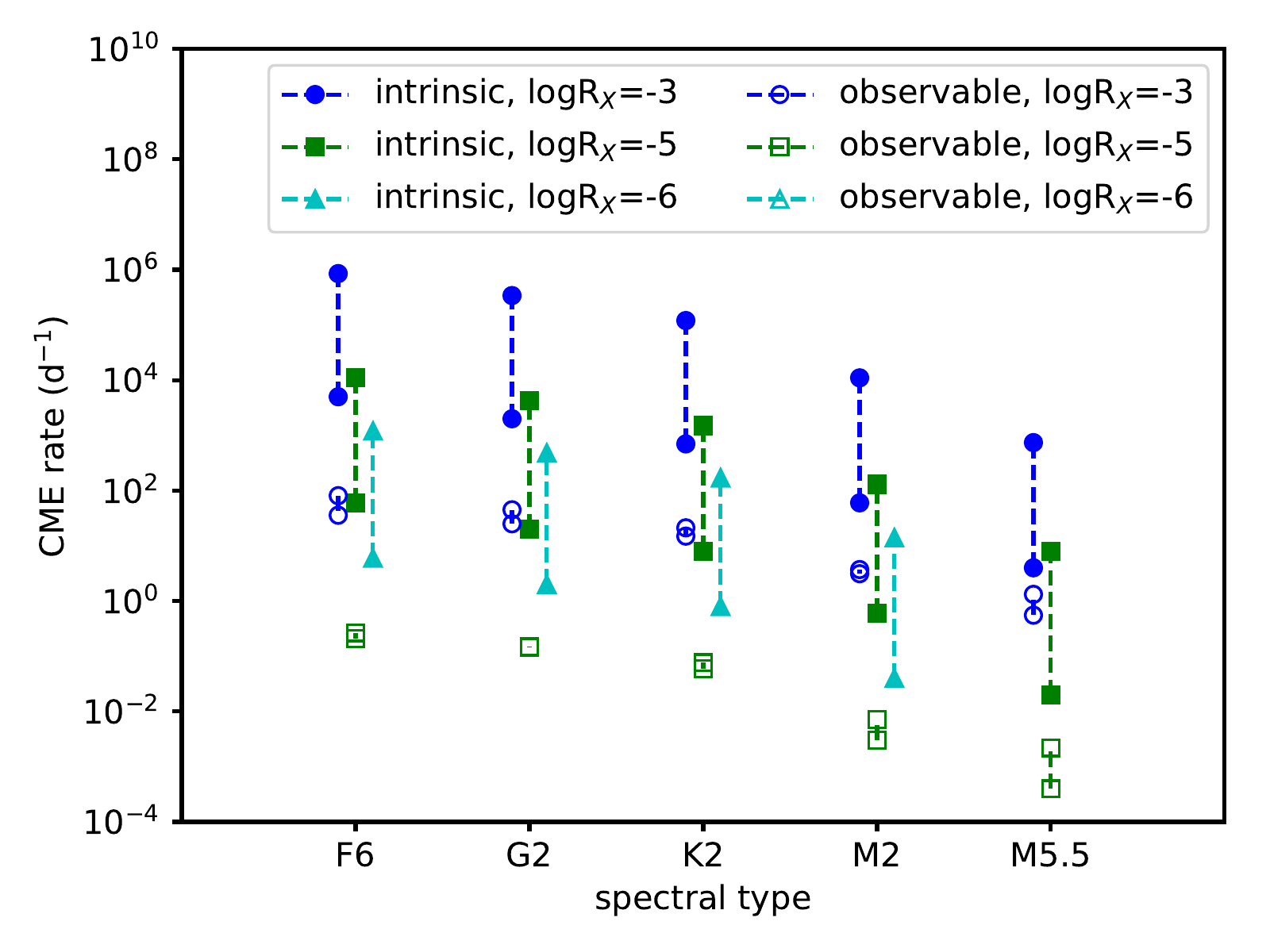}\hfill
	\includegraphics[width=\columnwidth]{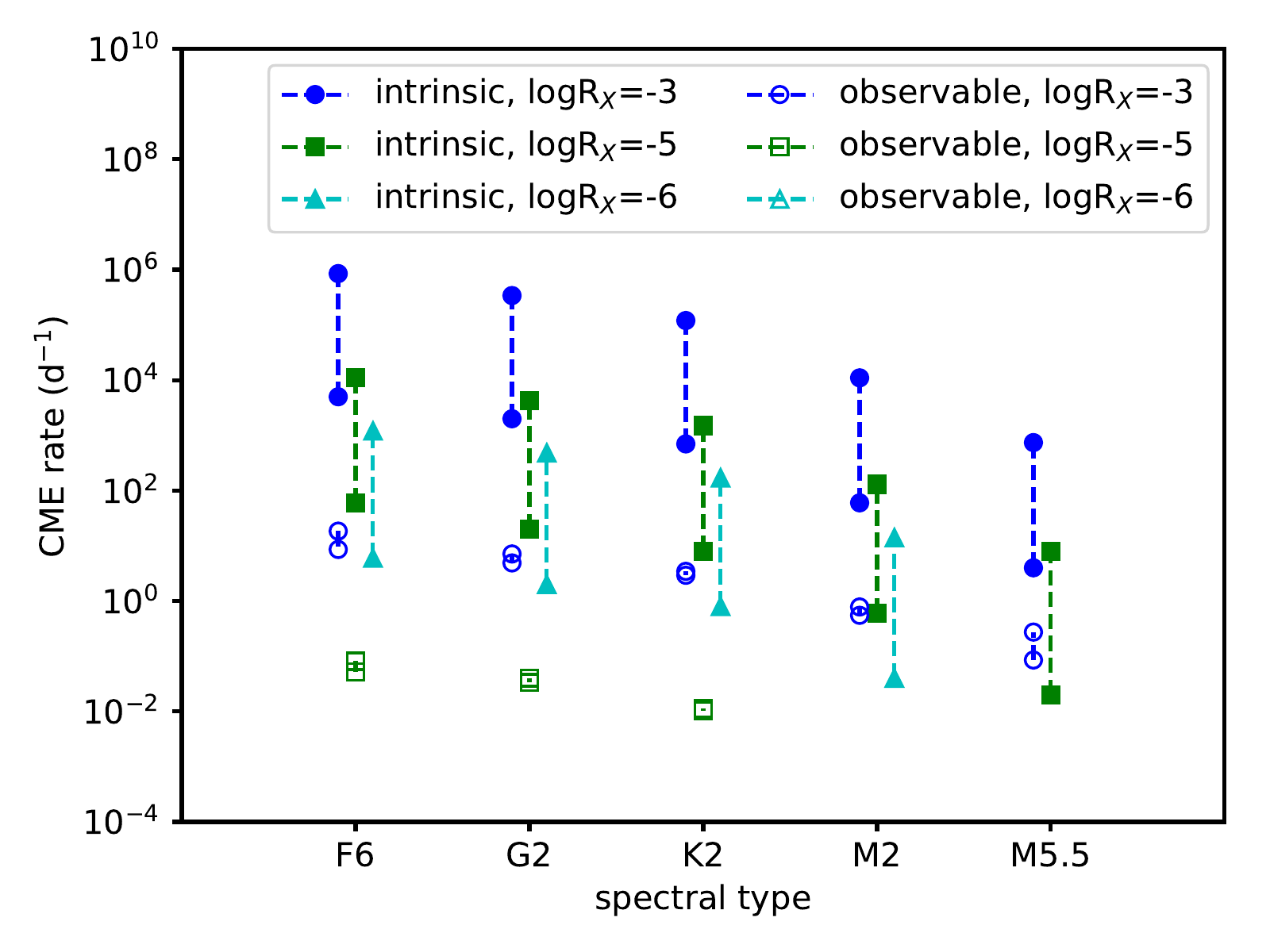}
	\includegraphics[width=\columnwidth]{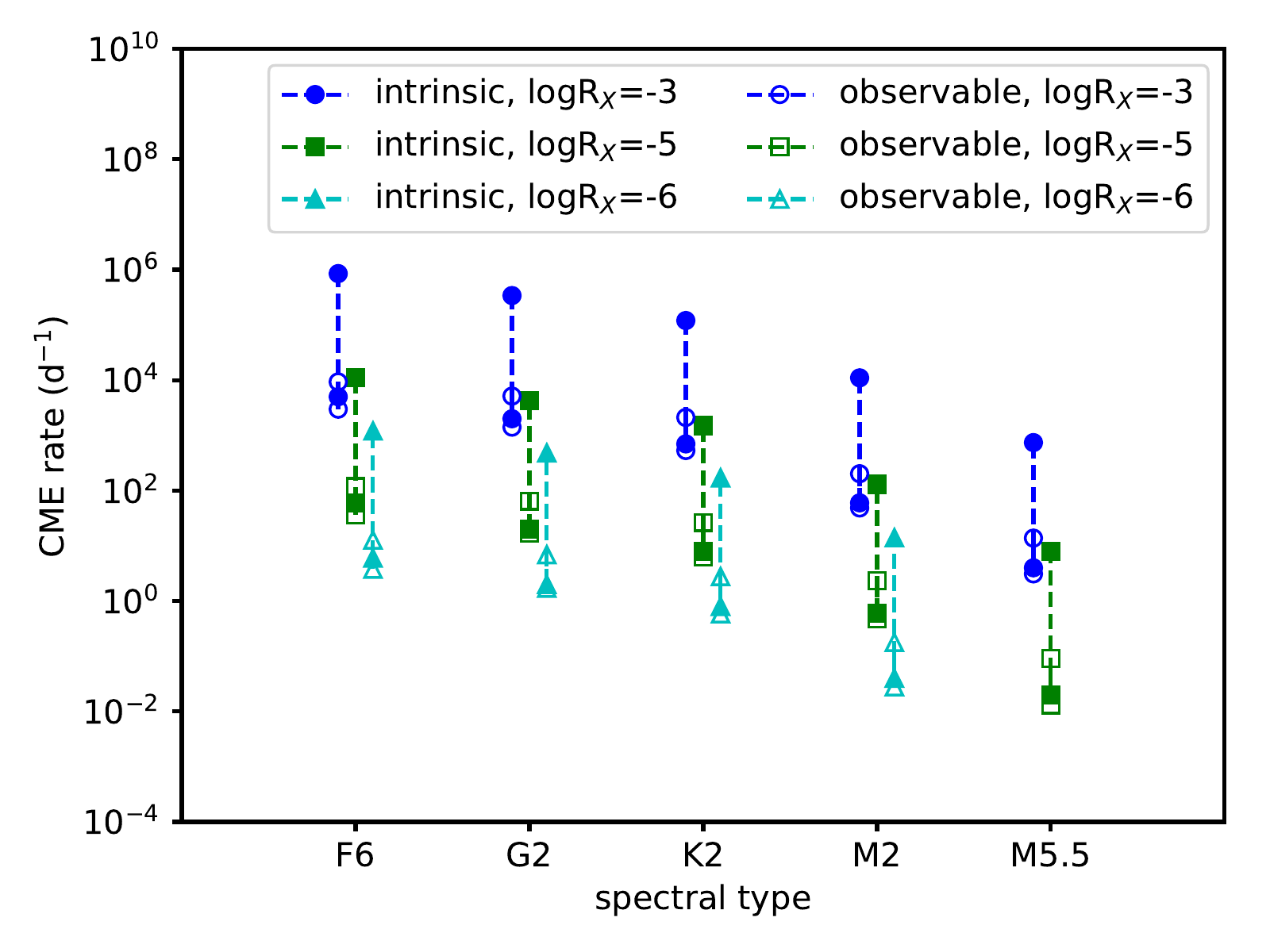}\hfill
	\includegraphics[width=\columnwidth]{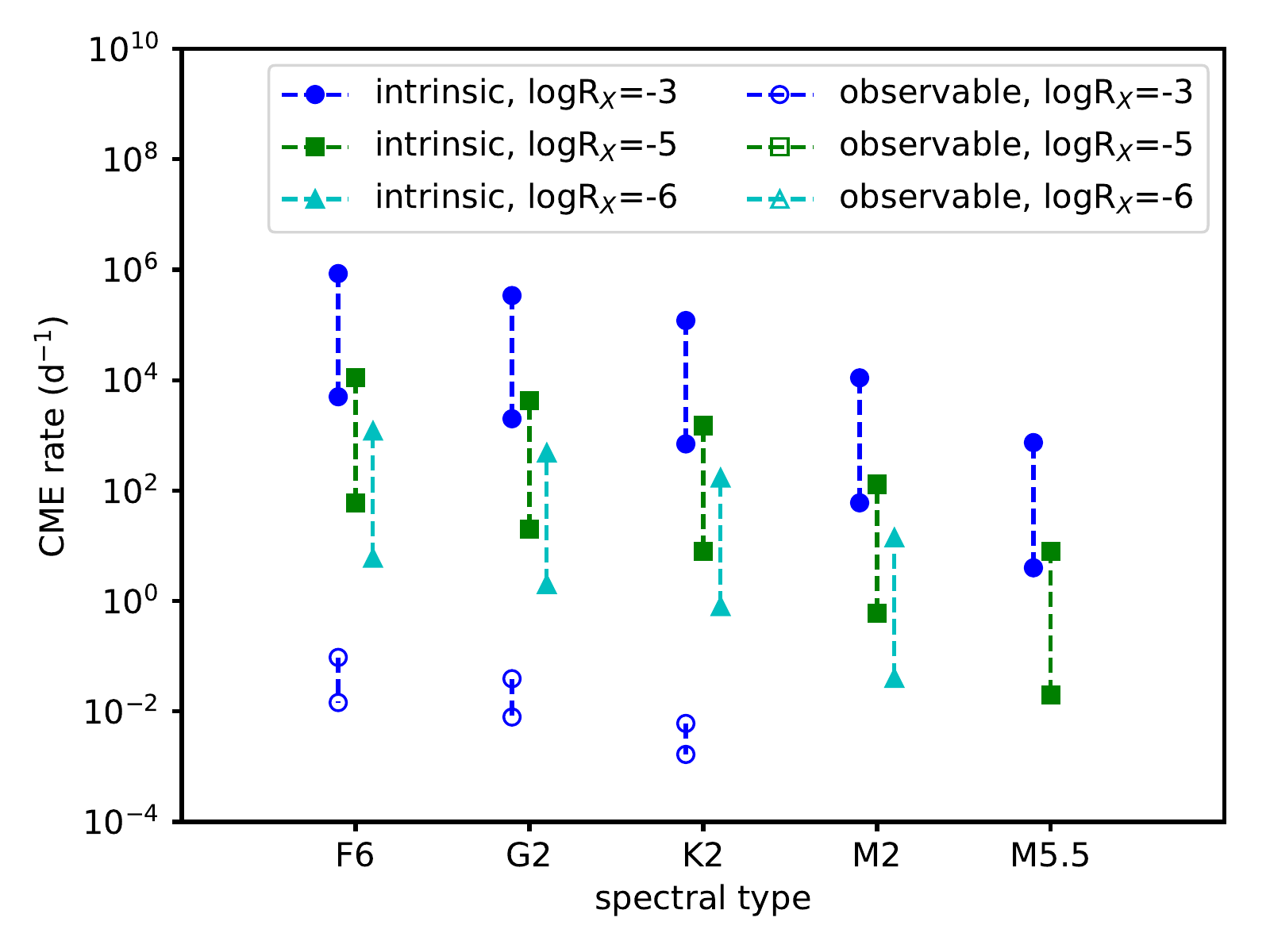}
	\contcaption{Upper row: reduction of CME speeds by factors of 5 (left) and 10 (right); middle row: spectral lines H$\beta$ (left) and H$\gamma$ (right); lower row: column densities of $10^{18}$ (left) and $10^{22}$\,cm$^{-3}$ (right). Missing symbols indicate observable and/or intrinsic rates of zero.}
\end{figure*}

\clearpage

\section{Minimum observing time}\label{app:minobs}
In Fig.~\ref{fig:app_minobs}, we show the minimum observing time to detect CMEs for several stellar spectral types. Relative to Fig.~\ref{fig:minobs}, we vary one of the default parameters.

\begin{figure*}
	\centering
	\includegraphics[width=\columnwidth]{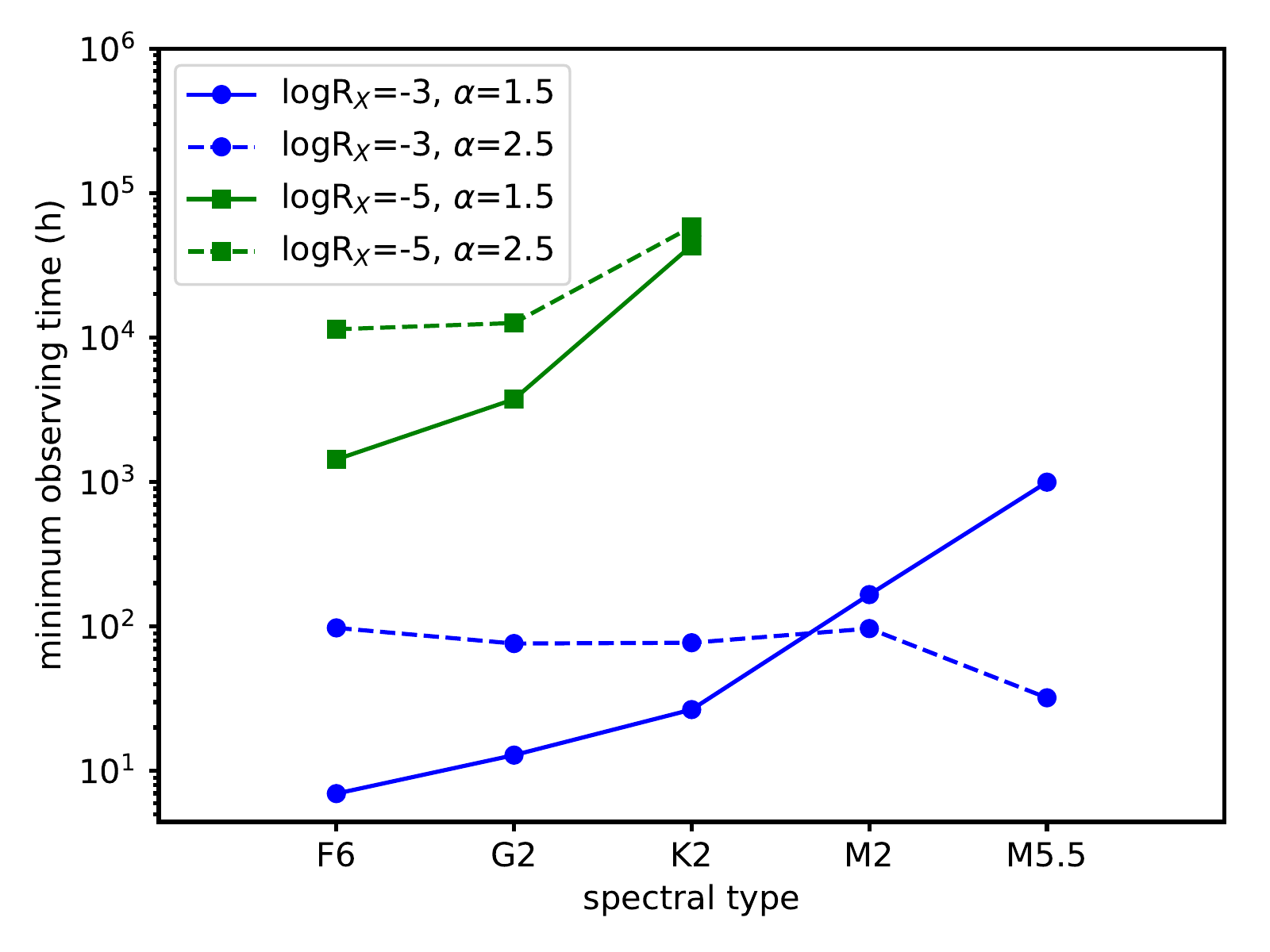}\hfill
	\includegraphics[width=\columnwidth]{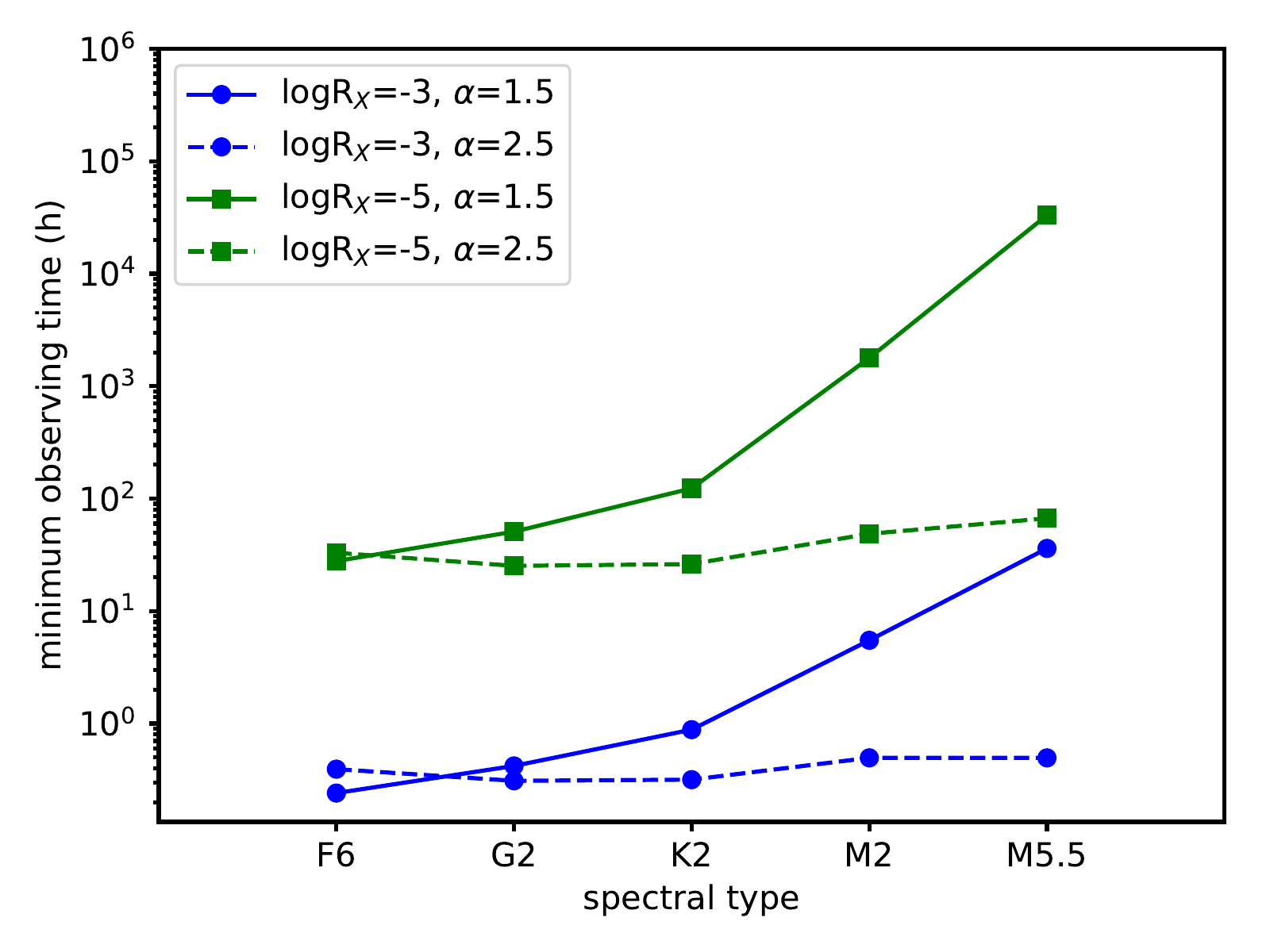}
	\includegraphics[width=\columnwidth]{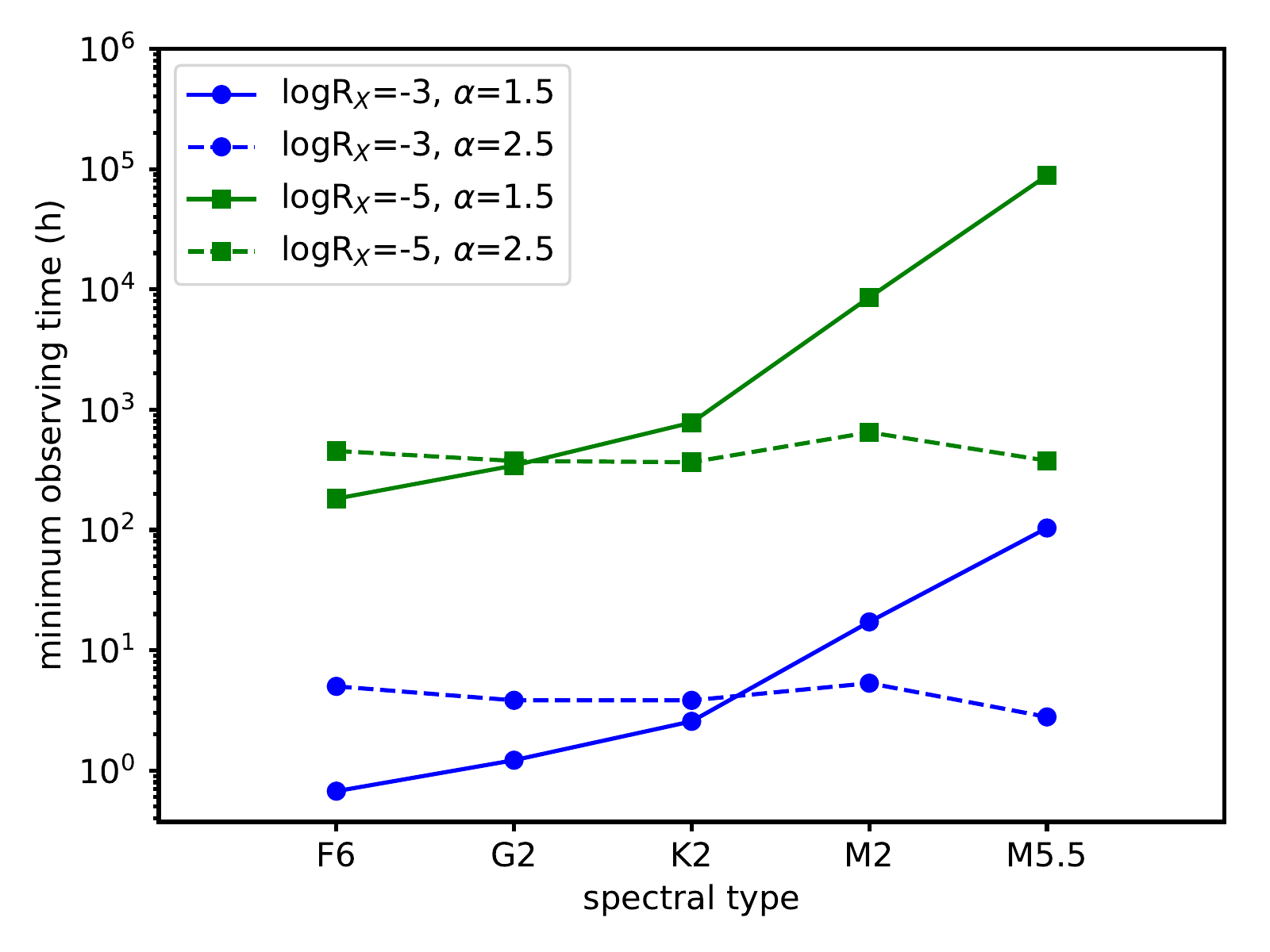}\hfill
	\includegraphics[width=\columnwidth]{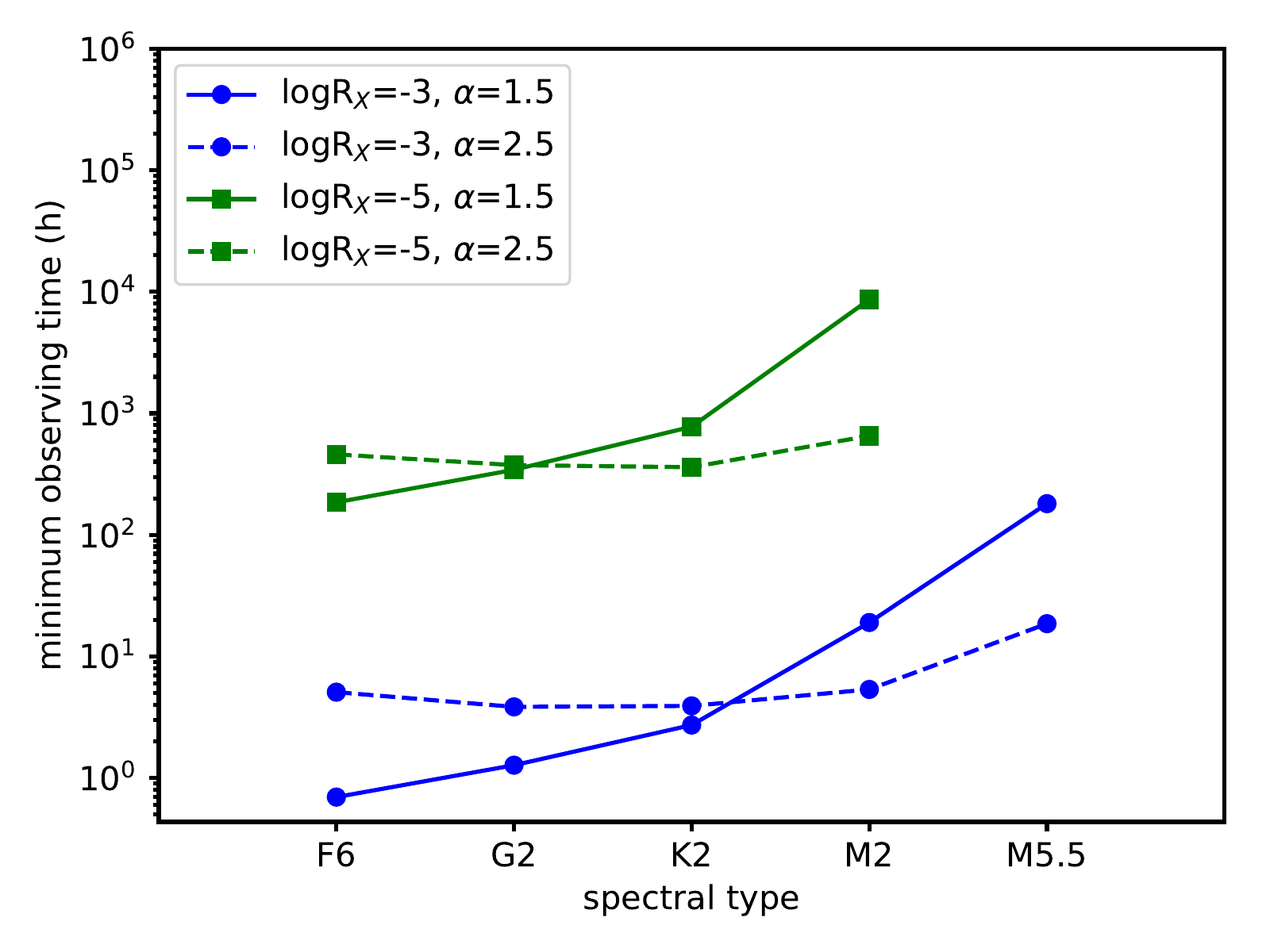}
	\includegraphics[width=\columnwidth]{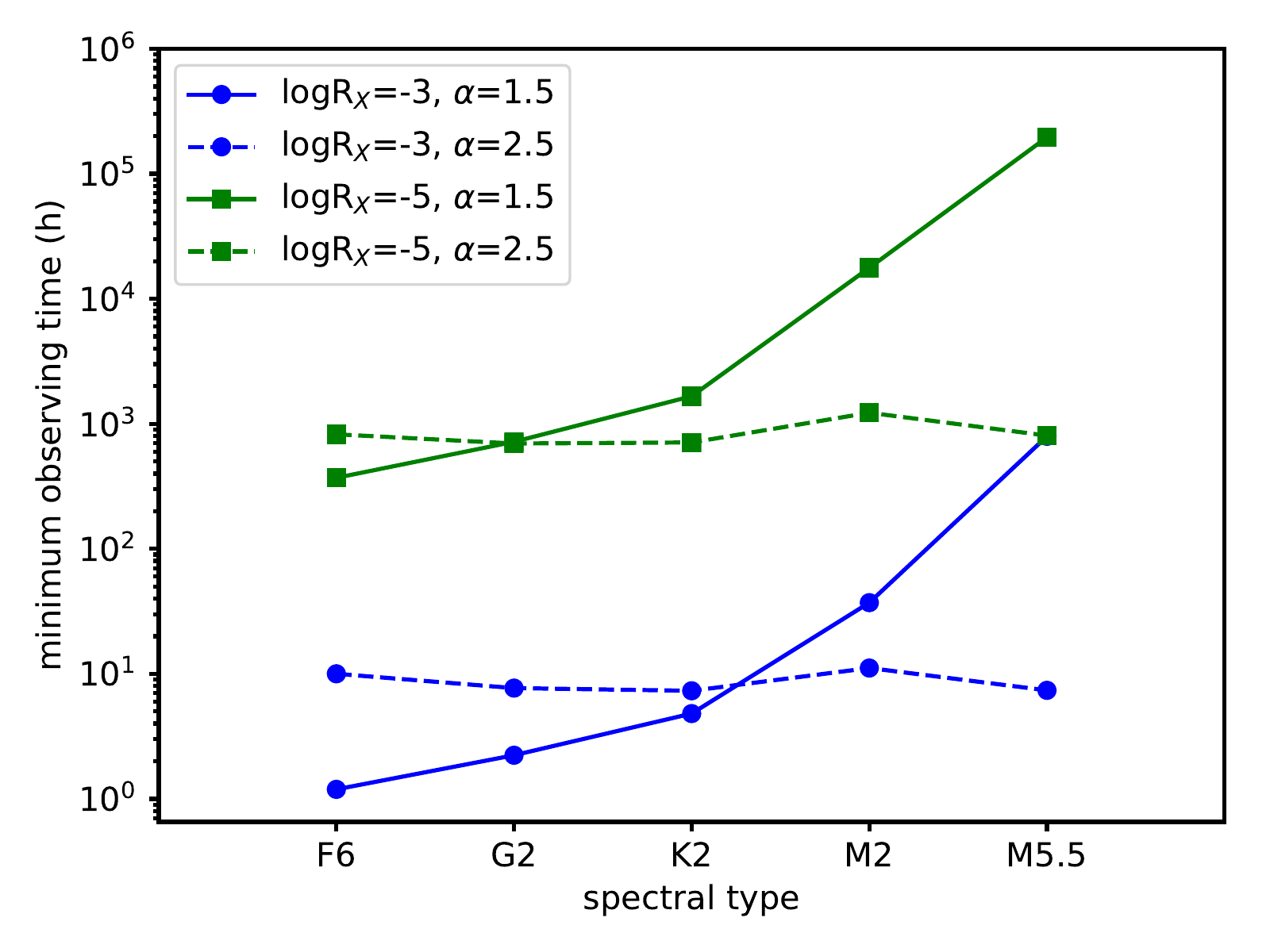}\hfill
	\includegraphics[width=\columnwidth]{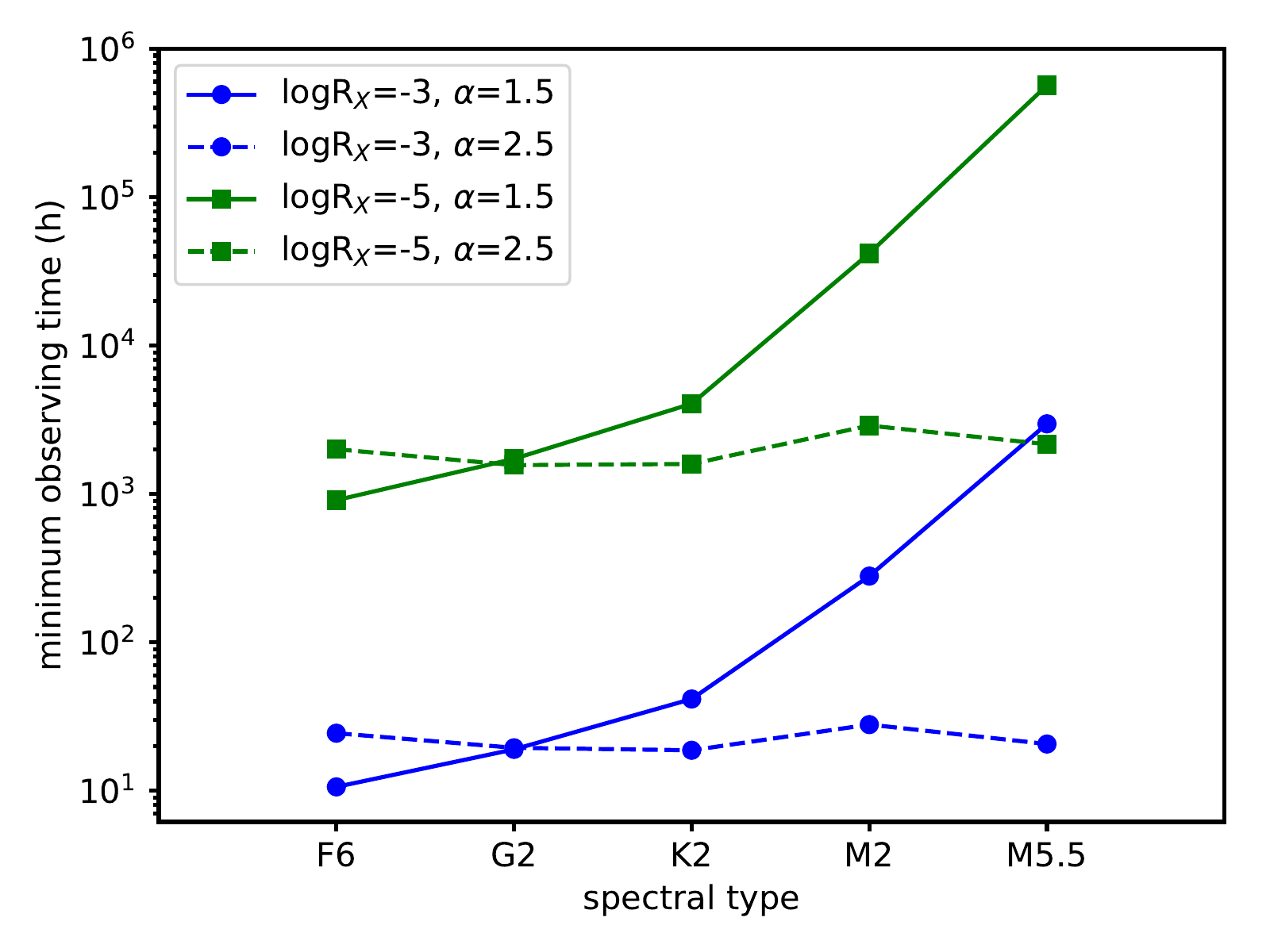}
	\caption{Minimum observing time - dependence on parameters. Upper row: SNR of 30 (left) and 300 (right); middle row: velocity limit of 300\,km\,s$^{-1}$ (left) and 600\,km\,s$^{-1}$ (right); lower row: exposure time of 10~min (left) and 20~min (right). Missing symbols indicate observable rates of zero (i.e. infinite observing time).}
	\label{fig:app_minobs}
\end{figure*}

\begin{figure*}
	\centering
	\includegraphics[width=\columnwidth]{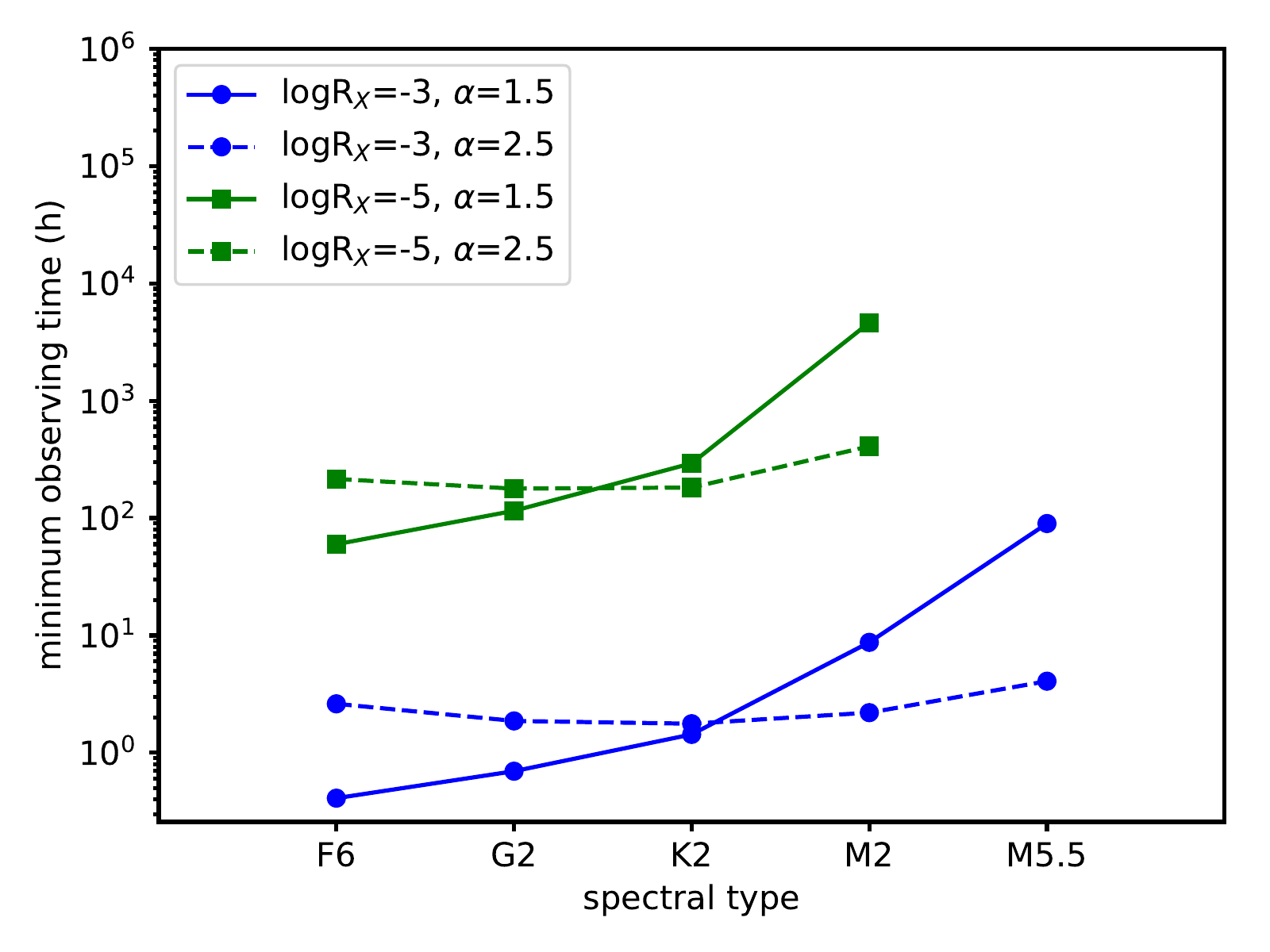}\hfill
	\includegraphics[width=\columnwidth]{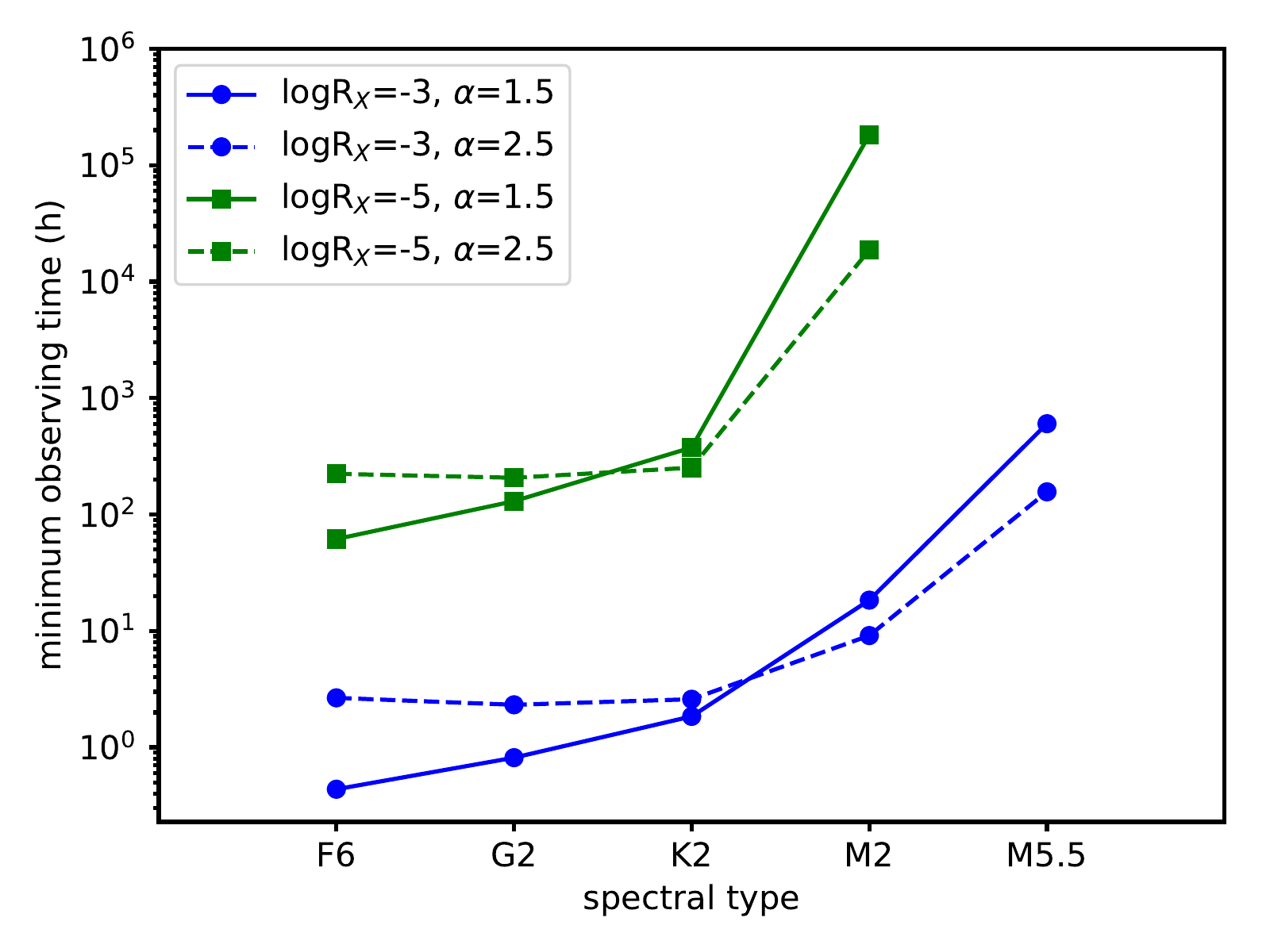}
	\includegraphics[width=\columnwidth]{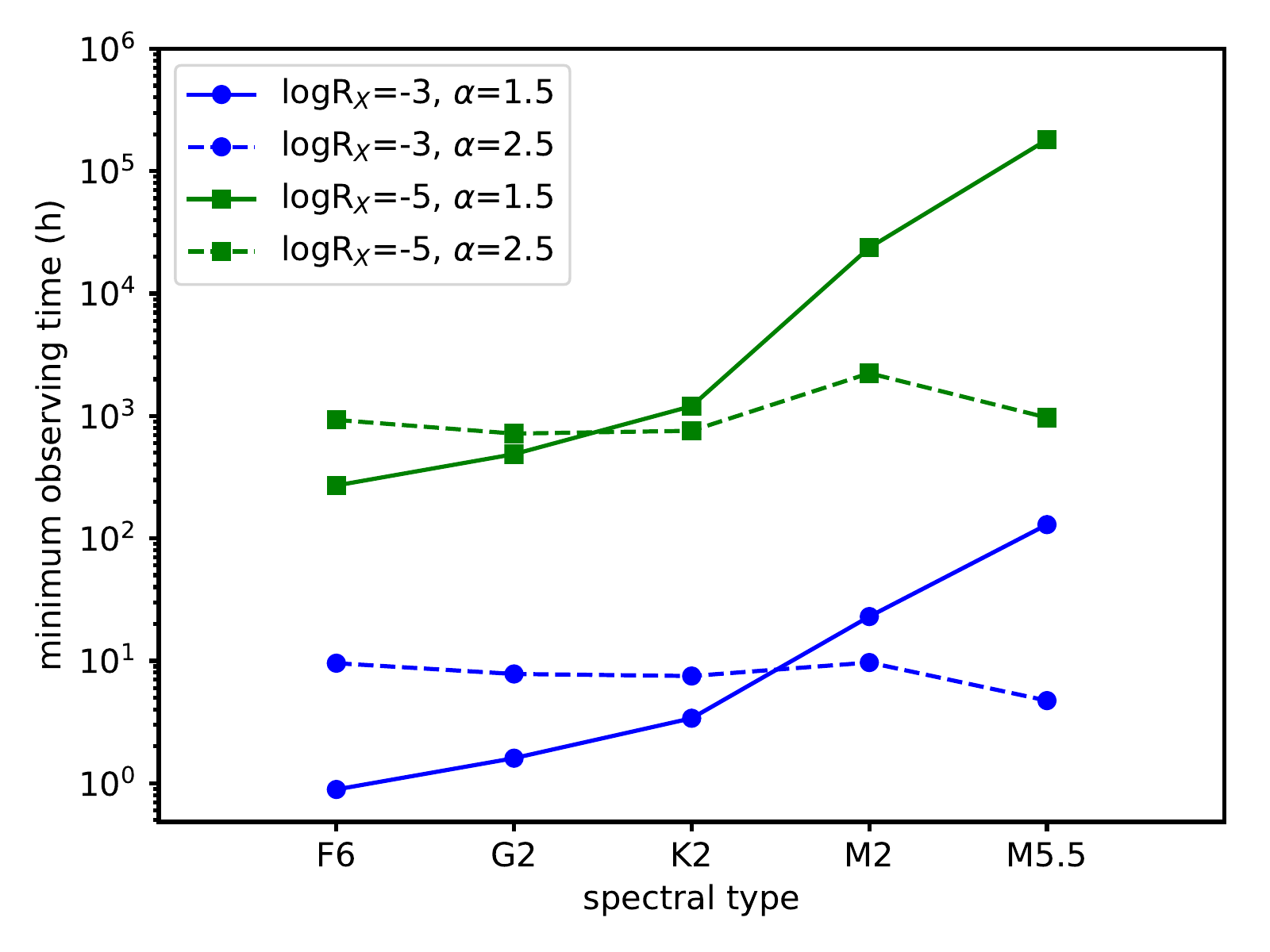}\hfill
	\includegraphics[width=\columnwidth]{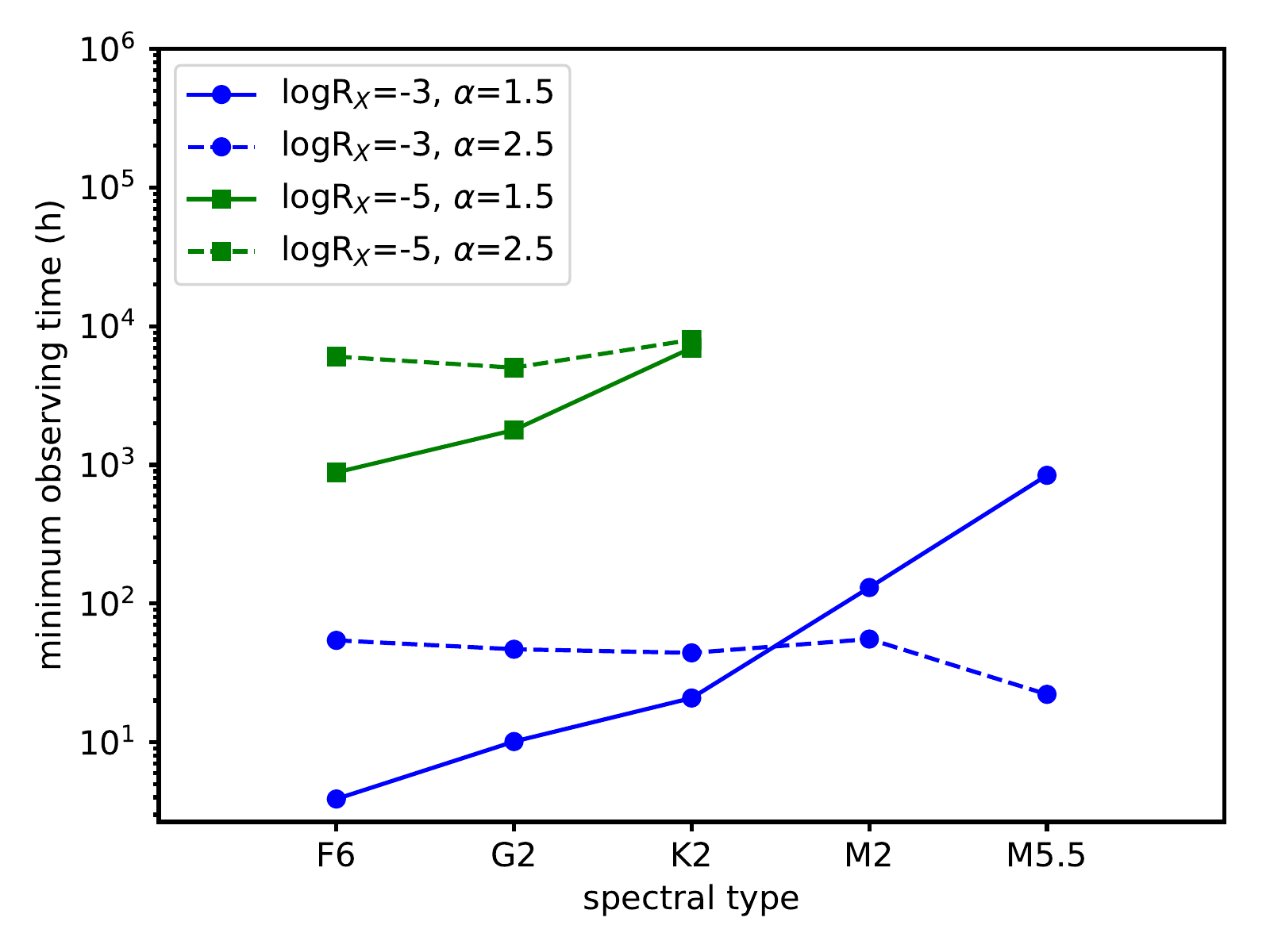}
	\includegraphics[width=\columnwidth]{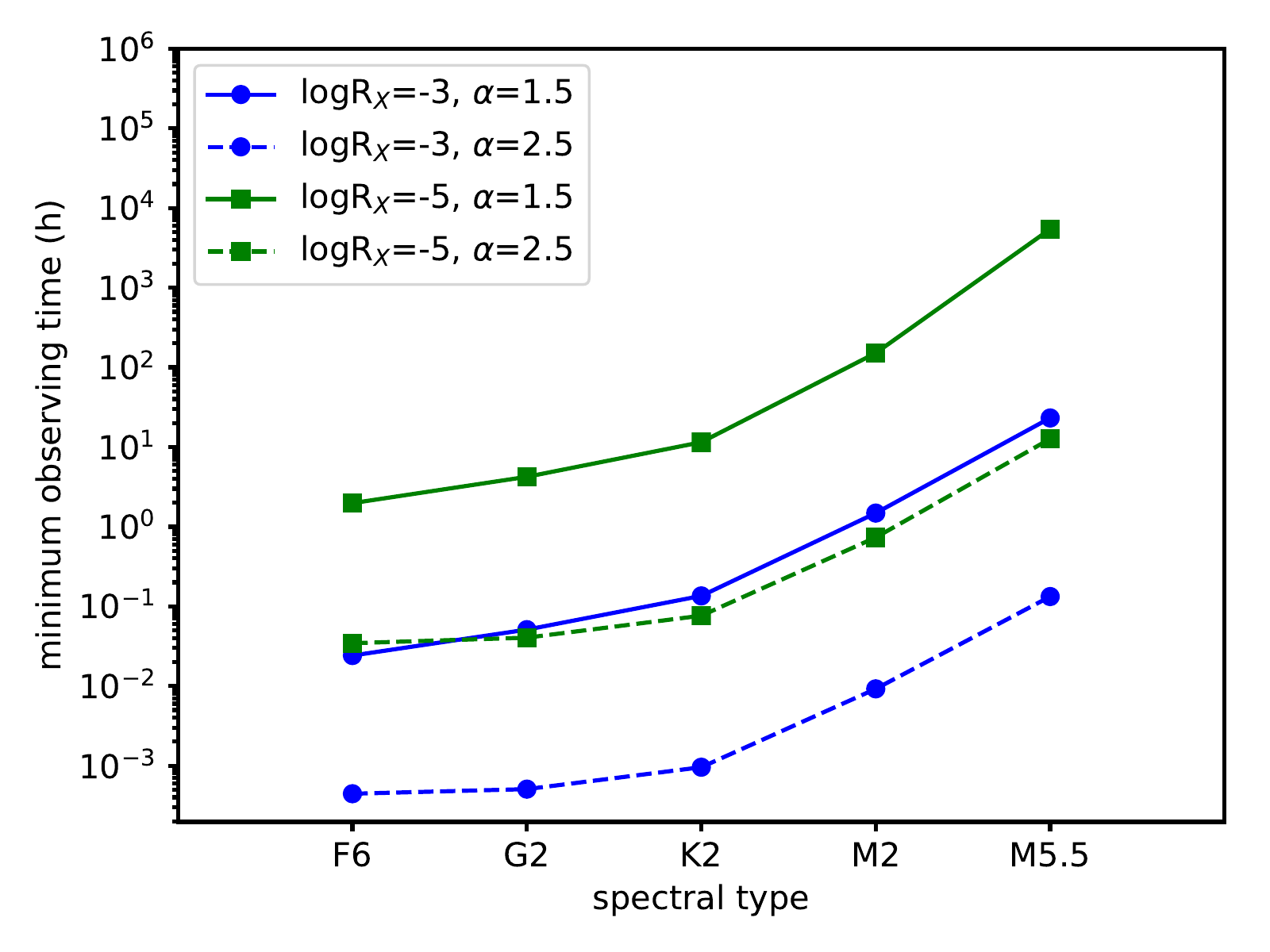}\hfill
	\includegraphics[width=\columnwidth]{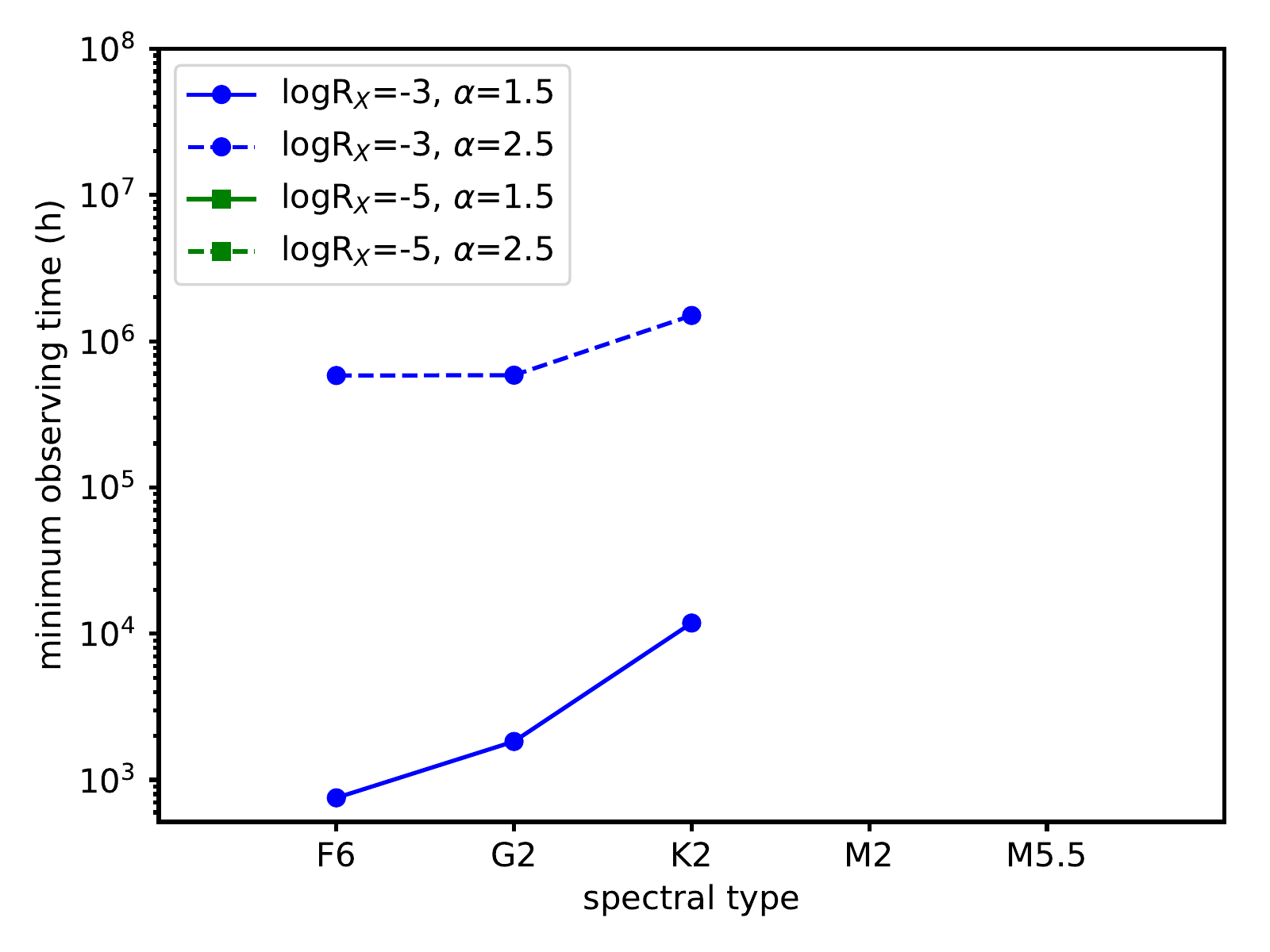}
	\contcaption{Upper row: reduction of CME speeds by factors of 5 (left) and 10 (right); middle row: spectral lines H$\beta$ (left) and H$\gamma$ (right); lower row: column densities of $10^{18}$ (left) and $10^{22}$\,cm$^{-3}$ (right). Missing symbols indicate observable rates of zero (i.e. infinite observing time).}
\end{figure*}

\clearpage

\section{Tables}
Parameters and estimated CME rates of the target stars from \citet{Leitzinger14} (Table~\ref{tab:blanco}) and \citet{Guenther97} (Table~\ref{tab:cmestarsg97}).

\begin{table*}
\centering
\caption{Member stars of the open cluster Blanco-1 observed by \citet{Leitzinger14}. Spectral types and X-ray luminosities are taken from \citet{Leitzinger14} and references therein. Stellar radii are estimated from the given spectral types using Table~12 of \citet{Boyajian12a}. $SNR$ values are determined from the individual spectra. CME rates and probabilities are calculated assuming a flare power law index range of 1.5--2.5, corresponding to the given number ranges. Intrinsic CME rates during $t_\mathrm{obs}=4.95$\,h are calculated using our empirical CME prediction model \citepalias{Odert17}. The maximum number of observable CMEs during $t_\mathrm{obs}$, $N_\mathrm{exp,max}$, is calculated as described in Section~\ref{sec:rates}. The last column gives the probability for a non-detection within the observing time.}
\label{tab:blanco}
\begin{tabular}{llllllll}
	\hline
2MASS ID & spectral & log$L_\mathrm{X}$ & $R_{\star}$  & $SNR$  & CME rate                         & $N_\mathrm{exp,max}$              & $P(0)$ \\
	     & type & (erg\,s$^{-1}$) & ($R_{\sun}$) &        & ($t_\mathrm{obs}^{-1}$) 			& ($t_\mathrm{obs}^{-1}$) 			& (\%) \\
\hline
00032088-3004472  & M4-5 & 28.71 & 0.16 & 18.02    & 6.2 - $1.1{\times}10^3$                & 0.129 - 1.59  & 87.9 - 20.4\\
00030027-3003215  & K4-7 & 29.62 & 0.67 & 190.29   & 46 - $8.3{\times}10^3$   				& 5.51 - 13.4   & 0.41 - $<$0.1\\
00031154-2958103  & M4-5 & 29.16 & 0.16 & 18.57    & 17 - $3{\times}10^3$                	& 0.362 - 4.53  & 69.6 - 1.07\\
00032417-2956229  & K7-M0& 28.77 & 0.59 & 85.68    & 7 - $1.3{\times}10^3$                	& 0.364 - 0.55  & 69.5 - 57.7\\
00032466-2955146  & K5-7 & 29.32 & 0.64 & 115.73   & 24 - $4.3{\times}10^3$ 				& 1.82 - 2.66   & 16.2 - 7.01\\
00032273-2953505  & M3-4 & 28.40 & 0.30 & 46.37    & 3.1 - $5.8{\times}10^2$                & 0.189 - 0.777 & 82.8 - 46.0\\
00021972-2956074  & M0-1 & 29.46 & 0.49 & 41.48    & 32 - $5.9{\times}10^3$ 				& 1.28 - 0.959  & 27.7 - 38.3\\
00022589-2952392  & M3-4 & 29.08 & 0.30 & 64.14    & 14 - $2.6{\times}10^3$                	& 1.61 - 7.15   & 20.0 - $<$0.1\\
00023545-3007019  & K4-5 & 29.46 & 0.71 & 218.93   & 32 - $5.9{\times}10^3$ 				& 3.93 - 9.93   & 1.95 - $<$0.1\\
00023482-3005255  & K5-7 & 30.00 & 0.64 & 134.43   & $1.1{\times}10^2$ - $1.9{\times}10^4$ 	& 10.3 - 16.9   & $<$0.1\\
00022819-3004435  & K4-5 & 29.76 & 0.71 & 188.72   & 63 - $1.1{\times}10^4$   				& 6.85 - 13.6   & 0.11 - $<$0.1\\
00022512-3004235  & M3-5 & 28.63 & 0.19 & 14.09    & 5.1 - $9.5{\times}10^2$                & 0.07 - 0.418  & 93.2 - 65.9\\
00022289-3002532  & M3-4 & 29.11 & 0.30 & 37.28    & 15 - $2.7{\times}10^3$                	& 0.723 - 2.21  & 48.5 - 11.0\\
\hline
\end{tabular}
\end{table*}

\begin{table*}
\caption{Same as Table~\ref{tab:blanco}, but for CTTS and WTTS in Chamaeleon \citep[taken from][]{Guenther97}. Average $SNR$ values are taken from the third night of the observations (Jan 29, 1995).}
\label{tab:cmestarsg97}
\begin{tabular}{llllllll}
	\hline
	type & ID & spectral & log$L_\mathrm{X}$     & SNR & CME rate                         	& $N_\mathrm{exp,max}$             	& $P(0)$ \\
	     &    & type     & (erg s$^{-1}$) &     & ($t_\mathrm{obs}^{-1}$) 			& ($t_\mathrm{obs}^{-1}$) 			& (\%) \\
	            \hline
	CTTS & VZ Cha       	& K6    & 29.28 & 25.81 & 61 - $1.1{\times}10^4$ 					& 0 				& 100 \\
	CTTS & VW Cha       	& K2    & 28.89 & 28.92 & 26 - $4.8{\times}10^3$                	& 0 				& 100 \\
	CTTS & VV Cha       	& M1.5  & 28.94 & 8.56  & 29 - $5.3{\times}10^3$                	& 0 				& 100 \\
	CTTS & HM 32        	& M0.5  & 29.67 & 13.29 & $1.5{\times}10^2$ - $2.6{\times}10^4$ 	& 0 				& 100 \\
	CTTS & WX Cha       	& K7-M0 & 28.64 & 14.42 & 15 - $2.8{\times}10^3$                	& 0 				& 100 \\
	CTTS & WY Cha       	& K7-M0 & 29.40 & 15.35 & 80 - $1.5{\times}10^4$ 					& 0 				& 100 \\
	CTTS & WW Cha       	& K5    & 29.02 & 11.75 & 35 - $6.3{\times}10^3$                	& 0 				& 100 \\
	CTTS & CS Cha       	& K5    & 29.75 & 39.48 & $1.7{\times}10^2$ - $3.1{\times}10^4$ 	& 0.34 - 0.015 		& 71.1 - 98.5 \\
	CTTS & TW Cha       	& M0    & 29.28 & 18.91 & 61 - $1.1{\times}10^4$ 					& 0 				& 100 \\
	CTTS & SY Cha       	& M0    & 29.09 & 16.79 & 40 - $7.4{\times}10^3$ 					& 0 				& 100 \\
	WTTS & J1150.9-7411 	& M4    & 30.19 & 9.76  & $4.5{\times}10^2$ - $8.2{\times}10^4$ 	& 0.129 - 0.0011 	& 87.9 - 99.9 \\
	WTTS & J1014.2-7636 	& M2    & 29.94 & 7.92  & $2.6{\times}10^2$ - $4.7{\times}10^4$ 	& 0 				& 100 \\
	WTTS & J1111.7-7620 	& K3    & 30.07 & 34.84 & $3.5{\times}10^2$ - $6.3{\times}10^4$  	& 0.715 - 0.024 	& 48.9 - 97.6 \\
	WTTS & J1202.8-7718 	& M0    & 30.09 & 7.10  & $3.6{\times}10^2$ - $6.6{\times}10^4$ 	& 0.004 - $3\times10^{-5}$ & 99.6 - 100 \\
	WTTS & J1204.6-7731 	& M3    & 30.14 & 11.71 & $4.1{\times}10^2$ - $7.3{\times}10^4$ 	& 0.158 - 0.002 	& 85.4 - 99.8 \\
	WTTS & J1149.8-7850$^*$ & M0    & 30.25 & 13.53 & $5.2{\times}10^2$ - $9.3{\times}10^4$ 	& 0.318 - 0.004 	& 72.8 - 99.6 \\
	WTTS & J1158.5-7754 	& M2    & 30.80 & 10.25 & $1.7{\times}10^3$ - $3.1{\times}10^5$  	& 1.01 - 0.007 		& 36.3 - 99.3 \\
	WTTS & J1108.8-7519 	& M2    & 29.93 & 20.04 & $2.6{\times}10^2$ - $4.6{\times}10^4$ 	& 0.198 - 0.004 	& 82.0 - 99.6 \\
	WTTS & J1044.6-7849 	& M2    & 29.91 & 12.96 & $2.5{\times}10^2$ - $4.4{\times}10^4$ 	& 0.061 - 0.001 	& 94.0 - 99.9 \\
	WTTS & SZ Cha       	& K0    & 29.22 & 67.44 & 54 - $9.8{\times}10^3$                	& 0.136 - 0.014 	& 87.2 - 98.6 \\
	WTTS & J1048.9-7655 	& K6    & 29.88 & 17.34 & $2.3{\times}10^2$ - $4.2{\times}10^4$ 	& 0.12 - 0.002 		& 88.7 - 99.8 \\
	WTTS & J1150.4-7704 	& K2    & 30.18 & 45.22 & $4.4{\times}10^2$ - $8{\times}10^4$ 		& 1.85 - 0.061 		& 15.7 - 94.1 \\
	WTTS & J1120.3-7828     &       & 30.68 & 36.77 & $1.3{\times}10^3$ - $2.4{\times}10^5$ 	& 6.53 - 0.115 		& 0.15 - 89.1\\
	WTTS & J1108.2-7728     & K6    & 29.81 & 19.21 & $2{\times}10^2$ - $3.6{\times}10^4$ 		& 0.114 - 0.002 	& 89.2 - 99.8 \\
	WTTS & J1159.7-7601 	& K2    & 30.25 & 18.25 & $5.2{\times}10^2$ - $9.3{\times}10^4$ 	& 0.5 - 0.008 		& 60.6 - 99.2 \\
	WTTS & J1106.3-7721     &       & 30.14 & 56.47 & $4.1{\times}10^2$ -  $7.3{\times}10^4$ 	& 2.43 - 0.091 		& 8.8 - 91.3 \\
	WTTS & J1039.5-7538     &       & 30.64 & 71.93 & $1.2{\times}10^3$ -  $2.2{\times}10^5$ 	& 13.6 - 0.503 		& $<$0.1 - 60.4 \\
	\hline
\end{tabular}\\
$^*$DZ Cha, one CME detected.
\end{table*}



\bsp	
\label{lastpage}
\end{document}